\documentclass[a4paper,11pt]{article}
\pdfoutput=1 

\usepackage{jcappub} 

\usepackage[T1]{fontenc} 
\usepackage{graphicx}
\usepackage{bm}
\usepackage{amssymb}
\usepackage{amsmath}
\usepackage{epsfig}
\usepackage{hyperref}
\usepackage{hhline}
\usepackage{color}
\usepackage{multirow}
\usepackage[normalem]{ulem}
\usepackage{adjustbox}

\definecolor{jd}{rgb}{0.858, 0.188, 0.478}

\def\lapp{\mathrel{\rlap{\raise.5ex\hbox{$<$}}
                    {\lower.5ex\hbox{$\sim$}}}}
\def\gapp{\mathrel{\rlap{\raise.5ex\hbox{$>$}}
                    {\lower.5ex\hbox{$\sim$}}}}

{
{

\def\gt{\zeta}

\newcommand{\bmt}{\begin{pmatrix}}
\newcommand{\emt}{\end{pmatrix}}
\newcommand{\ba}{\begin{array}{c}}
\newcommand{\ea}{\end{array}}
\newcommand{\be}{\begin{equation}}
\newcommand{\ee}{\end{equation}}
\newcommand{\bea}{\begin{eqnarray}}
\newcommand{\eea}{\end{eqnarray}}

\newcommand{\bi}{\begin{itemize}}
\newcommand{\ei}{\end{itemize}}

\newcommand{\baz}{\begin{array}{cc}}

\newcommand{\mathsym}[1]{{}}

\newcommand{\bt}{\begin{tabular}}
\newcommand{\et}{\end{tabular}}

\newcommand{\benu}{\begin{enumerate}}
\newcommand{\eenu}{\end{enumerate}}

\newcommand{\bav}{\begin{array}{cccc}}

\affiliation[a]{Department of Physics, Indian Institute of Technology Guwahati, North Guwahati, Assam- 781039, India}
\affiliation[b]{Department of Physics, Indian Institute of Technology Kanpur, Kanpur, Uttar Pradesh- 208016, India}

\author[a]{Basabendu Barman,}
\emailAdd{bb1988@iitg.ernet.in}
\author[a]{Subhaditya Bhattacharya,}
\emailAdd{subhab@iitg.ernet.in}
\author[a]{Sunando Kumar Patra,}
\emailAdd{sunando.patra@gmail.com}
\author[b]{and Joydeep Chakrabortty}
\emailAdd{joydeep@iitk.ac.in}

\abstract{
Vector boson dark matter (DM) appears in $SU(2)_N$ extension ($N$ stands for neutral) of Standard Model (SM) where an additional global $U(1)_P$ symmetry is assumed and results in a generalized lepton number defined as: $L=P+T_{3N}$. Breaking of $U(1)_P$ leads to the breaking of $L$ to $(-1)^L$, thus stabilizing DM through modified $R=(-1)^{3B+L+2J}$. This model, already discussed in literature, offers several novel features to elaborate upon. For example, $t$-channel annihilation and dominant $s$-channel direct search,  along with co-annihilation, helps the DM to evade stringent direct search bounds from LUX and XENON1T after satisfying relic density constraints. On the other hand, the exotic particles of the model can be produced at the Large Hadron Collider (LHC) yielding multilepton final states. Hadronically quiet four lepton signal with large missing energy, in specific, is shown to provide a smoking gun signature of such a framework. We study the details of $E(6) \to SM \otimes SU(2)_N$ breaking patterns (through D-parity odd/even cases) which yield important phenomenological consequences.}

\begin{document}

\renewcommand*{\thefootnote}{\fnsymbol{footnote}}
\title{\bf Non-Abelian Vector Boson Dark Matter, its Unified Route and signatures at the LHC\\}


\maketitle
\flushbottom


\setcounter{footnote}{0}
\renewcommand*{\thefootnote}{\arabic{footnote}}

\section{Introduction}

Astrophysical observations, such as rotation curve of spiral galaxies around the cluster \cite{zwicky,rubin}, inhomogeneity in cosmic microwave background radiation (CMBR) \cite{cmbr}, or more recent observations in Bullet cluster \cite{bullet} indicate towards the existence of dark matter (DM) in the universe. A particle description of DM is much sought after and lot of efforts are made (for a brief review, see for example, \cite{Bertone:2004pz, jungman}) to accommodate DM in extensions of standard model (SM). Thermal freeze-out of weakly interacting massive particles (WIMP) (see for example, \cite{Kolb:1990vq}) provides the most popular framework of DM phenomenology. While an experimental verification of DM is still awaited, a plethora of DM models are constructed assuming the DM to be a scalar, fermion or a vector boson.  

Scalar (see for example, ~\cite{Guo:2010hq, Feng:2014vea, Steele:2013fka, Cline:2013gha, subha1}) and fermionic (see for example,~\cite{Kim:2008pp,Okada:2013rha,Bai:2013iqa, subha2,subha3,subha4,subha5,subha6}) DMs can be realized with minimal extensions to SM and are discussed widely. However, vector boson DMs are rather difficult to come across as it requires an extension of the SM gauge group or even more exotic frameworks. Abelian vector boson DM appears in universal extra dimension models with conserved $KK$ parity \cite{ued1,ued2,ued3}, in little Higgs framework with $T$ parity \cite{lht1,lht2} or in $U(1)$ extensions of SM (for example, see~\cite{Baek:2013nr, Duch:2015jta}). In this paper we have analyzed the phenomenology of a non-abelian vector boson DM, which arises in an extension of SM by $SU(2)_N$ gauge group, that has been proposed in ~\cite{Farzan:2012kk, DiazCruz:2010dc, Bhattacharya:2011tr, Fraser:2014yga}. $SU(2)_N$ is broken subsequently and yields massive gauge bosons. This gauge group is {\em dark} in the sense that $SU(2)_N$ charges do not contribute to the hypercharge. Thus all three $SU(2)_N$ gauge bosons are electromagnetic charge neutral and the lightest one can in principle be a DM. The stability of the DM is ensured by an extra global symmetry $U(1)_P$, breaking of which explicitly breaks the generalized lepton number $L=P+T_{3N}$ to $(-1)^L$ leading to a discrete $R-$parity as in supersymmetric theory. One of the main motivations of such a framework is to envisage a unified theory, $E(6)$ \cite{DiazCruz:2010dc}, which subsequently breaks to $SM\otimes SU(2)_N$. In the process of $E(6) \to SM\otimes SU(2)_N$, D-parity is broken in presence of an $SU(2)_N$ scalar triplet which allows $g_N$ to be different from $SU(2)_L$ gauge coupling and serves as a parameter of the theory. VEV of this triplet scalar breaks the degeneracy of the dark gauge bosons and ensures a single component DM. Absence of the $SU(2)_N$ triplet (D-parity even case) leads to a constrained degenerate DM scenario, which is not viable from direct search constraints. Intermediate symmetry breaking takes care of the fact that hypercharge is determined without $SU(2)_N$ contribution, ensuring charge neutrality of the $SU(2)_N$ gauge bosons. 

Non-Abelian vector boson DM has been addressed in some other contexts as well (see for example, ~\cite{Hambye:2008bq,Hambye:2009fg,Arina:2009uq,Arcadi:2017kky,Kopp:2016yji}). DMs in context of $SO(10)$ Grand Unified Theory (GUT) has also been discussed (see for example~\cite{so10-dm1,so10-dm2}), but mostly for scalars and fermions.

 Phenomenology of the non-abelian vector boson DM is a major motivation of this paper and this is discussed more elaborately compared to the earlier analyses of this model~\cite{DiazCruz:2010dc, Bhattacharya:2011tr}. In particular, non observation of DM particle in direct search experiments is putting a huge constraint on the single component DM parameter space, eventually ruling out a large class of WIMP models from spin-independent direct search cross-sections. However, the model discussed here survives in a large region of relic density allowed parameter space from WMAP \cite{WMAP}/PLANCK~\cite{Ade:2015xua} data with current direct search bounds from  LUX~\cite{Akerib:2016vxi}. This is mainly due to its t-channel annihilation and dominant s-channel direct search interaction along with co-annihilation features with heavier gauge boson. This has been shown through elaborate parameter space scan, assuming $SU(2)_N$ coupling as an independent parameter (and obeying unification prescription), unlike the previous analyses.


Extended gauge group demands the presence of exotic fermions which can be produced at the Large Hadron Collider (LHC). They subsequently decay to DM providing a variety of signatures of multilepton final states that can testify such a framework in future runs of LHC. Two most promising signals of this model are discussed in this work as `opposite sign dilepton' and `hadronically quiet four lepton' with large missing energy. It has been explicitly shown, by analyzing three different benchmark points, that  `hadronically quiet four lepton' signal is a smoking-gun signature for this model, typically because of less SM background. This potentially interesting feature was also overlooked in the earlier analyses.

The paper is organized as follows: the model and the basic formalism is discussed in Sec. \ref{sec:model}; DM phenomenology is discussed elaborately in Sec. \ref{sec:VBDM} including relic density and direct search constraints. In Sec. \ref{sec:unified}, we study unified framework under $E_6$ and subsequent constraints on the model parameters. Some benchmark points, identified thereafter, are studied for collider signatures in Sec. \ref{sec:collider}. We finally conclude in Sec. \ref{sec:summary}.

\section{Dark Vector multiplet and exotic particles}
\label{sec:model}
The theory under consideration is $SU(2)_N \otimes SU(2)_L \otimes U(1)_Y \otimes SU(3)_C$ gauge symmetric. An additional global abelian symmetry $U(1)_P$ has also been introduced. This redefines the lepton number of the particles as $L=P+T_{3N}$. Explicit breaking of $U(1)_P$ breaks $L$ to $(-1)^L$ and results in conservation of a discrete symmetry $R=(-1)^{3B+L+2J}$ (very similar to $R$-parity in supersymmetry). This stabilizes the lightest particle odd under $R$-parity and serves as DM. $SU(2)_N$ charges do not contribute to the hypercharge ($Y$) and thus to electromagnetic charge, defined as $Q=T_{3L}+Y$. Hence, $SU(2)_N$ gauge bosons $X_{1,2,3}$ are electromagnetic charge neutral and the lightest of them, $X_1$ (odd under $R$), aptly fits into the criterion of a DM. Once $SU(2)_N$ symmetry is broken completely, the gauge bosons acquire masses of the scale of TeV. We adopt the same particle configuration as in \cite{DiazCruz:2010dc,Bhattacharya:2011tr} to realise its high-scale origin, DM constraints, and collider signatures through elaborate analysis. 
Following is the particle content of the model, where the quantum numbers are mentioned under $SU(2)_L \otimes SU(2)_N \otimes U(1)_Y \otimes SU(3)_C\otimes U(1)_P$:
\vskip 0.3cm
\noindent
{\underline{{\it Fermions:}}}\;\;
\begin{center}
$\begin{pmatrix}
   u \\
   d
  \end{pmatrix}
= [2,1,1/6,3;0],~u^c= [1,1,-2/3,\bar{3};0],~(h_{q}^c\;~d^c)=[1,2,1/3,\bar{3};-1/2], h_q=[1,1,-1/3,3;1],$\\ \vskip 0.2cm
$\quad 
\begin{pmatrix}
N & \nu \\
E & e
\end{pmatrix}
\quad
= [2,2,-1/2,1;1/2],~(E^c\;\; N^c)=[2,1,1/2,1;0],~
e^c=[1,1,1,1;-1],~(\nu^c\;\; n^c)=[1,2,0,1;-1/2]$, 
\end{center}
\vskip 0.4cm
{\underline{{\it Scalars:}}}\;\;
\begin{center}
$\quad
\begin{pmatrix}
\phi_1^0 & \phi_3^0 \\
\phi_1^- & \phi_3^-
\end{pmatrix}
\quad= [2,2,-1/2,1;1/2],~
(\chi_1^0\;\;\chi_2^0)=[1,2,0,1;-1/2]$,
$\begin{pmatrix}
   \phi_2^+ \\
   \phi_2^0
  \end{pmatrix}
= [2,1,1/2,1;0]$,\\
$\quad
\begin{pmatrix}
\Delta_2^0/\sqrt{2} & \Delta_3^0 \\
\Delta_1^0 & -\Delta_2^0/\sqrt{2}
\end{pmatrix}
\quad= [1,3,0,1;1].$
\end{center}

Vertical parentheses indicate doublet under $SU(2)_L$ and the horizontal ones indicate doublet under $SU(2)_N$. For brevity, we mention one family of the particle spectrum and assume that gauge and Yukawa interactions are flavour diagonal. The fermion sector is augmented by an exotic quark $h_q$, an exotic electron $E$ and two exotic neutrinos $N,n$. The scalar sector consists of one $SU(2)_L$ doublet, one $SU(2)_N$ doublet, a bi-doublet and an $SU(2)_N$ triplet. $SU(2)_N$ gauge bosons $X_{1,2,3}$, are not assigned any global $U(1)_P$ charge as they transform under adjoint representation. Their respective $T_{3N}$ quantum numbers are $[ 1,0,-1 ]$, leading to $R$-charges for $X_{1,2}$ as (-1), and $X_3$ as (+1) (since $J=1$). The odd $R$-charged dark gauge bosons are stable as they cannot decay to a pair of SM particles or to other heavier (by construction) exotic particles \footnote {Other odd $R-$charged particles are not electromagnetic charge neutral and hence do not qualify as DM. See \cite{Bhattacharya:2011tr}.}. Thus, in our scenario, the lightest of $X_{1,2}$ qualifies to be DM and in case of a degeneracy, both may serve as DM candidates. Scalar potential of the model and the allowed Yukawa couplings are discussed in Appendix \ref{sec:potential} and Appendix \ref{sec:yukawa} respectively.


\section{Vector Boson as Dark Matter}
\label{sec:VBDM}


This section reveals the allowed parameter space of the vector boson DM through elaborate scan. For the ease of discussion, this section is divided into several subsections as follows: DM-SM interactions and annihilation cross-sections are elaborated in subsection \ref{sec:annihilation}, Boltzmann equation and thermal freeze out in \ref{sec:BEQ}, relic density and allowed parameter space in \ref{sec:relic-density}, direct search cross-section and constraints in \ref{sec:DDC} and finally effects of co-annihilation is pointed out in subsection \ref{sec:coann}.

In our scenario, the neutral scalars that acquire vacuum expectation value (VEV) are $\chi_{1,2}^0,\phi_{1,2}^0$, and $\Delta_{1,3}^0$. $SU(2)_N$ and electroweak symmetries are spontaneously broken through the VEV of $\chi_2^0$ ($\kappa_2$) and $\phi_{1,2}^0$ ($v_{1,2}$) respectively. Due to the presence of a bi-doublet and an $SU(2)_L$ Higgs doublet, the physical Higgs field ($h$) can be written as a linear combination of CP-even neutral components ($\phi_{1,2}$) ($h=(v_1\;\phi_1^0\;+v_2\;\phi_2^0)/v$; where $v=\sqrt{v_1^2+v_2^2}$ = 246 GeV is the electroweak symmetry breaking scale). The model behaves as a two-Higgs doublet model of type II \cite{Branco:2011iw,Bhattacharyya:2015nca}, where part of the bi-doublet $\begin{pmatrix}
   \phi_1^0 \\
   \phi_1^-
  \end{pmatrix}$ couples to up type quarks and $\begin{pmatrix}
   \phi_2^+ \\
   \phi_2^0
  \end{pmatrix}$ couples to down type quarks. Physical scalar fields and their masses after spontaneous symmetry breaking of the model has been elaborated in~\cite{Bhattacharya:2011tr}. The scalar potential can be found in the appendix \ref{sec:potential}. 

\begin{figure}[htb!]
$$
\includegraphics[height=8cm]{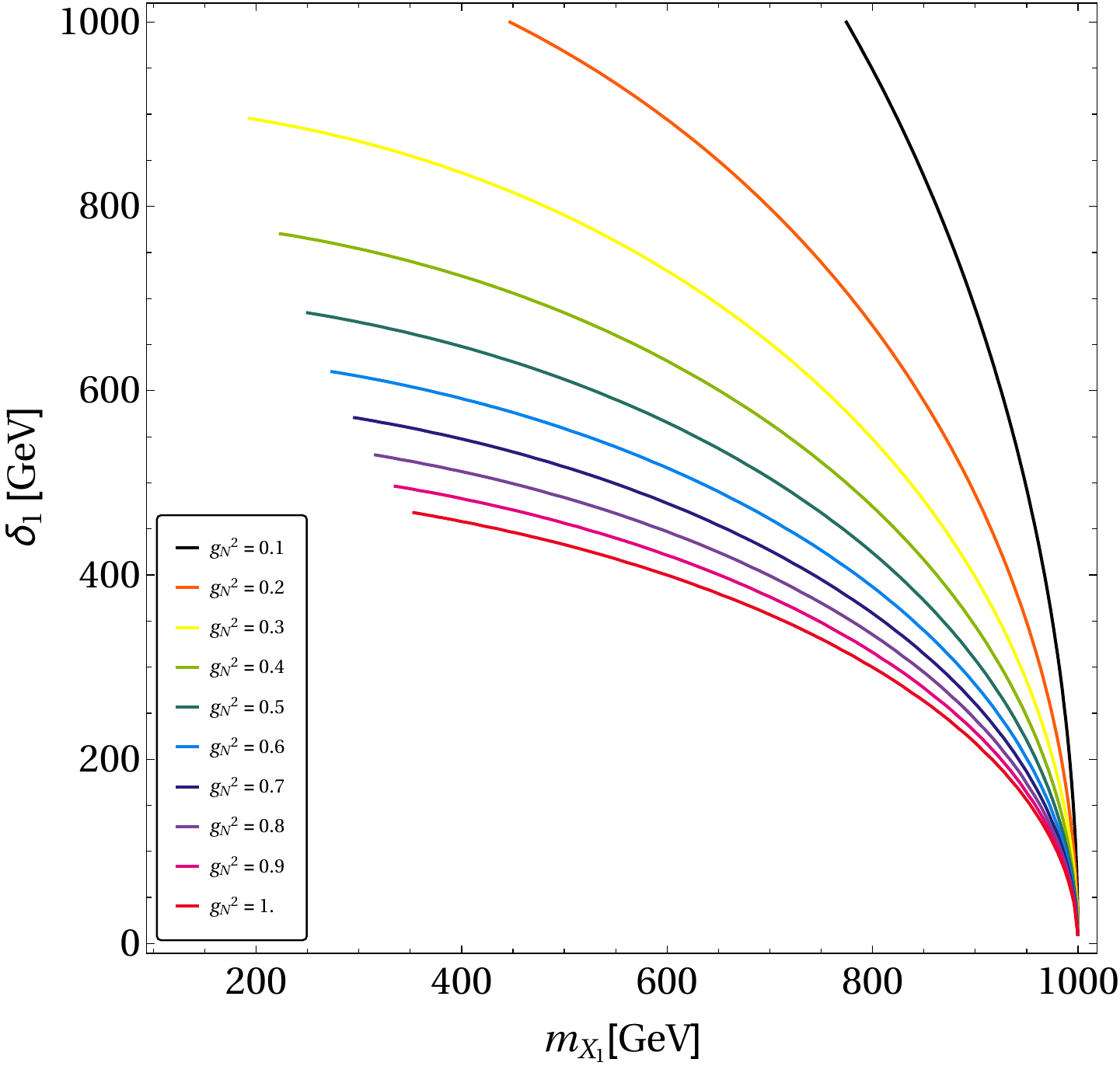}
$$
\caption{Contours of triplet VEV $\delta_1$ with DM mass $m_{X_1}$ is shown for $m_{X_3}=1$ TeV for different choices of $SU(2)_N$ coupling ($g_N$) in the maximum splitting scenario ($\delta_1=\delta_2$). All the points below these lines are ruled out as they predict $X_3$ lighter than TeV.}
\label{fig: u1-mx}
\end{figure}

VEV of the $SU(2)_N$ triplet components $\Delta_{1,3}^0$ ($\delta_{1,2}$) contribute to the $SU(2)_N$ gauge boson masses and causes the mass splitting between two lighter dark gauge bosons as \cite{DiazCruz:2010dc,Bhattacharya:2011tr}: 
\bea
m_{X_{1,2}}^2 &=& \frac{1}{2}g_N^2\left[\kappa_2^2+v_1^2+2(\delta_1\mp \delta_2)^2\right].
\label{eq:mx12}
\eea
Therefore, the absence of the triplets (with $\delta_1=\delta_2=0$) will make $X_{1,2}$ degenerate: $m_{X_{1}}^2=m_{X_{2}}^2=\frac{1}{2}g_N^2\left[\kappa_2^2+v_1^2\right]$. The other gauge boson $X_3$ (even under $R$) mixes with the usual SM neutral gauge boson $Z$ through \cite{DiazCruz:2010dc,Bhattacharya:2011tr}:

\bea
m_{Z,X_3}^2= \frac{1}{2}
\begin{pmatrix}
\left(g_1^2+g_2^2\right)\left(v_1^2+v_2^2\right) &  -g_N\sqrt{g_1^2+g_2^2}~v_1^2\\ \\
-g_N\sqrt{g_1^2+g_2^2}~v_1^2  &  g_N^2\left[\kappa_2^2+v_1^2+4\left(\delta_1^2+\delta_2^2\right)\right]
\end{pmatrix}\,.
\label{eq:mzx3}
\eea\\

The mass term for $X_3$ can then be assumed as: $m_{X_3}^2 \approx \frac{1}{2}g_N^2\left[\kappa_2^2+v_1^2+4(\delta_1^2 + \delta_2^2)\right]$ as $Z-X_3$ mixing is very much constrained \cite{Andreev:2014fwa}. Now, it is also evident from from Eqn.~(\ref{eq:mx12}) that the maximum splitting between $X_1, ~ X_2$ can be achieved for $\delta_1=\delta_2=\delta$, where 
\bea
m_{X_{1}}^2=\frac{1}{2}g_N^2\left[\kappa_2^2+v_1^2\right], ~~m_{X_{2}}^2=m_{X_{3}}^2=\frac{1}{2}g_N^2\left[\kappa_2^2+v_1^2+8\delta^2\right]
\label{eq:maxsplit}.
\eea

In this framework, using the exclusion limit of $m_{Z^{'}} \sim \mathcal{O}(\rm TeV)$ from heavy neutral gauge boson search \cite{gb-lhc}, we can put tentative lower limits on DM mass ($m_{X_1}$) depending on the choice of $\delta_1$. For different values of $SU(2)_N$ gauge coupling ($g_N$), the lower limit on $m_{X_1}$ is shown in Fig.~\ref{fig: u1-mx}. For example, with $g_N^2=0.4$ (light green line), $m_{X_1}$ can be 400 GeV for $\delta_1 \sim 750 $ GeV. The smaller is the value of $\delta_1$, the larger is the allowed $m_{X_1}$. DM mass is equal to the heavy neutral gauge boson mass ($m_{X_1}=m_{X_3}=1 ~{\rm TeV}$) in the limit $\delta_1=\delta_2=0$, independent of the choice of $g_N$. As $X_3$ masses can be larger than 1 TeV, points above the contours are allowed while the ones below are disfavored. It is also quite apparent from Eq.~\ref{eq:maxsplit}, presence of the triplets gives rise to a single component DM in terms of $X_1$. When the triplet is absent or the vev is zero, all three gauge bosons are degenerate, yielding a multipartite DM framework, which necessarily requires to be heavier than $\sim$ 1 TeV as $m_{Z^{'}} \gtrsim 1 ~{\rm TeV}$.

\subsection{Annihilation channels and cross-section}
\label{sec:annihilation}
\begin{figure}[htb!]
$$
\includegraphics[height=4.0cm]{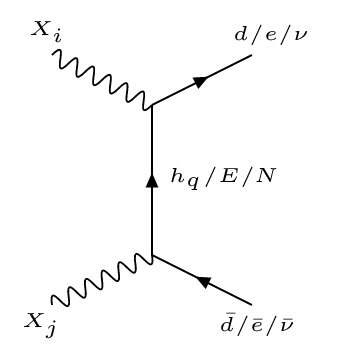}
$$
 \caption{Annihilations (Co-annihilations with $i\neq j, ~ \{i,j\}=\{1,2\}$) to SM fermion pairs by exotic quark exchange.}
\label{fig:fd-ann1}
\end{figure}

\begin{figure}[htb!]
$$
\includegraphics[height=3.5cm]{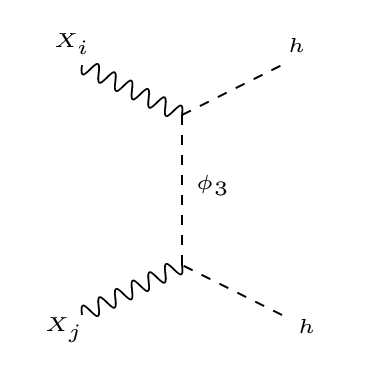}
\includegraphics[height=3.5cm]{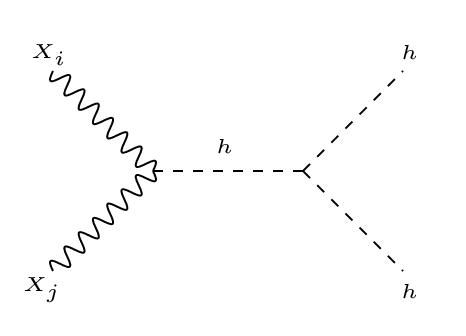}
\includegraphics[height=3.5cm]{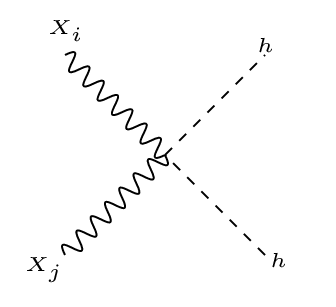}
 $$
 \caption{Annihilations (Co-annihilations with $i\neq j, ~\{i,j\}=\{1,2\}$) to SM Higgs pair.}
\label{fig:fd-ann2}
\end{figure}

\begin{figure}[htb!]
$$
 \includegraphics[height=3.5cm]{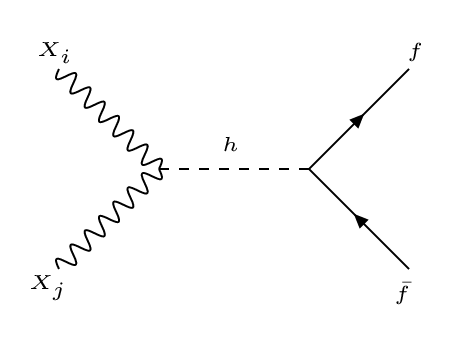}
 \includegraphics[height=3.5cm]{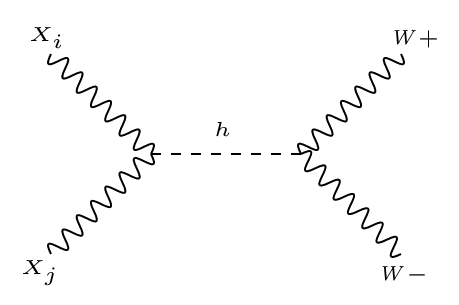}
 \includegraphics[height=3.5cm]{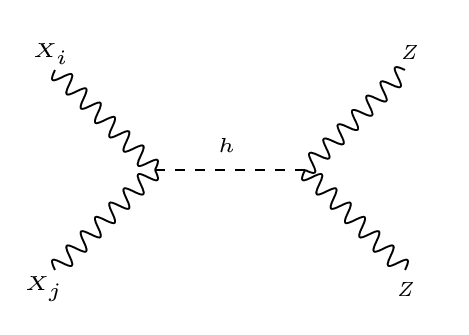}
 $$
 \caption{Annihilations (Co-annihilations with $i\neq j, ~\{i,j\}=\{1,2\}$) to SM fermions and gauge bosons through Higgs exchange}
\label{fig:fd-ann3}
\end{figure}

Relic density of DM is evaluated through thermal freeze out, governed by the number changing processes of the DM. $X_1$ being odd under $R$-charge, cannot directly couple to a pair of SM particles, but can connect to them via the exotic fermions odd under $R$. For example, one can have couplings like $\bar d ~h_q~ X_1$ or $\bar E ~e~X_1 $. Hence, there are different channels through which $X_1$ can annihilate to SM particles which eventually help the DM to freeze out. Here, we have categorized them systematically: 

\begin{itemize}
\item  Annihilations to fermion pairs $d\bar{d};\;e\bar{e};\;\nu \bar{\nu}$ by exchanging $h_q,\;E,\;N$ respectively as shown in Fig.~\ref{fig:fd-ann1},
\item  Annihilation to pair of Higgs by exchanging $\phi_3,\;h$ and through four point contact term shown in Fig.~\ref{fig:fd-ann2},
\item  Annihilation to pair of SM fermions and gauge bosons through Higgs exchange as in Fig.~\ref{fig:fd-ann3}. 
\end{itemize}

It is also important to note that $X_2$ being heavier, can contribute to co-annihilation of $X_1$, which in turn will alter the relic density constraint of the model. We will discuss this in details later. For annihilation cross-section, we compute only the $s$-wave contribution, where $s_0=4m_{X_1}^2$. Therefore, we can approximate: $\langle \sigma v \rangle \sim \sigma v|_{s_0}$, which simplifies relic density calculations significantly.  This approximation is justified as all the diagrams (except $\phi_3$ mediated one) will have non-zero $s$-wave contribution; higher order terms are therefore negligible and will not cause any visible change in the allowed parameter space of the model. Total annihilation cross-section, computed from diagrams in Figs.~(\ref{fig:fd-ann1}), (\ref{fig:fd-ann2}) and (\ref{fig:fd-ann3}), can be written as \cite{peskin}:

\begin{equation}\label{eq:sigma-v2}
\begin{split}
(\sigma \rm{v})_{X_1 X_1 \to SM}|_{s_0=4m_{X_1}^2}& = {g_N^4 m_{X_1}^2 \over 72 \pi}\Bigg\{\sum_{h_q} {N_c \over (m_{h_q}^2
+ m_{X_1}^2)^2} + \sum_E {1 \over (m_E^2 + m_X^2)^2} +  \sum_N {1 \over (m_N^2 + m_{X_1}^2)^2}\Bigg\}\\ &
+ {g_N^2  \over 2 \pi}\left(\frac{v_1}{v}\right)^{2}\Bigg\{\frac{1}{6}\frac{m_f^{2}}{(4m_{X_1}^{2}-m_h^{2})^{2}
+\Gamma_h^2 m_h^2}\left(1-\frac{m_f^{2}}{m_{X_1}^{2}}\right)^{\frac{3}{2}}\\ & + \frac{1}{4m_{X_1}^{2}}\frac{m_Z^{4}}{(4 m_{X_1}^{2}-m_{h}^{2})^{2}+\Gamma_h^2 m_h^2}\left(1-\frac{m_Z^{2}}{m_{X_1}^{2}}\right)^{\frac{1}{2}}\\ &+
\frac{1}{4m_{X_1}^{2}}\frac{m_W^{4}}{(4 m_{X_1}^{2}-m_{h}^{2})^{2}+\Gamma_h^2 m_h^2}\left(1-\frac{m_W^{2}}{m_{X_1}^{2}}\right)^{\frac{1}{2}}\Bigg\}\\ &	
+ \frac{1}{32\pi m_{X_1}^2}\sqrt{1-\frac{m_{h}^2}{m_{X_1}^2}}\Bigg\{\frac{1}{3}~g_N^4\left(\frac{v_1}{v}\right)^4+\frac{3~g_N^2 m_h^4}{\left(4 m_{X_1}^2-m_h^2\right)^2+\Gamma_h^2 m_h^2}\left(\frac{v_1}{v}\right)^2 \\ &+\frac{2 g_N^3 m_h^2 \left(4m_{X_1}^2-m_h^2\right)}{{(4m_{X_1}^2-m_h^2)^2+\Gamma_h^2 m_h^2}}\left(\frac{v_1}{v}\right)^3\Bigg\}.
\end{split}
\end{equation}

First three terms in above equation (Eq.~\ref{eq:sigma-v2}) comes from annihilation to SM fermions through $t-$ channel graphs as shown in Fig.~\ref{fig:fd-ann1} \cite{DiazCruz:2010dc,Bhattacharya:2011tr}. $N_c=3$ is the colour factor which appears only for the colour fermion ($h_q$) exchange. Next three terms, proportional to $(v_1/v)^2$, are collective contributions from Fig.~\ref{fig:fd-ann2}, where $X_1$ is annihilated through SM Higgs exchange to a pair of SM fermions, neutral and charged gauge bosons respectively.  Last three terms are contributions from Fig.~\ref{fig:fd-ann2}, where DM annihilates to a pair of SM Higgs. All Higgs exchange diagrams are suppressed by $(v_1/v)^2$ with $v_1<<v$ to adjust $X_3-Z$ mixing and were ignored in earlier analyses \cite{DiazCruz:2010dc,Bhattacharya:2011tr}.

\begin{figure}[htb!]
$$
\includegraphics[height=4.8cm]{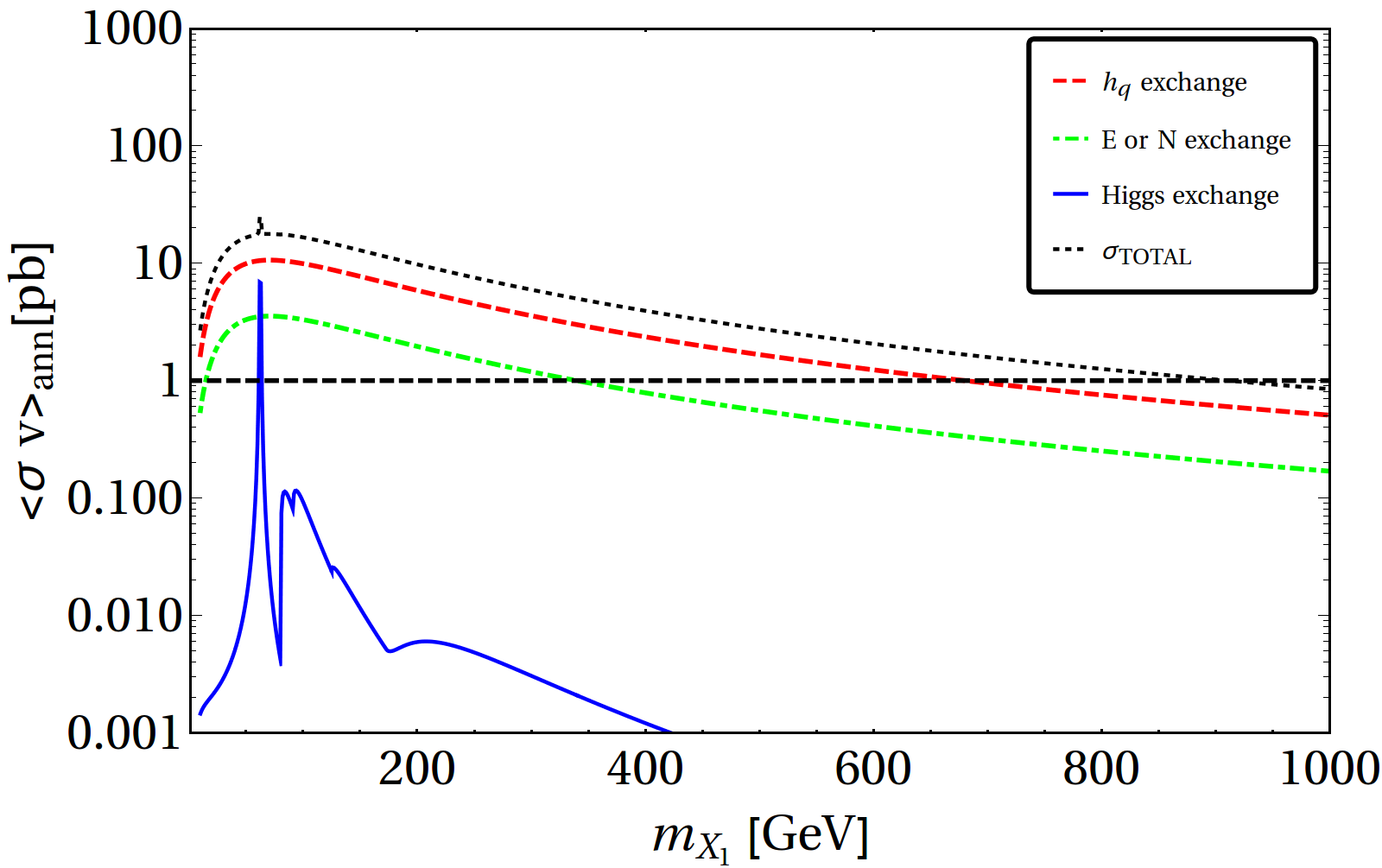}
\includegraphics[height=4.8cm]{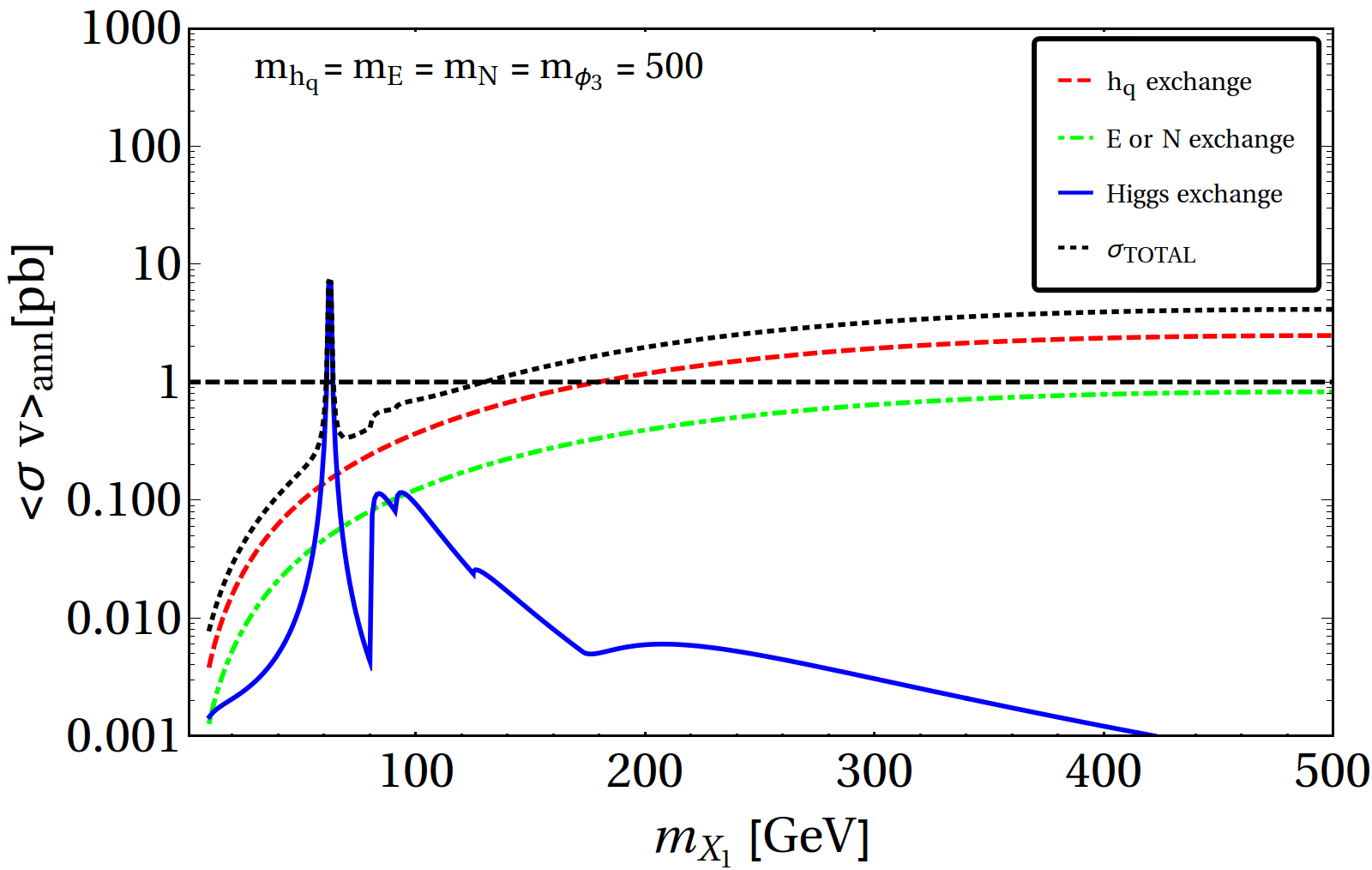}
 $$
 \caption{Variations of total annihilation cross-sections $\langle \sigma v \rangle \sim \sigma v|_{s_0=4m_{X_1}^2}$ and contributions from different channels is shown as a function of DM mass ($m_{X_1}$) for two hypothesis.  Left panel: mass of the exotic fermions are assumed to be ($m_{X_1}+100$) GeV;  Right panel: mass of the exotic fermions are fixed at 500 GeV. For both cases we set $g_N^2=0.4$ for illustration.}
 \label{fig:sigmav-compare}
 \end{figure}

Independent parameters which control the phenomenology of the model are: $SU(2)_N$ gauge coupling \footnote{ This, in principle, is not equal to SM $SU(2)_L$ coupling $g_L$ (except for D-parity conserving scenario as discussed later). The freedom of choosing $g_N$ was not considered in earlier analysis \cite{DiazCruz:2010dc,Bhattacharya:2011tr}.} ($g_N$) and the masses of the exotic non-SM particles. This yields an effective four dimensional parameter space characterizing the DM sector of the model:
 \be
 \{g_N, m_{X_1}, m_{h_q},m\},
\label{eq:parameters}
 \ee

where we simplify the situation by assuming the uncoloured exotic masses to be the same: $m_E=m_N=m_{\phi_3}=m$. This still keeps phenomenological implications of the model intact. Different contributions to total annihilation cross section as a function of DM mass is shown in Fig.~\ref{fig:sigmav-compare}. Here we have adopted two scenarios: (i) all exotic fermions are degenerate and heavier than the DM by 100 GeV, shown in the left panel and (ii) all exotic fermion masses are set to 500 GeV and shown in the right panel of Fig.~\ref{fig:sigmav-compare}.

From both the plots in Fig.~\ref{fig:sigmav-compare}, contribution from $h_q$ exchange is visibly high due to its colour interaction. Significant enhancement in the cross-section is seen at $m_{X_1}\simeq m_h/2$ due to resonance through Higgs exchange. Smaller peaks at $m_{X_1}\simeq m_{W}$ and $m_{X_1}\simeq m_{Z}$ marks the opening of $WW$ and $ZZ$ final states. The patterns of the individual as well as total annihilation cross-section are different in left and right panels of Fig.~\ref{fig:sigmav-compare}. In the left panel, the propagators become heavier with increase in DM mass, causing depletion in cross-section. In the right panel, the propagator masses are kept fixed at 500 GeV. Thus, when $m_{X_1}$ increases and approaches the masses of the exotic fermions, final state phase space gradually increases, increasing the cross-section. Note that, in both the plots, contribution of the Higgs exchange diagram (polynomial in $\frac{v_1}{v}$) is small since $v_1<<v$ to satisfy $Z-X_3$ mixing \cite{Andreev:2014fwa}. The horizontal black dotted line denotes the  annihilation cross-section required to satisfy correct relic density. Regions above denote under-closure (larger cross section) while the region below with smaller cross-section is ruled out by over closure (larger relic density).

\subsection{Boltzmann Equation and thermal freeze out}
\label{sec:BEQ}

To compute thermal freeze out of the DM, we need to solve Boltzmann's equation (BEQ)~\cite{Kolb:1990vq}:
\bea
&& \dot{n}_{X_1}+3H n_{X_1} =
 -\langle \sigma \;{\rm v} \rangle_{_{X_1 X_1 \to SM\;SM}} \left[ (n_{X_1})^2- (n_{X_1}^{EQ})^2\right], 
\label{eq:BEQ}
\eea
where $n_{X_1}$ is DM number density, $H$ is the Hubble constant. $n_{X_1}^{(EQ)}$ is DM number density in equilibrium given by: 
\be
n_{X_1}^{(EQ)} =  \int \frac{\gt_{X_1} d^3 p}{(2 \pi)^3 2E}  \tilde{f}_{X_1}^{(EQ)},    \hspace{.4 cm} {\rm with\; equilibrium\; density} \hspace{.2 cm} \tilde{f}_{X_1}^{EQ}  = \frac{1}{e^{E/k_B T}-1}\;.
\label{eq:no-density}
\ee

$\langle \sigma {\rm v} \rangle_{X_1 X_1 \to SM\;SM}$ depicts thermal averaged annihilation cross-section of the DM defined as~\cite{Kolb:1990vq,gondolo}:

\bea
\langle \sigma \; {\rm v} \rangle_{X_1 X_1 \to SM\;SM} 
 &= &  \int_{\hat{s}_0}^{\infty}d\hat{s}~ \frac{\gt_{X_1}^2 \hat{s}\sqrt{\hat{s}-4m_{X_1}^2}~K_1(\frac{\sqrt{\hat{s}}}{T})~(\sigma {\rm v})_{_{X_1 X_1 \to SM\; SM}}}{16 ~m_{X_1}^4 T~K_2(\frac{m_{X_1}}{T})^2}\,, 
\label{eq:sigma-v1}
\eea
where  $K_{1,2}$ are the modified Bessel's functions, $\gt_{X_1}$ is the internal degrees of freedom associated with $X_1$, which is 3 for its vectorial nature, and $\hat{s}=(p_{X_1}+p^{'}_{X_1})^2$. We will use Eq.~\ref{eq:sigma-v2} for computing the thermal averaged cross-section. 

In order to identify the freeze-out of the DM, we will recast BEQ in terms of co-moving density $(Y)$ as: 
\be
 Y=\frac{n_{X_1}}{s},
 \label{eq:y}
 \ee
Here, the entropy density ($s$) of the universe is given by~\cite{Kolb:1990vq}
\bea
s = \frac{2 \pi^2}{45} g_s (T) T^3 \,; \quad {\rm with}\;
g_s(T) =  r_k g_k \left (\frac{T_k}{T} \right )^3\theta (T-m_k) \,.
\eea
where the repeated index $k \in {\rm}$\{all particles\}, is summed over.
Here, $T_k$ and $g_k$ are the temperature and internal DOF of $k^{th}$ particle 
with $ r_k=1\,(7/8)$ for boson(fermion).

Now we can rewrite Eq.~\ref{eq:BEQ} in terms of $Y$ and $x=\frac{m_{X_1}}{T}$ as:
\bea
\frac{dY}{dx}&=& -\frac{x \langle \sigma {\rm v} \rangle s}{H(m)} \left[ ( \sigma\; {\rm v} )_{X_1X_1\rightarrow SM\;SM} (Y^2-{Y^{EQ}}^2)\right],  
\label{eq:beq3}
\eea
where
\be
H(m)= 1.66 \sqrt{g_{*}}\frac{m^2}{m_{Pl}}; ~ g_{*}=\sum_{i} \chi_i g_i (\frac{T_i}{T})^4,
\ee
where $i$ runs over bosons (fermions) with $\chi_i=1\;(7/8)$ for bosons(fermions),
and $m_{Pl}=\frac{1}{\sqrt {8\pi G}}=2.43 \times 10^{18} ~{\rm GeV}$.

We also have the equilibrium co-moving number density:
\be
Y^{EQ}=0.145 \frac{g}{g_{*s}} (\frac{m_{X_1}}{T})^{\frac{3}{2}}e^{-\frac{m_{X_1}}{T}}.
\label{eq:Yeq}
\ee

\begin{figure}[htb!]
$$	
\includegraphics[height=4.8cm]{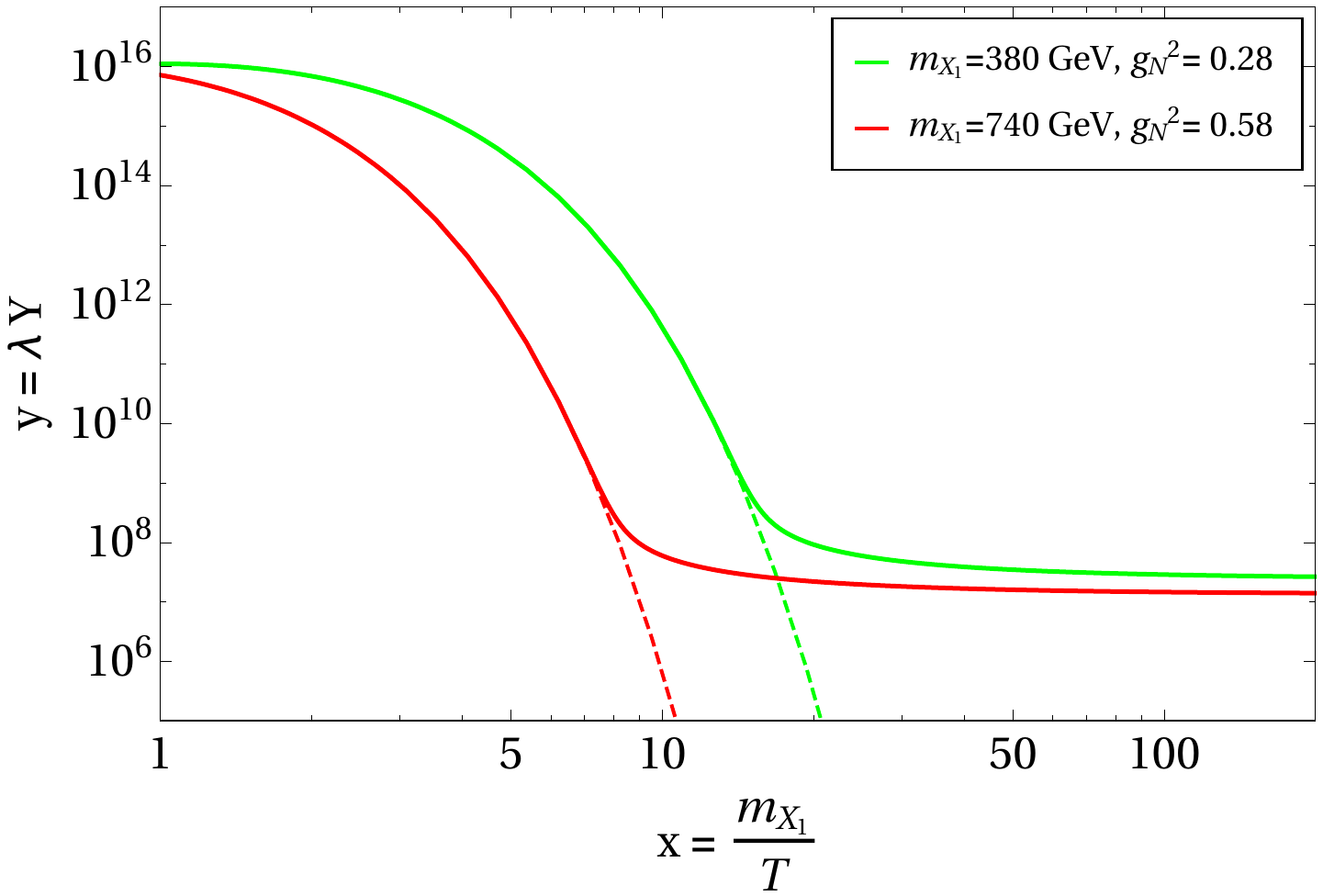}
\includegraphics[height=4.8cm]{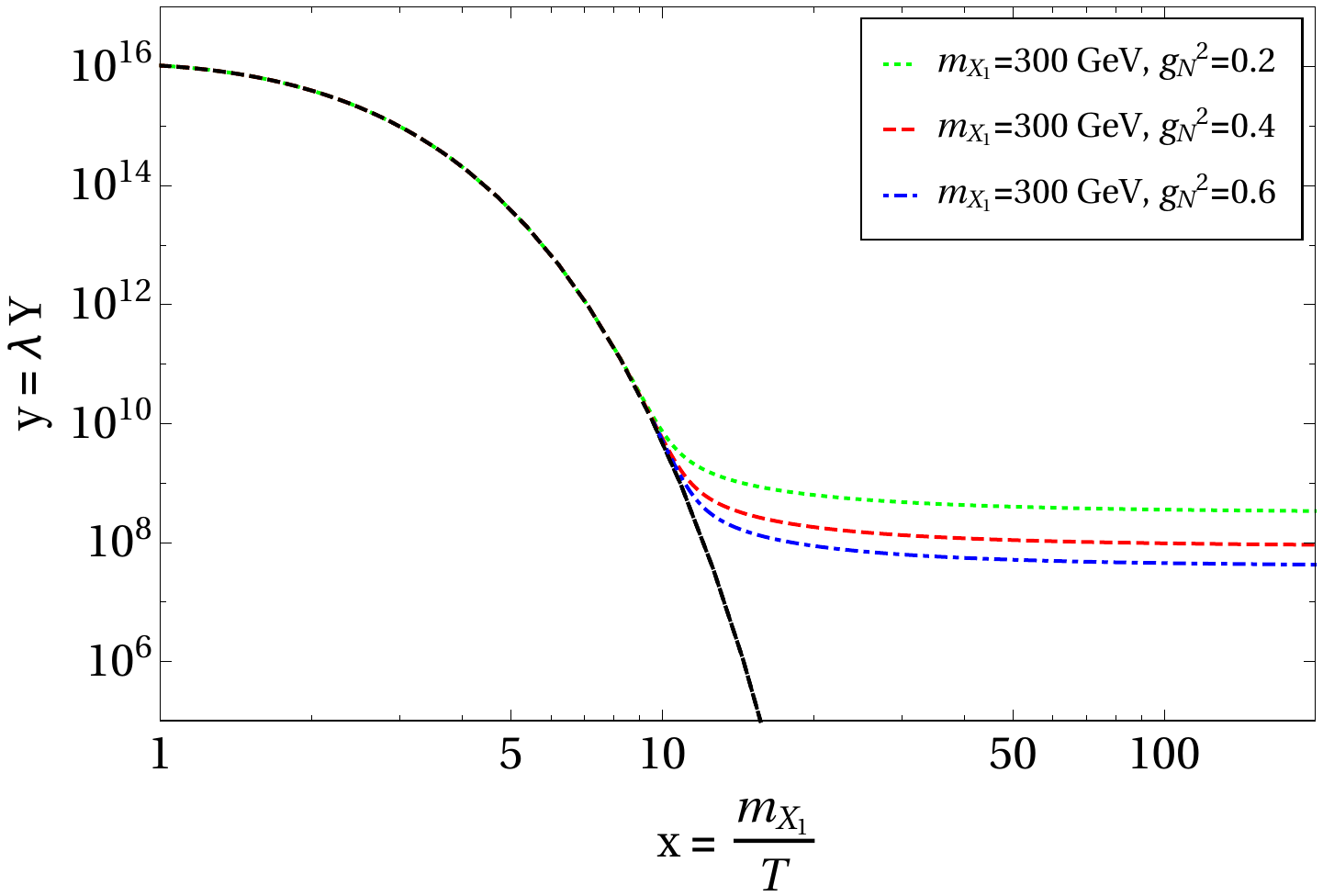}
$$
 \caption{Patterns of thermal freeze out of $X_1$ has been framed in $y=\lambda~ Y$ vs $x=m_{X_1}/T$ plane. Left-panel: $\{m_{X_1},~g_N\}=\{380 ~{\rm GeV}, 0.28\},~\{ 740 ~{\rm GeV}, 0.58 \}$ are chosen as illustration. Dashed lines show the respective equilibrium distributions. Right-panel: different values of $g_N=\{0.2,0.4,0.6\}$ are chosen for DM mass $m_{X_1}=300$ GeV.}
 \label{fig:freeze-out}
 \end{figure}

Now the Boltzmann equation can be further simplified to a compact form as:
\be
\frac{dy}{dx}= -\frac{m_{X_1}}{x^2} \left[\sigma _0(y^2-{y^{EQ}}^2)\right],  
\label{eq:beq4}
\ee
where $y=\lambda Y$ with $\lambda={(0.264 ~ m_{Pl} \frac{g_{*s}}{\sqrt{g_*}})}$, $\sigma_0=(\sigma\rm{v})_{_{X_1X_1\rightarrow SM}}$ given in Eq.~\ref{eq:sigma-v2}.

We demonstrate the patterns of DM freeze out in Fig.~\ref{fig:freeze-out}. Left panel shows two different combinations of \{$m_{X_1},{g_N}^2\}=\{380 ~{\rm GeV}, 0.28\},~\{740 ~{\rm GeV}, 0.58\}$; right panel shows freeze out for three different values of ${g_N}^2=\{0.2,0.4,0.6\}$ for same DM mass $m_{X_1}=300$ GeV.  In the left plot, due to different choices of the DM masses, the equilibrium distributions (shown by dashed lines) are different, lighter ($m_{X_1}=380$ GeV shown by the green line) appears above the heavier one ($m_{X_1}=740$ GeV shown by the red line).  Decoupling for these two cases occur at $x=20 ~(T\sim 19 ~{\rm GeV})$ and $x=7 ~(T\sim 105 ~{\rm GeV})$. Heavier component still has a smaller yield due to significantly larger coupling (${g_N}^2=0.58$), thanks to larger annihilation cross section. In the right panel, we illustrate the same for a fixed DM mass. Choices of the parameters here will be explained shortly.

\subsection{Relic Density and allowed parameter space}
\label{sec:relic-density}

\begin{figure}[htb!]
$$
\includegraphics[height=4.8cm]{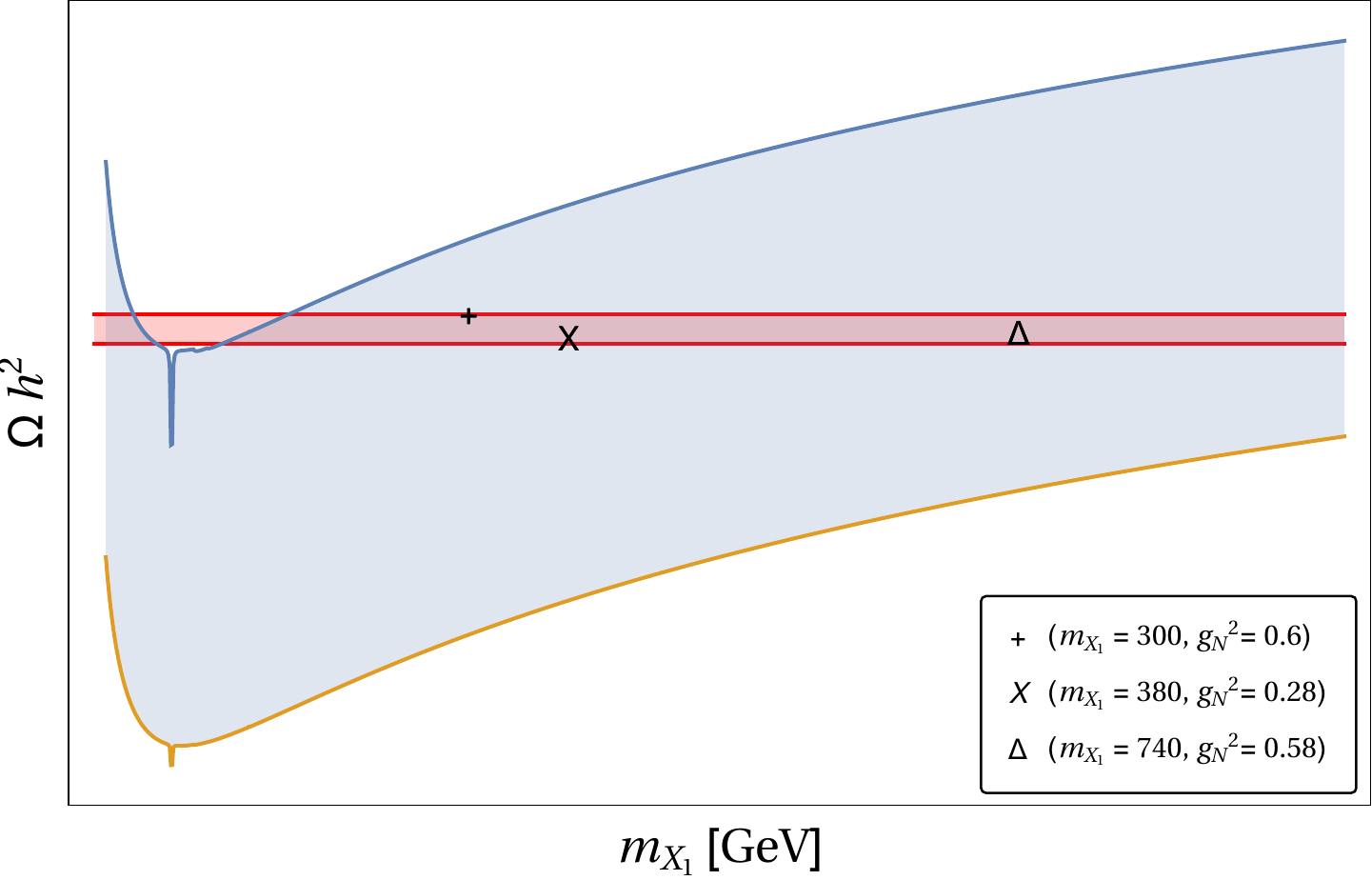}
\includegraphics[height=4.8cm]{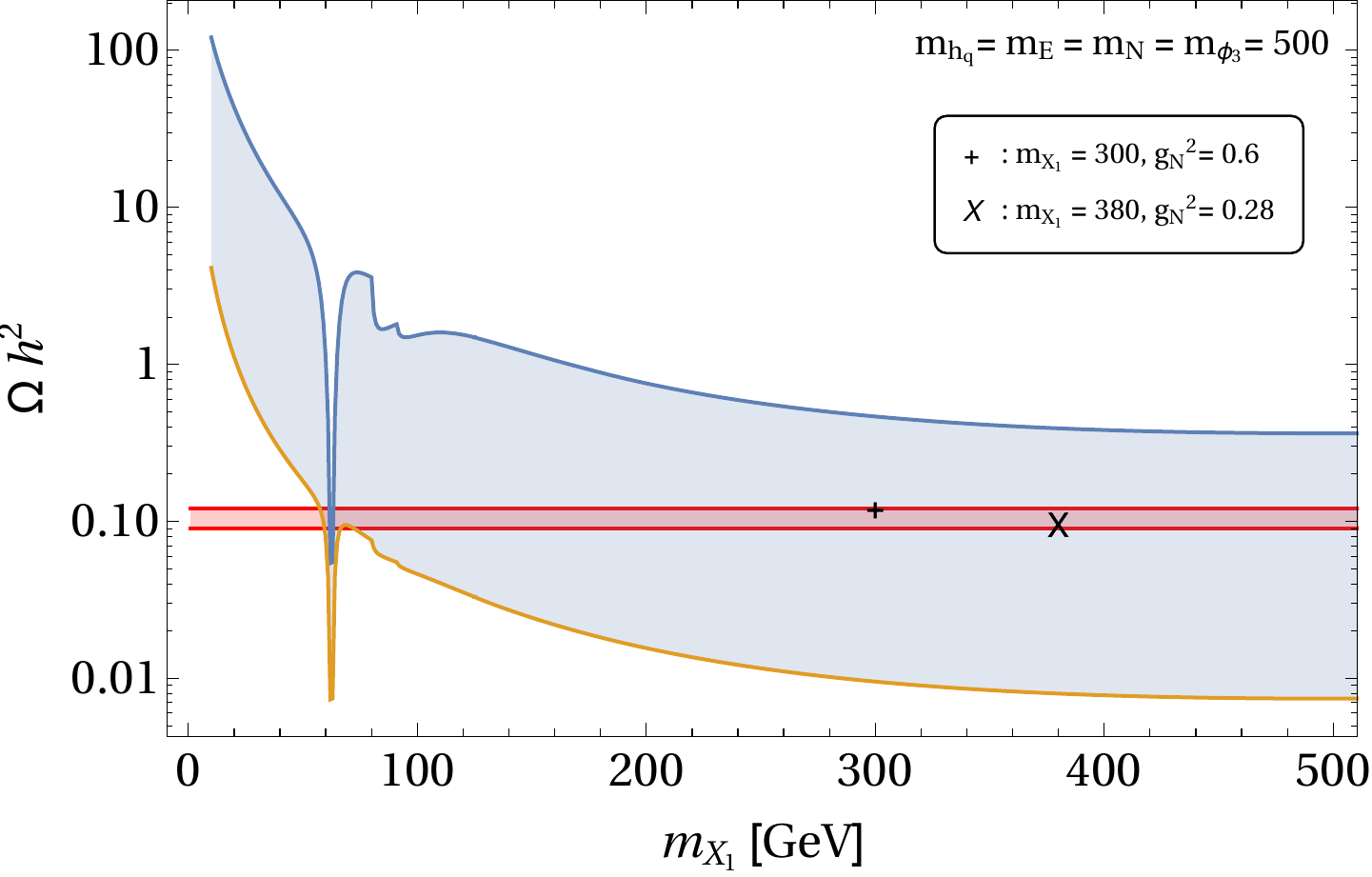}
$$
\caption{Variation in relic density is shown as a function of DM mass for two hypothesis: On left panel, masses of the exotic fermions are proposed to be ($m_{X_1}+100$) GeV; and on right-panel they are fixed at 500 GeV. The $SU(2)_N$ coupling $g_N^2$ is also varied within the range of $\{0.32:0.78\}$ for each values of $m_{X_1}$. Correct relic density window is shown by red band. Freeze-out points of Fig.~\ref{fig:freeze-out} are also indicated.}
\label{fig:Omega-m}
\end {figure}

After freeze-out, relic density for the DM is obtained from the yield as: 
 \be
\Omega h^2 = \frac{m_{X_1}s_0\sqrt{g_{*}}}{3H_0^2~m_{pl}^3~ 0.26 g_{*s}}y(x_\infty),
\ee
where $y(x_\infty)$ is the solution of Eq.~\ref{eq:beq4} at large values of $x$. This can be written in terms of annihilation cross-section under $s-$ wave approximation as \cite {Kolb:1990vq}: 
\be
\Omega h^2  \simeq \frac{2.4 \times 10^{-10}{\rm GeV^{-2}}}{( \sigma \; {\rm v})_{_{X_1 X_1 \to SM \;SM}}}.
\label{eq:Omega-sigmav}
\ee
We will therefore adopt Eq. ~\ref{eq:Omega-sigmav} as approximate analytical solution for evaluating relic density of the DM. The variation of relic density with DM mass is shown in Fig.~\ref{fig:Omega-m}. On the left hand side, we vary all the exotic fermion masses as $(m_{X_1}+100)$ GeV while on the right hand side, we keep them fixed at 500 GeV and vary DM mass upto that (similar to Fig.~\ref{fig:sigmav-compare}). In both the plots a sharp drop in relic density is observed due to resonant enhancement in the Higgs exchange diagram at $m_{X_1}\simeq m_h/2$. Relic density varies over a wide range (blue band) for a particular DM mass  due to a large span of $SU(2)_N$ gauge coupling $g_N$: $\{0.32-0.78\}$. The horizontal red band depicts the allowed range of relic density following WMAP in $3\sigma$ range \footnote{PLANCK data dictates relic density in a similar range $\Omega_{\rm DM} h^2 = 0.1196\pm 0.0031$ \cite{Ade:2015xua}. Though a little more restrictive, would not have altered the main outcome obtained in the analysis.}
\bea
0.09\leq \Omega_{\rm DM} h^2 \leq 0.12 \,.
\label{eq:wmap.region}
\eea

The points which were chosen to demonstrate thermal freeze-out in Fig.~\ref{fig:freeze-out}, satisfy relic density, as shown in Fig.~\ref{fig:Omega-m} by ($\times$), ($+$) and ($\triangle$).



\begin{figure}[htb!]
$$
\includegraphics[height=4.5cm]{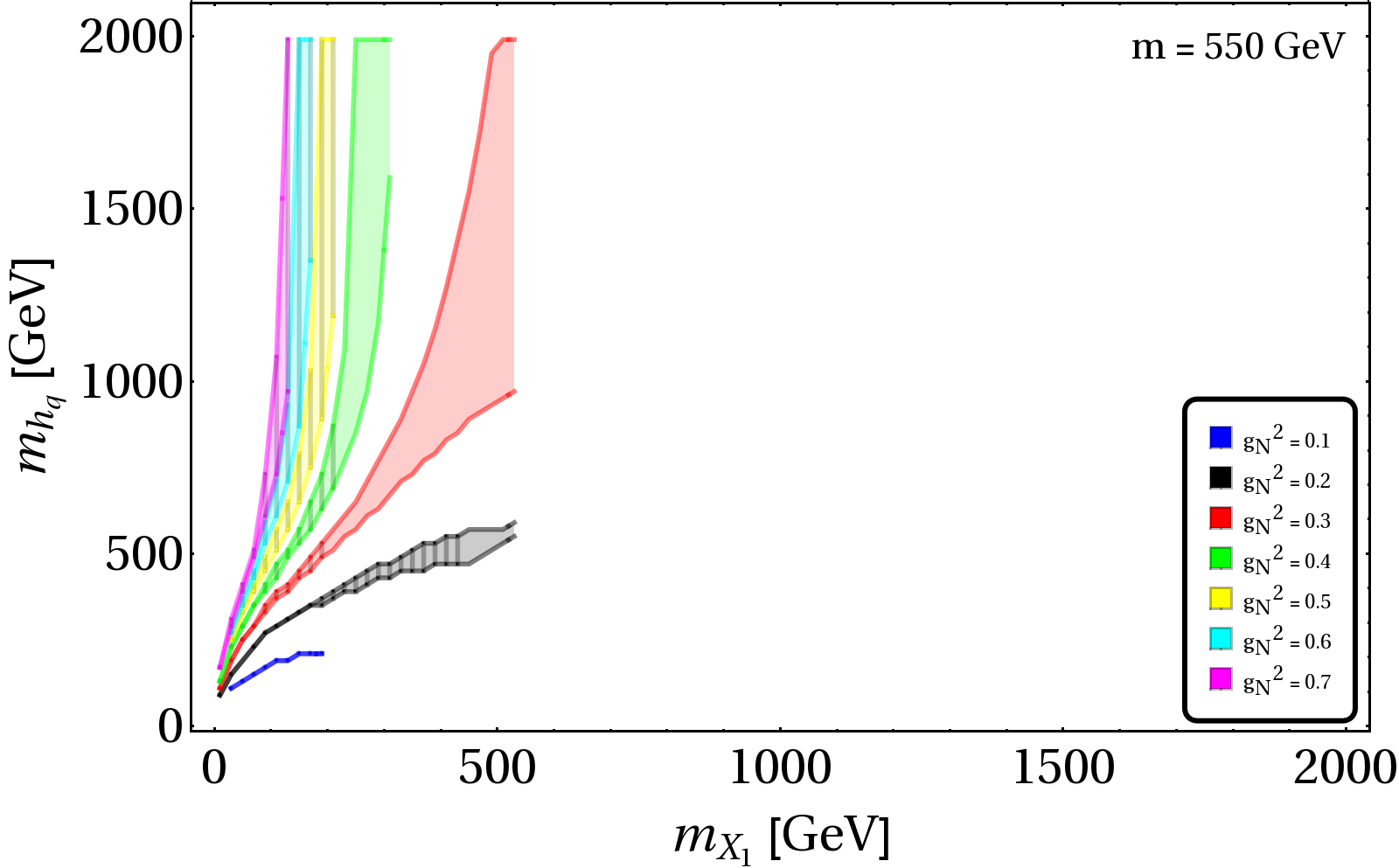}
\includegraphics[height=4.5cm]{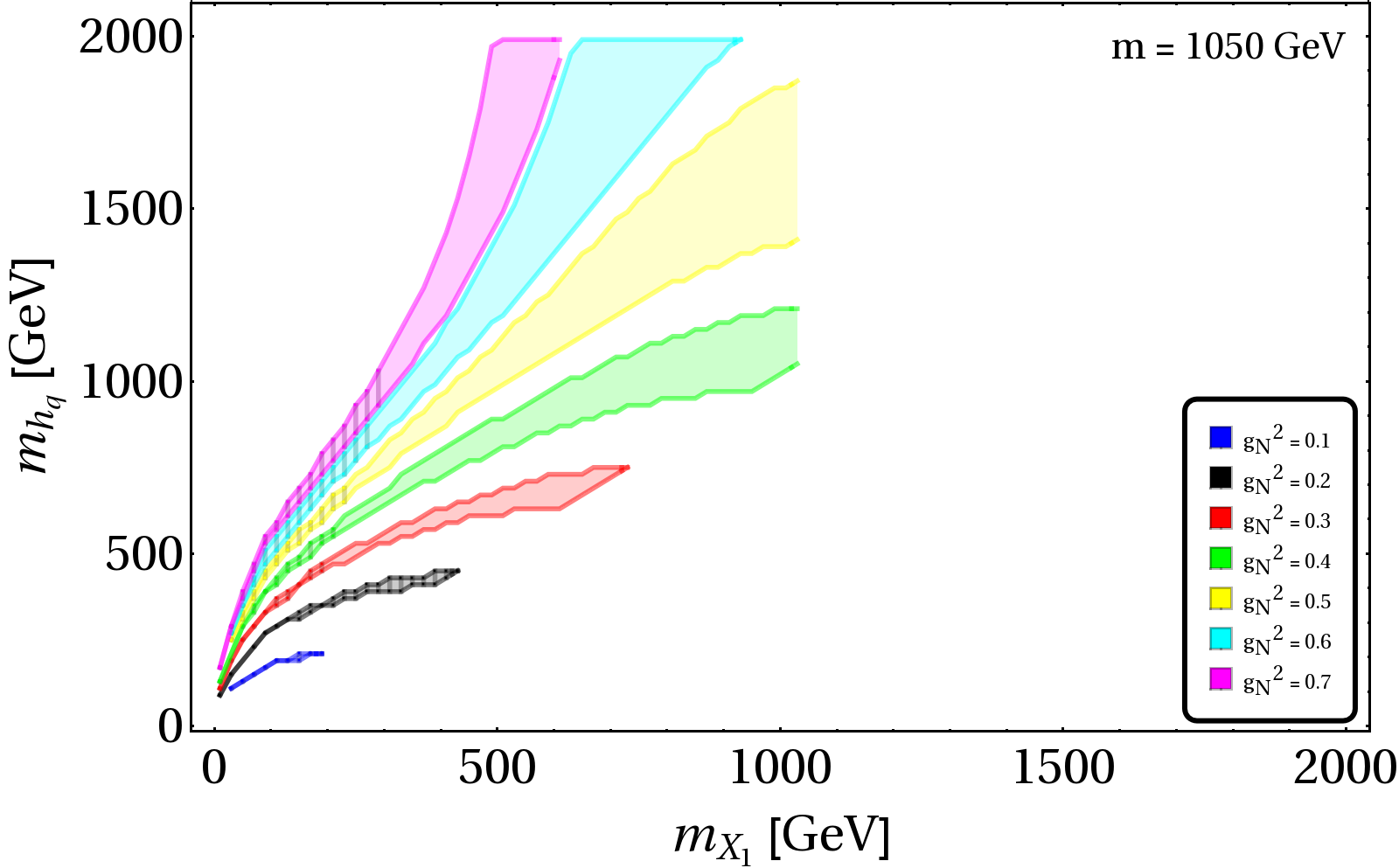}
$$
$$
\includegraphics[height=4.5cm]{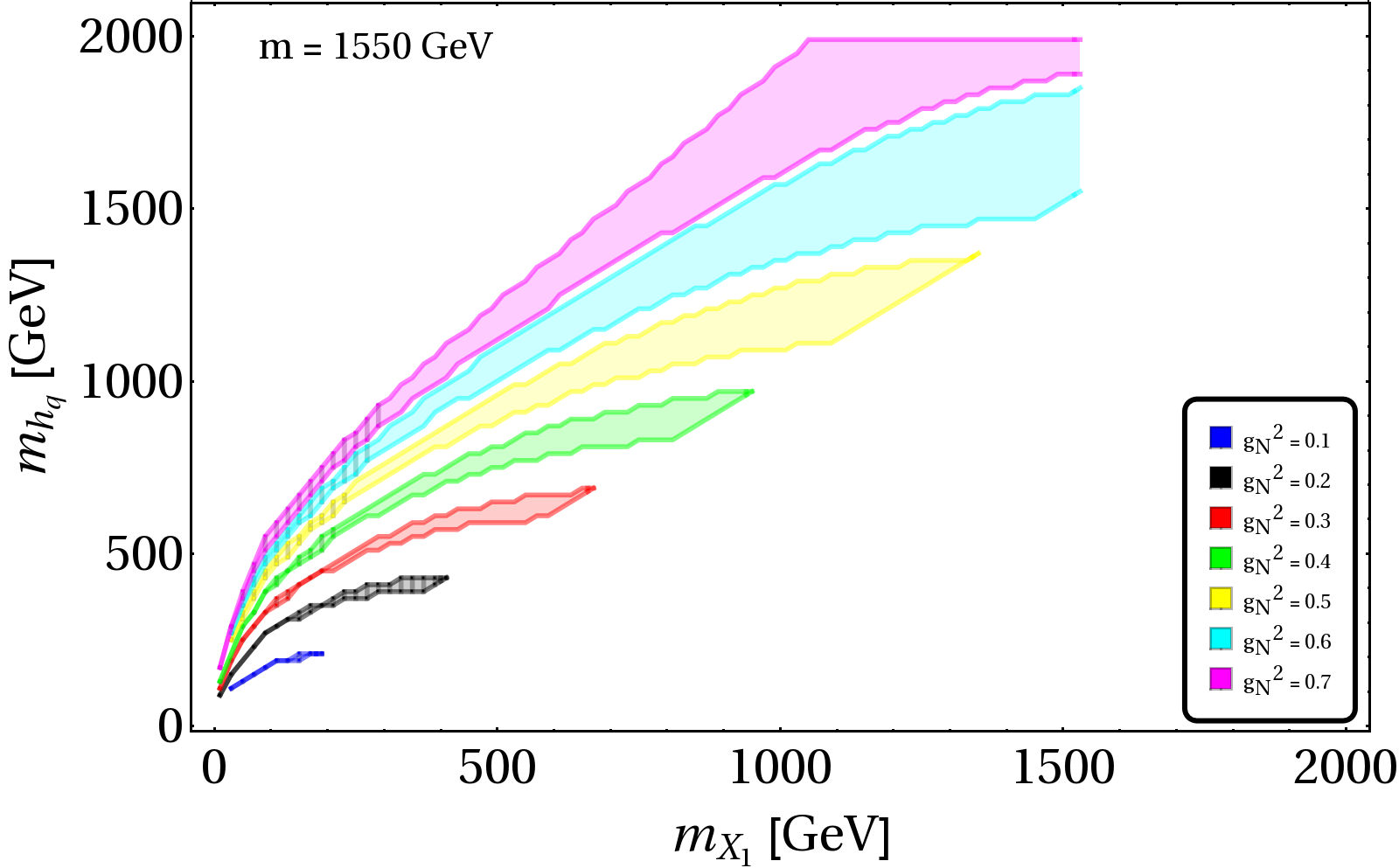}
\includegraphics[height=4.5cm]{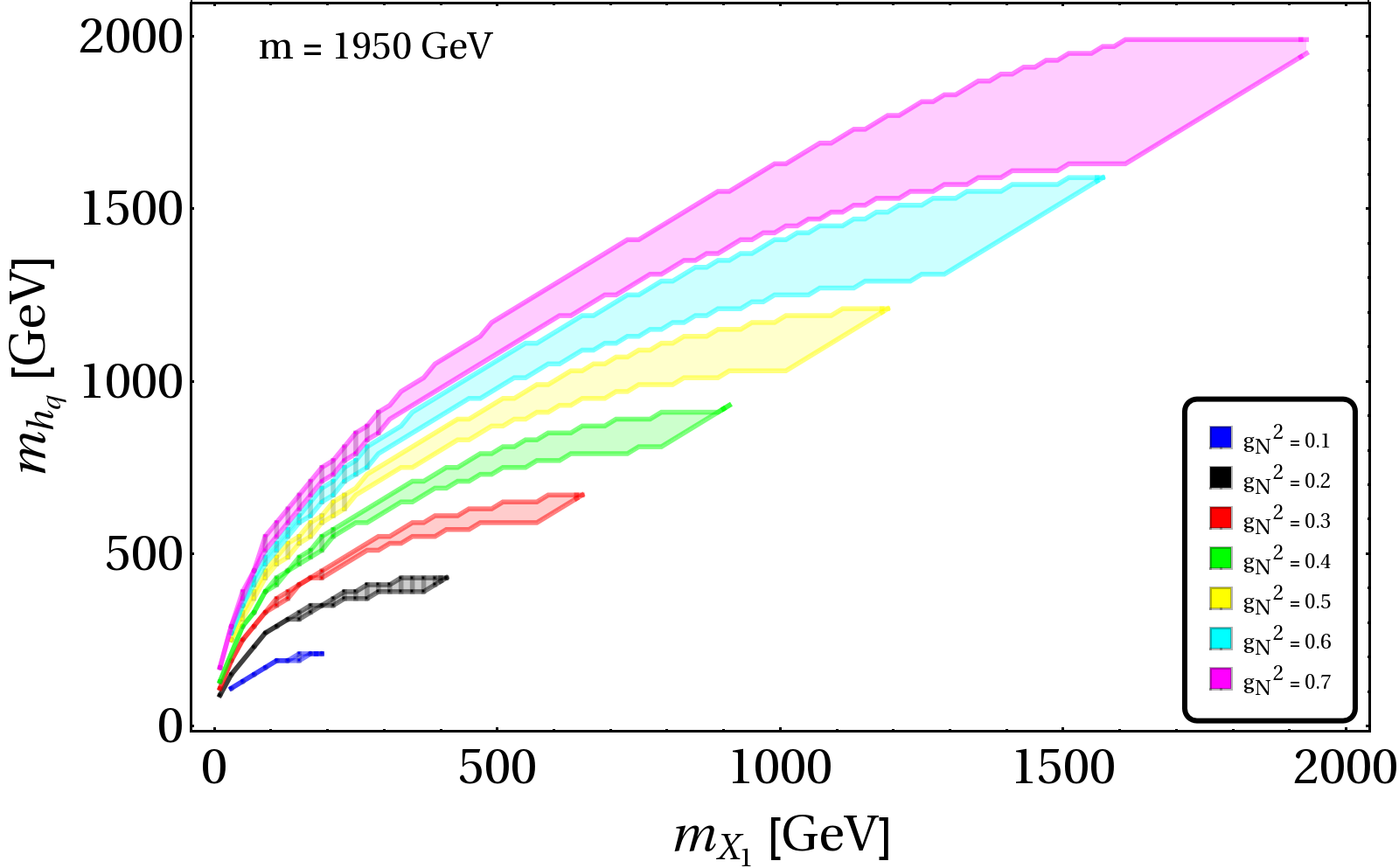}
$$
\caption{Relic density allowed parameter space in $m_{X_{1}}-m_{h_q}$ plane for four different values of $m=m_E=m_N$: 550 GeV (top left), 1050 GeV (top right), 1550 GeV (bottom left), 1950 GeV (bottom right) for different values of $g_N^2\to\{0.1-0.7\}$. The hatched region is excluded by the VEV constraints.}
\label{fig:m1-mhq1}
\end{figure}

\begin{figure}[htb!]
$$
\includegraphics[height=7cm]{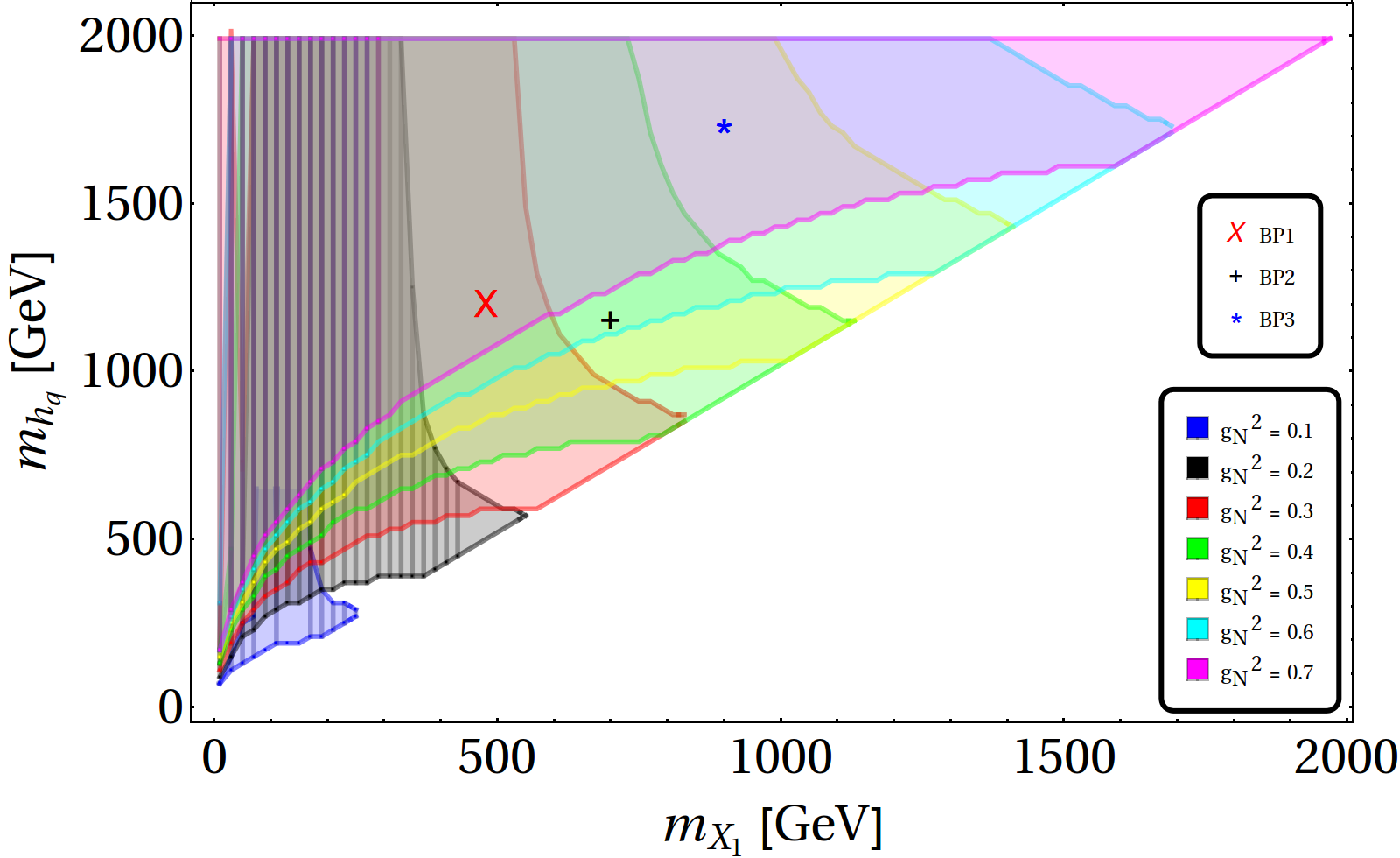}
$$
\caption{Relic density allowed parameter space in $m_{X_{1}}$-$m_{h_q}$ plane when $m=m_N=m_E$ is varied between $\{20-2000\}$ GeV for different $g_{N}^{2}\to\{0.1-0.7\}$. The hatched region is discarded by VEV limit. Benchmark points (BP1, BP2, BP3) as discussed in Table \ref{tab:BP} of the next section are also indicated in the plot.}
\label{fig:m1-mhq1-allm}
\end{figure}

To find the allowed region of parameter space satisfying Eq.~\ref{eq:wmap.region}, we vary the parameters $\{g_N, m_{h_q},m_N = m_E = m,m_{X_1}\}$ as discussed earlier. In $m_{X_1}$-$m_{h_q}$ plane, the allowed parameter space is shown in Fig.~\ref{fig:m1-mhq1} for different $g_N^2$ and $m$. Each figure is for a specific value of $m$ and contains different colour coded regions corresponding to different values of $g_N^2$. For each of these regions, upper and lower boundaries correspond to maximum and minimum allowed values of $m_{h_q}$. This pattern is expected as relic density is closed from both sides. Annihilation cross-section decreases for larger $m_{h_q}$, causing more contribution to the relic density. Thus, for a given $g_N^2$, a cut-off in $m_{h_q}^{max}$ corresponds to the maximum allowed relic density. Similarly, the lower cut-off in $m_{h_q}$ corresponds to the maximum annihilation cross-section i.e, minimum relic density. From different figures in Fig.~\ref{fig:m1-mhq1}, we observe that for small values of $m$ only a small part of $m_{X_1}$ is allowed. This is a consequence of the constraint: $m_{X_1}\le m$. Now, there exists a lower limit of $m_{X_1}$ for a given value of $g_N$ allowed by the VEV constraints as depicted in  Fig.~\ref{fig: u1-mx}. The hatched regions on the left hand side of each figure in Fig.~\ref{fig:m1-mhq1} shows the parameter space disallowed by VEV. In addition, using PLANCK limit shrinks the allowed region but keeps the interpretations same. 

Fig.~\ref{fig:m1-mhq1-allm} summarizes the allowed parameter space in $m_{X_1}$ vs $m_{h_q}$ plane when $m$ is varied upto 2 TeV. Three benchmark points (BP1, BP2, BP3), identified after imposing all constraints, are indicated in this plot and listed in Table \ref{tab:BP} in subsection \ref{sec:summary-dm}.

\begin{figure}[htb!]
$$
\includegraphics[height=7cm]{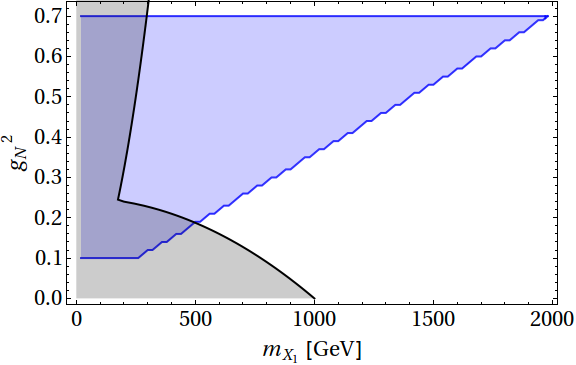}
$$
\caption{Relic density compatible parameter space in $m_{X_{1}}-g_N^2$ plane has been described by the blue region. The grey shaded region depicts the VEV disallowance and thus excluded.}
\label{fig:m1-gn}
\end{figure}

The correlation between DM mass and coupling, satisfying right relic density, is shown in Fig.~\ref{fig:m1-gn} by blue shaded region. VEV exclusion limit is shown by translucent grey region on the left hand side constraining low DM mass.

\subsection{Direct detection constraints}
\label{sec:DDC}


\begin{figure}[htb!]
\includegraphics[height=4.2cm]{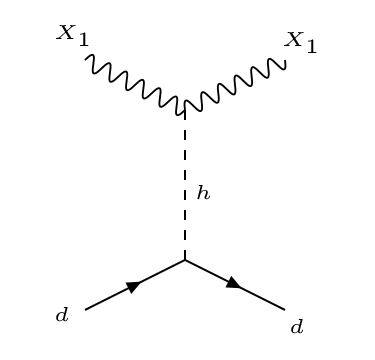}~\includegraphics[height=4.2cm]{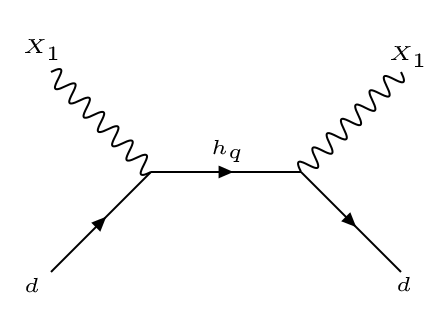}~\includegraphics[height=4.2cm]{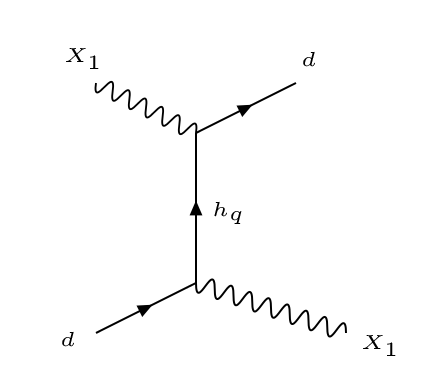}
\caption{Relevant interactions of  dark matter ($X_{1}$) with quarks (nucleons) for direct-search experiments.}
\label{fig:fd-direct}
\end{figure}

\begin{figure}[htb!]
$$
\includegraphics[height=6.2cm]{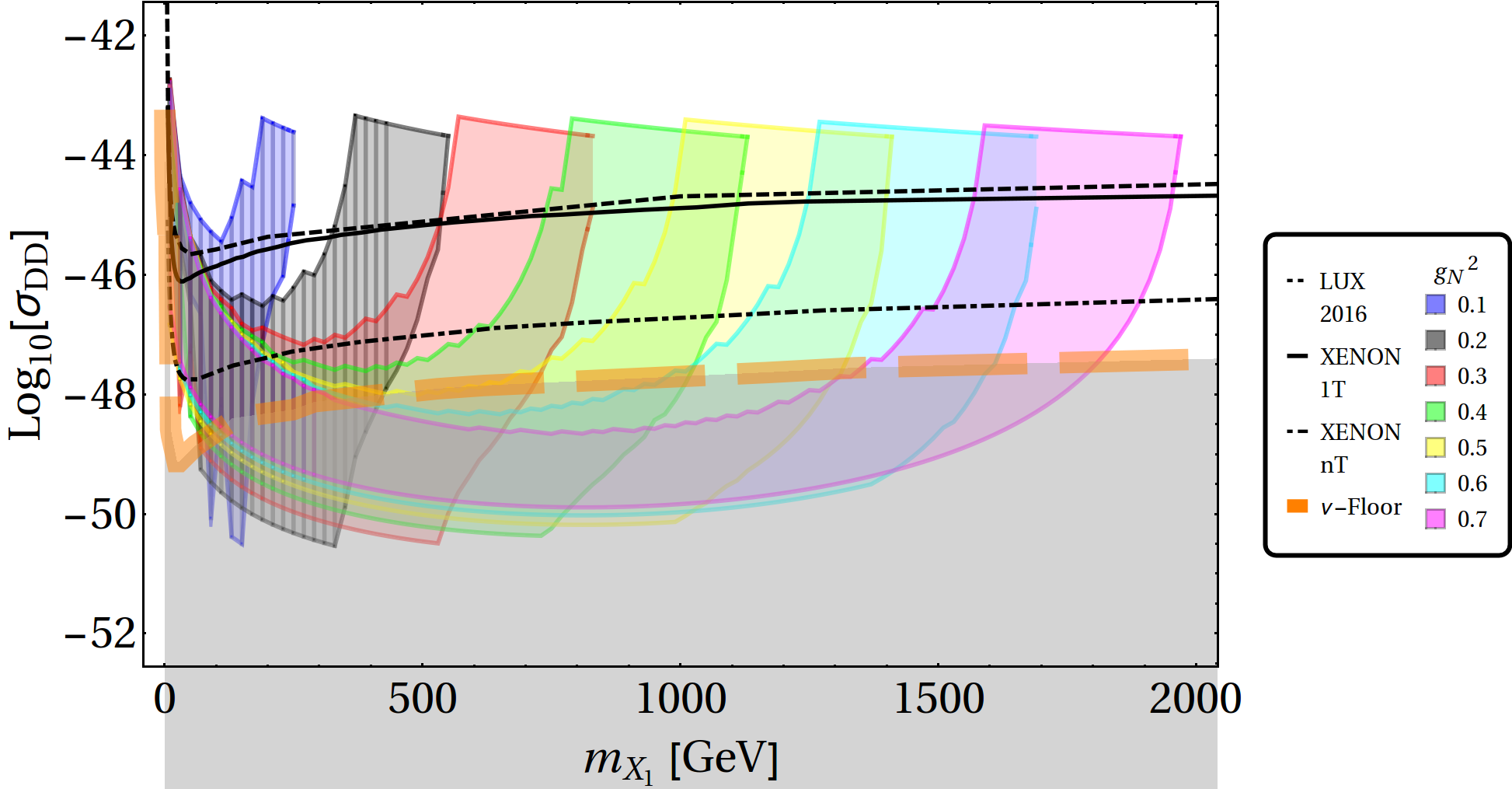}
$$
$$
\includegraphics[height=6.2cm]{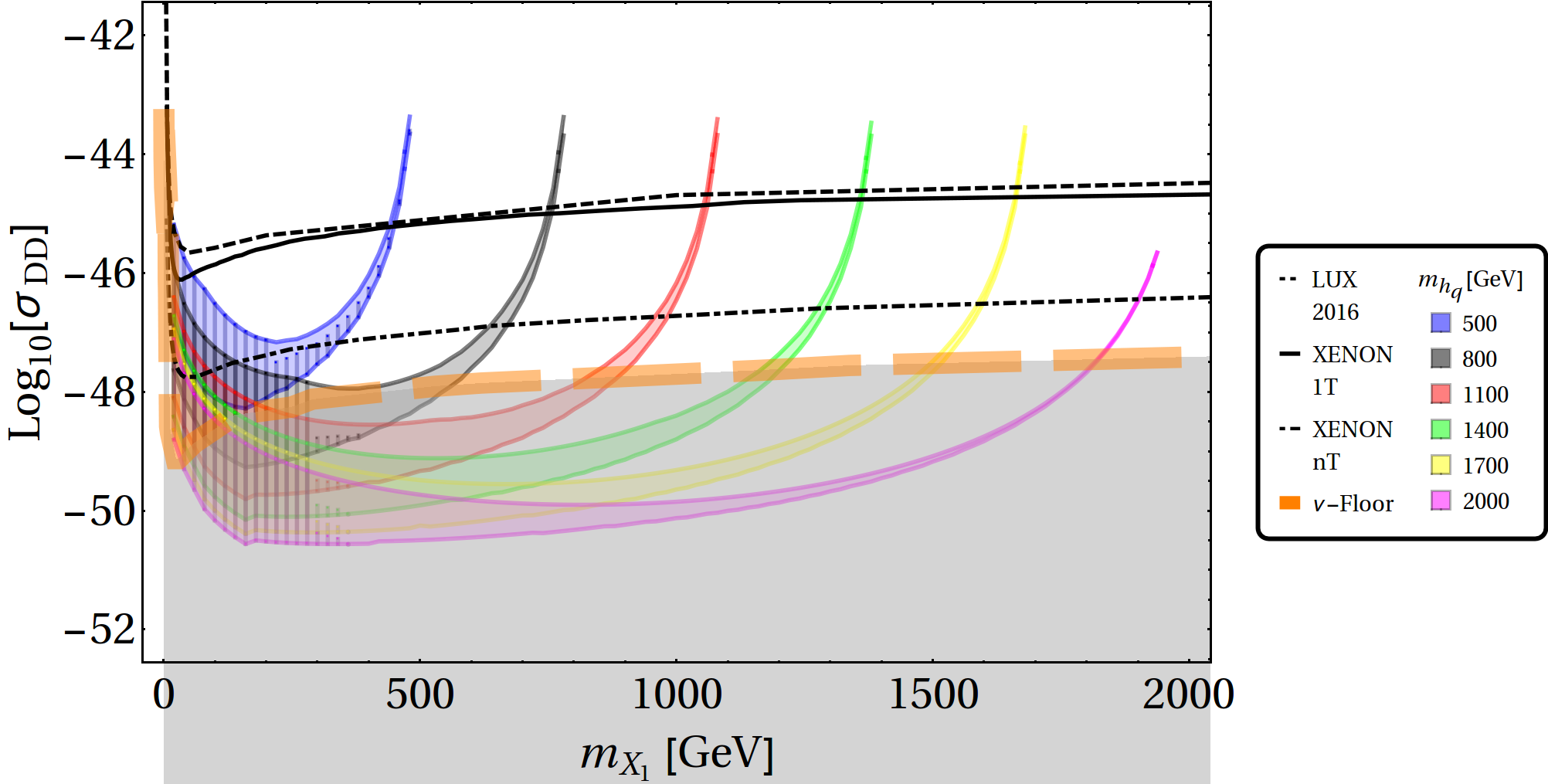}
$$
\caption{Spin independent direct search cross-section for the vector boson DM $X_1$ with respect to DM mass ($m_{X_1}$) for relic density allowed parameter space. Top: Different values of $g_N^2$ is shown by different colours where $m_{h_q}$ is varied appropriately to yield correct relic density. Bottom: Different $m_{h_q}$ values are shown by different coloured regions where $g_N^2$ is varied to yield correct relic density. Limits from LUX2016, XENON1T and XENONnT predictions are depicted by the dashed, solid and dot-dashed black lines respectively. The orange thick dashed line shows the background limit from solar, atmospheric and diffuse supernovae neutrinos  while the grey shaded region below shows the neutrino floor. VEV constraints are shown by hatched regions.}
\label{fig:m1-sigma}
\end{figure}

Spin-independent direct search cross-section for vector boson DM scattering off nuclei can be expressed as \cite{Hisano:2010yh,Hisano:2015bma}: 

\begin{equation}
\sigma_{DD}^{SI} = {1 \over \pi} \left( {m_{nu} \over m_{X_1}+m_{nu}} \right)^2 \left| 
{Z f_p + (A-Z) f_n \over A} \right|^2,
\label{eq:sigma-DD}
\end{equation}
where $Z$  and $(A-Z)$ are the number of protons and neutrons in Xe nucleus respectively. The mass of nucleus is given as $m_{nu}=Am_p+(A-Z)m_n$. Here, $m_{p(n)}$ and $f_{p(n)}$ are the mass and form factor for proton (neutron).

The ratios of the form factors of proton and neutron {\it w.\;r.\;t.} their respective masses are computed using the diagrams in Fig.~\ref{fig:fd-direct} by incorporating the gluonic contributions along with the twist-2 operators. Here also the sub-dominant t-channel contribution due to Higgs mediation has been taken into account which was ignored in earlier analysis \cite{Bhattacharya:2011tr}. The ratios therefore read:
\bea
{f_{i} \over m_{i}} &=& (I_i) \left[ - {g_N^2 \over 4 m_h^2}\left(v_1 \over v\right)^2 - {g_N^2 \over 16} 
{m_{h_q}^2 \over (m_{h_q}^2-m_{X_1}^2)^2} \right] + {3 \over 4} (J_i) \left[ - 
{g_N^2 \over 4} {m_{X_1}^2 \over (m_{h_q}^2 - m_{X_1}^2)^2} \right] \nonumber \\ 
&-& (K_i) \left( (1.19){g_N^2 \over 54 m_h^2}\left(v_1 \over v\right)^2 + {g_N^2 \over 36} \left[ 
(1.19) {m_{h_q}^2 \over 6 (m_{h_q}^2-m_{X_1}^2)^2} + {1 \over 3(m_{h_q}^2-m_{X_1}^2)} \right] 
\right)
\label{eq:fp-mp}
\eea
where $i$ stands either for proton ($p$) or neutron ($n$) and their respective form factors are $I_{p(n)}=0.052\;(0.061); \;J_{p(n)}=0.222\;(0.330);\; K_{p(n)}=0.925\;(0.922)$.

\begin{figure}[htb!]
$$
\includegraphics[height=4.6cm]{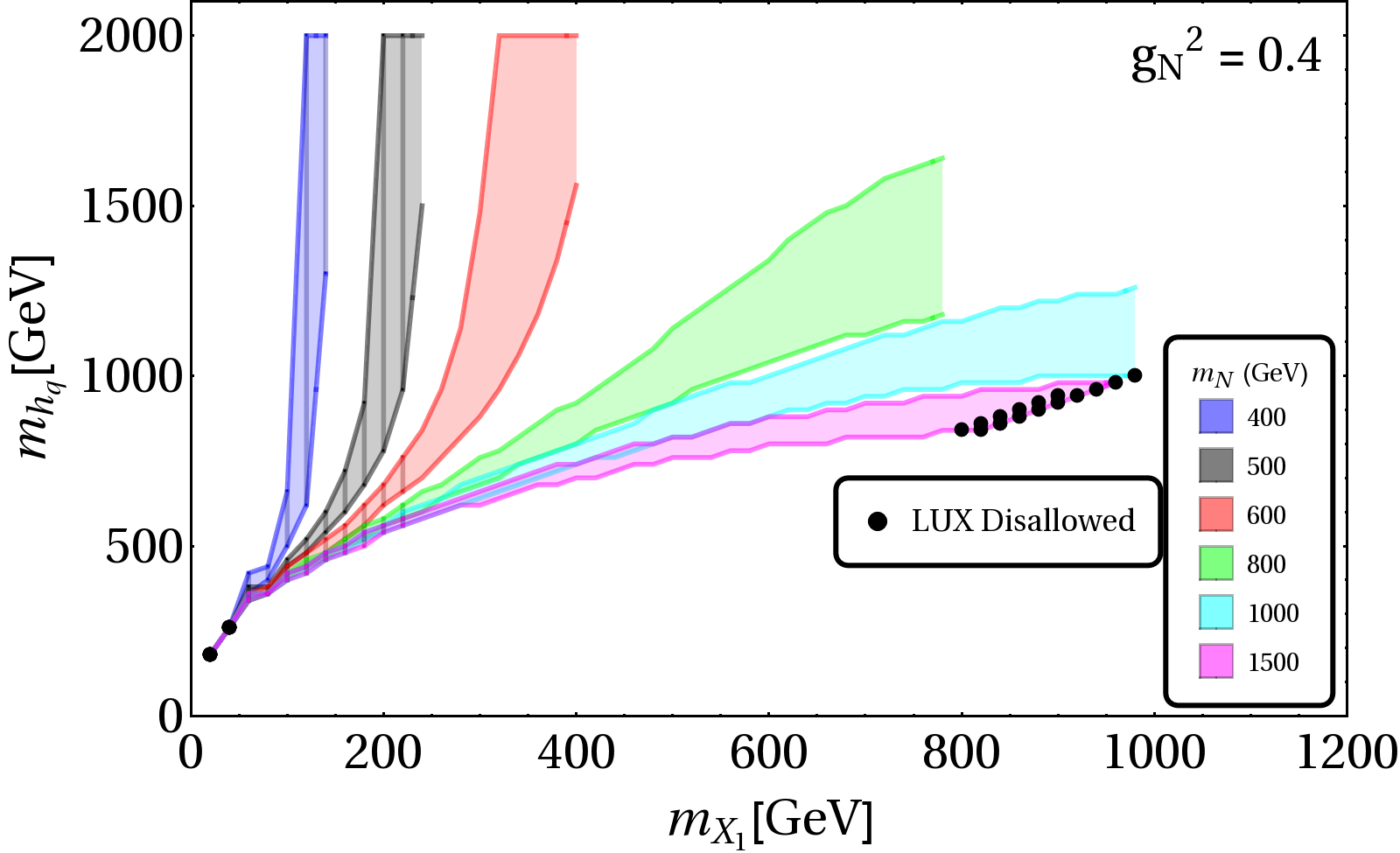}
\includegraphics[height=4.6cm]{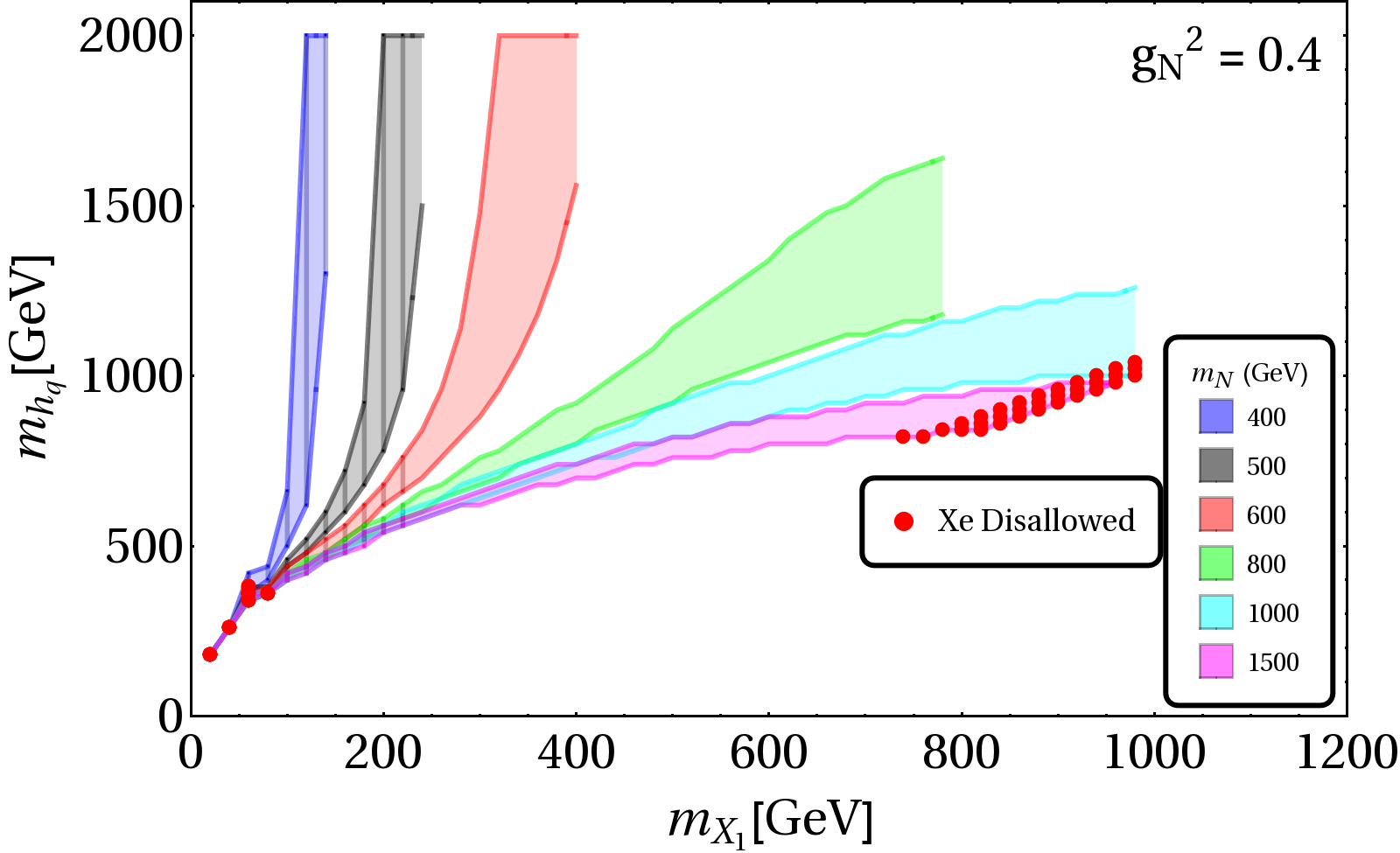}
$$
\caption{Direct search exclusion limits on relic density allowed parameter space of $m_{X_{1}}-m_{h_{q}}$ plane from LUX on the left hand side and from XENON nT on the right hand side. $g_N^2$=0.4 is chosen for illustration. Different allowed values of $m_N$ from relic density constraint are shown in different colours. VEV constraints are shown by hatched regions.}
\label{fig:DD}
\end{figure}

Variation of $\sigma_{DD}^{SI}$ with $m_{X_1}$ is summarized in Fig.~\ref{fig:m1-sigma}. In the upper panel, different colour coded regions correspond to different $g_N^2$ scanned over $m_{h_q}$ upto 2 TeV, allowed by relic density. Again, the maximum and minimum $m_{h_q}$ boundaries of each $g_N^2$ region correspond to maximum and minimum relic density for a given DM mass. $\sigma_{DD}$ has $s$-channel contribution from $h_q$ as shown in Fig.~\ref{fig:fd-direct}, where $m_{h_q}$ appears in the propagator and therefore, smaller $\sigma_{DD}$ indicates larger $m_{h_q}$. The t-channel Higgs mediation yields sub-dominant contribution to the direct search cross-section. The VEV disallowed points are shown by hatched regions which almost entirely discard low DM masses for small $g_N^2=\{0.1,0.2\}$. In the bottom-panel, different coloured regions depict relic density allowed parameter space for fixed $m_{h_q}$ and for $g_N^2:0.1-0.6$. LUX and XENON1T exclusion limits as well as the future prediction from XENONnT are shown in both the figure.  The main outcome of this analysis is to observe a huge region of parameter space still allowed by direct search limits. This can be attributed to the direct search cross-section of the DM guided by $m_{h_q}$ mediation in $s$-channel.\footnote{Recall that the t-channel Higgs mediation here is smaller due to $Z-Z^{'}$ mixing. This is in contrast to a large class of DM frameworks where direct search cross sections are dominated by large contributions from $t$-channel diagrams and therefore fall within the present exclusion bound.  On top of this, relic density for this particular DM is also obtained through exotic leptonic contribution (illustrated in earlier section), which do not take part in direct search process allowing the model to further survive direct search cuts.} 

As $\sigma_{DD}$ goes well below XENON1T and even XENONnT limit, it is important to look into the current projection of solar, atmospheric and diffuse supernovae neutrino background. The orange thick dotted line in Fig.~\ref{fig:m1-sigma} indicates the presence of 3-10 neutrino events \cite{neutrino-floor1, neutrino-floor2}. The grey shaded region below indicates the neutrino floor where DM signal will have to be separated from neutrino background. 


In Fig.~\ref{fig:DD} we show the $m_{h_q}$-$m_{X_1}$ relic density allowed parameter space for $g_N^2=0.4$, while highlighting the direct search exclusion bounds (LUX on left and XENON 1T on right). Direct search disallowed region corresponds to $s$-channel resonance ($m_{X_1}\sim m_{h_q}$), where the cross-section for the DM rises over the present exclusion limit.

\subsection{Co-annihilation of $X_1$ with $X_2$}
\label{sec:coann}

One of the important aspects of the DM under consideration is the co-annihilation effect, which further helps it to evade direct search bounds. There are two R-charge odd gauge bosons ($X_{1,2}$) present in this model. With $m_{X_2} > m_{X_1}$, the lightest one ($X_1$) is stable and serves as DM candidate as discussed so far. The heavier one ($X_2$) can contribute to the relic density of $X_1$ by co-annihilation to a pair of SM particles. The possible co-annihilation channels are already depicted in Figs.~\ref{fig:fd-ann1},~\ref{fig:fd-ann2},~\ref{fig:fd-ann3} (with $i=1,j=2$). Total effective cross-section, including co-annihilation, can be written as:

\be
\langle\sigma \; {\rm v}\rangle_{eff} = (\sigma\; {\rm v} )_{X_1X_1\rightarrow SM\;SM}~+~(\sigma\; {\rm v} )_{X_1X_2\rightarrow SM\;SM}\left(1+\frac{\Delta m}{m_{X_1}}\right)^{\frac{3}{2}}exp({-\Delta m}/{m_{X_1}}),
\label{eq:coann}
\ee
where $\Delta m=m_{X_2}-m_{X_1}$. Co-annihilation cross sections are significant for small $\Delta m$ $\left(m_{X_1} \sim m_{X_2}\right)$, or has a Boltzmann suppression otherwise (as evident from the last term of Eq.~\ref{eq:coann}). Following previous arguments, co-annihilation cross-sections are also computed at the threshold $s_0=(m_{X_1}+m_{X_2})^2$ considering only the dominant $s$-wave contribution. $(\sigma\; {\rm v} )_{X_1X_2\rightarrow SM\;SM}$ is obtained from Eq.~\ref{eq:sigma-v2}, by replacing $4m_{X_1}^2=(m_{X_1}+m_{X_2})^2$.

\begin{figure}[htb!]
$$
\includegraphics[height=8cm]{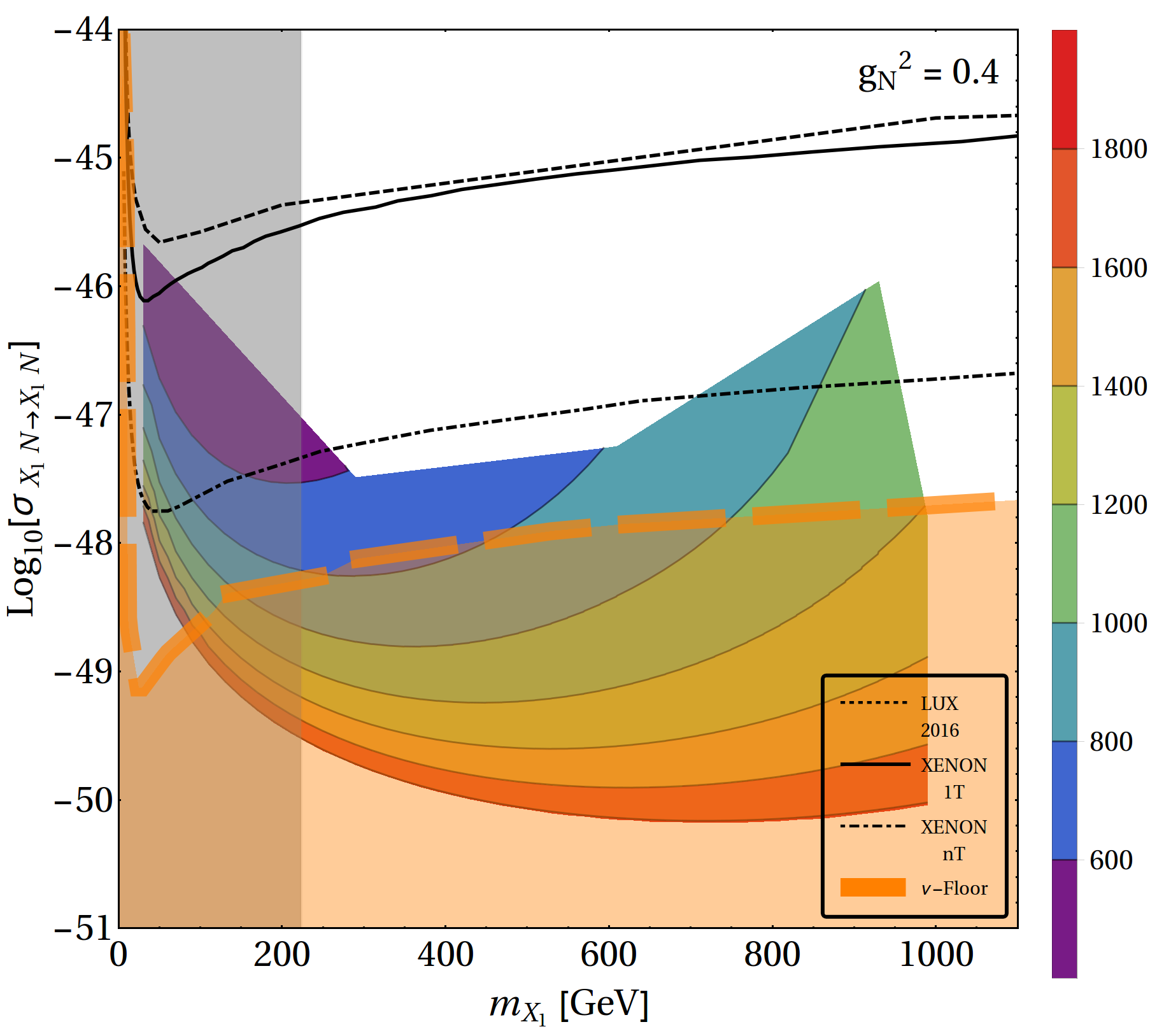}
$$
\caption{Direct search (spin independent) cross-section for relic density allowed parameter space for $X_1$ with co-annihilation taken into account. Different colour codes indicate different values of $m_{h_q}$. LUX 2016 and XENON 1T exclusion limits are shown along with XENONnT sensitivity by the dashed, solid and dot-dashed black lines respectively. Orange thick dashed line shows the background limit estimated from solar, atmospheric and diffuse supernovae neutrinos while the orange shaded region below indicates the neutrino floor.  VEV exclusion limit is represented by the grey region. }
\label{fig:coparamspace}
\end{figure}

 A scan of the relic density allowed parameter space of the model after incorporating co-annihilation effects is shown in Fig.~\ref{fig:coparamspace}. Here we have chosen a specific $g_N^2(=0.4)$ for illustration. Different values of $m_{h_q}$ are shown by different coloured regions. VEV exclusion limit is shown in the left hand side by the grey band, which puts a limit on the low DM mass $\sim$ 200 GeV. The orange thick dotted line points to the neutrino background (3-10 neutrino events), with the shaded region below depicting the neutrino floor. Noteworthy feature of Fig.~\ref{fig:coparamspace} is that the relic-density allowed parameter space lies safely much below the current direct search exclusion limits, a significant part of it going even lower than the XENONnT proposed limit. As has already been stated, this is a generic feature of DM models with co-annihilation, where such contributions (as $X_1X_2 \to ~SM$) help the DM to thermally freeze out, but do not take part in direct search processes.  To summarize, the model predicts that direct search of this DM is going to be difficult and direct search exclusions will not be able to rule-out a significant part of this model. This is one of the major outcomes of this analysis and was not pointed out in earlier analyses \cite{DiazCruz:2010dc,Bhattacharya:2011tr}.  



\section{Unified framework under $E(6)$}
\label{sec:unified}

Here we have analyzed in detail how the gauge symmetry of our interest, $SU(2)_L \otimes SU(2)_N \otimes U(1)_Y \otimes SU(3)_C$, can be realized within an unified scenario. To understand this, we have adopted $E(6)$ as the unified group that has already been outlined in \cite{DiazCruz:2010dc,Bhattacharya:2011tr}. We consider the following breaking patterns to achieve the low scale theory:

\begin{tabular}{lll}
$E(6)\xrightarrow{M_U} SU(3)_{L}\otimes SU(3)_{R}\otimes SU(3)_{C}$
$\xrightarrow{M_{I}^{'}}SU(2)_{L}\otimes U(1)_{L}\otimes SU(2)_{R}\otimes U(1)_{R}
\otimes SU(3)_{c}$ \\ \hskip 1 cm$ \xrightarrow {M_{I}}$ $SU(2)_{L}\otimes U(1)_{Y}\otimes SU(2)_{N}
\otimes SU(3)_{c}$  $ \xrightarrow{M_{I}^{N}} SU(2)_{L}\otimes U(1)_{Y}\otimes SU(3)_{c}$\\ \hskip 1 cm
$\xrightarrow{EWSB} SU(3)_{c}\otimes U(1)_{EM}$
\end{tabular}\\ 

In this section, we will first discuss the spontaneous symmetry breaking of $E(6)$ to the SM through multiple intermediate scales. We have computed the  beta-function coefficients~\cite{Chakrabortty:2009xm} for gauge coupling evolutions\footnote{ We have considered the running of gauge couplings only, and ignored the contributions from Yukawa couplings.} for different intermediate scales. In the process, we have also used the extended survival hypothesis (ESH) \cite{delAguila:1980qag} to allow  the minimal fine tuning in the scalar potential. All contributing field representations and the beta coefficients for different scales are appended below:


\begin{itemize}
\item {\bf From $M_{I^{'}} \to M_U$:}

$E(6)$ can be spontaneously broken to  $SU(3)_{L}\otimes SU(3)_{R}\otimes SU(3)_{C}\equiv \mathcal{G}_{333}$ through the vacuum expectation value (VEV) of  $650_H~(650^{'}_H)$ scalar keeping D-parity intact (broken). The scalar and fermion fields that contribute to the running of the gauge couplings $\{g_{3L},g_{3R},g_{3C}\}$, associated with $SU(3)_{L}, SU(3)_{R}, SU(3)_{C}$ respectively, are\footnote{We have relied on extended survival hypothesis (ESH) which tells that only those scalars are light which are participating in the symmetry breaking. Thus, even if we start with an unified $E(6)$ group, only relevant submultiplets are considered.}:
\begin{align*}
\begin{tabular}{lll}
$27_{F}= [\bar{3},3,1]+[3,1,3]+[1,\bar{3},3]$, \;\;
$650^{'}_{H}\supset[8,8,1]+[1,8,1]$,\;\;\\
$27_{H}\supset [\bar{3},3,1]$.
\end{tabular}
\end{align*}
Here we would like to mention, when this breaking occurs via VEV of $650_H$, then only $[8,8,1]$ contributes to the beta-functions and D-parity remains conserved. This can be understood intuitively as one can see, the fields contributing to the beta-coefficients in the running of $g_{3L}$ and $g_{3R}$ are identical. Thus $g_{3L}=g_{3R}$ will be maintained as long as  $\mathcal{G}_{333}$ is unbroken. We compute the  beta-coefficients as:

\begin{center}
$b_{3L}=7/2$,
$b_{3R}=7/2~(9/2)$,
$b_{3c}=-5$,
\end{center}
where the value of $b_{3R}$ in parenthesis denote the D-odd case.

\item {\bf From $M_{I^{}} \to M_{I^{'}}$:}

$SU(3)_{L}\otimes SU(3)_{R}\otimes SU(3)_{C}$ is further broken to 
$SU(2)_{L}\otimes U(1)_{L}\otimes SU(2)_{R}\otimes U(1)_{R} \otimes SU(3)_{C}(\equiv\mathcal{G}_{21213})$ through VEV of $(\bar{3},3,1)$.
The particles that contribute in the beta-functions are:
\begin{align*}
\begin{tabular}{lll}
$27_{F}$=$ [2,-1/2\sqrt{3},1,-1/\sqrt{3},1]+[2,1/2\sqrt{3},1,0,3]+[1,1/\sqrt{3},2,1/2\sqrt{3},1]
+[1,0,2,-1/2\sqrt{3},\bar{3}]$\\ +$[2,-1/2\sqrt{3},2,1/\sqrt{3},1]+[1,0,1,1/\sqrt{3},\bar{3}]+
[1,-1/\sqrt{3},1,0,3]+[1,1/\sqrt{3},1,-1/\sqrt{3},1]$ \\ [0.2cm]
$27_{H}\supset [2,-1/2\sqrt{3},1,-1/\sqrt{3},1]+[1,1/\sqrt{3},2,1/2\sqrt{3},1]+[2,-1/2\sqrt{3},2,1/2\sqrt{3},1]$\\[0.2cm]
$650_{H}\supset[1,0,3,0,1],$
\end{tabular}
\end{align*}
under the symmetry group $\mathcal{G}_{21213}$. The corresponding beta-coefficients are given by:
\begin{center}
$b_{2L}=-5/6$,
$b_{2R}=-5/6~(1/6)$,
$b_{LL}=b_{RR}=115/18$,
$\tilde{b}_{RL}=\tilde{b}_{LR}=1/9$,
$b_{3c}=-5$.
\end{center} 

\item {\bf From $M_{I}^{N} \to M_I$:}

$\mathcal{G}_{21213}$ is broken to $SM \otimes SU(2)_N$ in such a way that the hypercharge generator is originated as one of the linear combination of $U(1)_L$ and $U(1)_R$ generators. The fermion and scalar fields that contribute to the evolutions of the gauge couplings, $\{g_{2L},g_{2N},g_{1},g_{3C}\}$, are:

\begin{align*}
\begin{tabular}{lll}
$27_{F}= [2,1,1/6,3]+[1,1,-2/3,\bar{3}]+[1,2,1/3,\bar{3}]+[1,1,-1/3,3]+[2,2,-1/2,1]$\\+$[2,1,1/2,1]
+[1,1,1,1]+[1,2,0,1]$ \\ [0.2cm]
$27_{H} \supset [2,2,-1/2,1]+[2,1,1/2,1]+[1,2,0,1]$\\ [0.2cm]
$650^{'}_H \supset [1,3,0,1],$
\end{tabular}
\end{align*}

under the symmetry group $SU(2)_L \otimes SU(2)_N \otimes U(1)_Y \otimes SU(3)_C$. Corresponding beta-coefficients are:

\begin{center}
$b_{2L}=-5/6$,
$b_{2R}=1/6$,
$b_{1Y}=21/2$,
$b_{3c}=-5$,
\end{center}
respectively.

The sub-multiplet $ 650^{'}_H \supset [1,8,1] \supset [1,3,0,1]$ is considered for D-parity breaking case only. This $SU(2)_N $ triplet $\Delta\equiv [1,3,0,1]$ plays a crucial role in this analysis. The VEV of this triplet breaks the mass degeneracy of the $SU(2)_N$ gauge bosons, leading to one component dark matter scenario. Otherwise, for D-parity conserving case where this triplet is absent, this framework will have two degenerate dark matter candidates.

\item {\bf From $M_{EW} \to M_{I}^{N}$:}

In the next step, $[1,2,0,1]\in 27_{H}$, singlet under $SU(2)_{L}\otimes U(1)_{Y}\otimes SU(3)_{c}$, acquires VEV and causes spontaneous braking of $SU(2)_{N}$ leading to SM gauge symmetry. The fermion and scalar fields which participate in the evolution of gauge couplings are (under $SU(2)_{L},U(1)_{Y},SU(3)_{c}$):

\begin{align*}
\begin{tabular}{lll}
$27_{H}\supset$ $[2,-1/2,1]+[2,1/2,1]$\\
$27_{F}\supset$ $[2,1/6,3]+[1,-2/3,\bar{3}]+[1,1/3,\bar{3}]+[2,-1/2,1]+[1,1,1]$,
\end{tabular}
\end{align*}
and the corresponding beta-coefficients are:
\bea
\nonumber
b_{2L}=-3,
b_{1Y}=21/5,
b_{3c}=-7,
\eea
associated with $\{g_{2L},g_{1Y},g_{3C}\}$ respectively\footnote{$g_1$ is GUT normalized $U(1)_Y$ coupling and normalization factor 3/5 is used to fit within this framework.}. To summarize, the scalar and fermion representations, which participate in the symmetry breaking at different stages, are listed in Table~\ref{tab:gutparticles}.
\end{itemize}

\begin{table}[htb!]
\begin{center}
\begin{tabular}{|c| c| c| c|c|c|}
\hline
Particles & $E(6)$                         & $\left(SU(3)\right)^3$       & $\mathcal{G}_{21213}$              & $SM\otimes SU(2)_N$ & SM  \\ [0.5ex] 
\hline
	  &&&&&\\
          & $650^{'}_{H}$                  & $\left[8,8,1\right]$                                 & $\left[1,0,3,0,1\right]$     & $\left[1,3,0,1\right]$ & \\
          
          &                                &   $\left[1,8,1\right]$                               &                              &                         &\\
          & & & & & \\
          \cline{2-6}
          Scalar&&&&&\\
              & $27_{H}$                       & $\left[\bar{3},3,1\right]$      & $\left[2,-1/2\sqrt{3},1,-1/\sqrt{3},1\right]$     & $\left[2,2,-1/2,1\right]$ &
$\left[2,-1/2,1\right]$  \\
          &                                &                                 &    $\left[1,1\sqrt{3},2,1/2\sqrt{3},1\right]$                                                  & $\left[2,1,1/2,1\right]$                                         & $\left[2,1/2,1\right]$ \\
          
          &                                &                                 &   $\left[2,-1/2\sqrt{3},2,1/2\sqrt{3},1\right]$                                                 & $\left[1,2,0,1\right]$ & \\
          
          &                                &                                 &                                                    &    & \\         
\hline\hline

          &                                &                                 &                                                   &      & \\

          & $27_F$                         &  $\left[\bar{3},3,1\right]$     & $\left[2,-1/2\sqrt{3},1,-1/\sqrt{3},1\right]$                                                              &    $\left[2,1,1/6,3\right]$              & $\left[2,1/6,3\right]$\\

	  &   &  $\left[3,1,3\right]$       &  $\left[2,1/2\sqrt{3},1,0,3\right]$                                              &   $\left[1,1,-2/3,\bar{3}\right]$        & $\left[1,-2/3,\bar{3}\right]$ \\
	 
	  &   &  $\left[1,3,\bar{3}\right]$ &  $\left[1,1/\sqrt{3},2,1/2\sqrt{3},1\right]$                                              &  $\left[1,2,1/3,\bar{3}\right]$          & $\left[1,1/3,\bar{3}\right]$ \\

          Fermion&                          &      &  $\left[1,0,2,-1/2\sqrt{3},\bar{3}\right]$                                                 &   $\left[1,1,-1/3,3\right]$          & $\left[2,-1/2,1\right]$ \\
        
          &                          &      & $\left[2,-1/2\sqrt{3},2,1/\sqrt{3},1\right]$                                        &  $\left[2,2,-1/2,1\right]$        & $\left[1,1,1\right]$ \\
        
          &                          &      &  $\left[1,0,1,1/\sqrt{3},\bar{3}\right]$                                            & $\left[1,1,1,1\right]$       & \\
                
          &                          &      & $\left[1,-1/\sqrt{3},1,0,3\right]$                                         & $\left[1,2,0,1\right]$        & \\

          &                          &      & $\left[1,1/\sqrt{3},1,-1/\sqrt{3},1\right]$                                            &   & \\
          &&&&&\\
       
\hline 
\end{tabular}
\end{center}
\caption {The representations of scalar and fermion fields that contribute in the renormalization group evolutions of gauge couplings under different intermediate symmetries.} 
\label{tab:gutparticles}
\end{table}

\begin{figure}[htb!]
$$
\includegraphics[height=7cm]{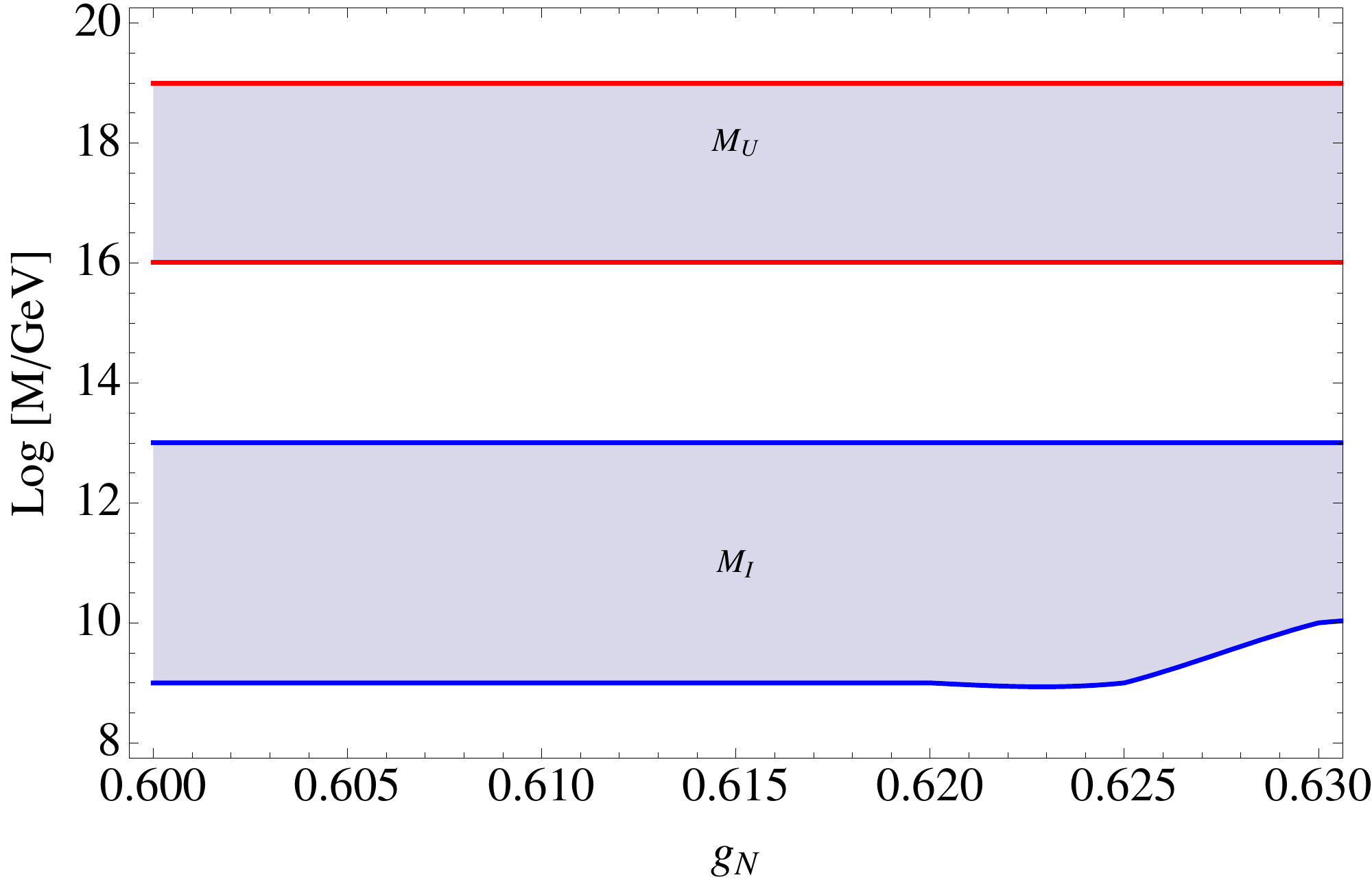}
$$
\caption{The unification within D-parity odd scenario has been encapsulated. The range of $SU(2)_N$ gauge coupling, $g_N$, and one of the intermediate scales ($M_I$)   have been explored with allowed unification scale ($M_U$), i.e., within $\{10^{18}-10^{19}\}$ GeV. Here, $M_{I}^{'}$ is considered to be degenerate with $M_I$, and $M_I^{N}$ is chosen to be TeV scale.} 
\label{fig:gnuni}
\end{figure}

\begin{figure}[htb!]
$$
\includegraphics[height=8cm]{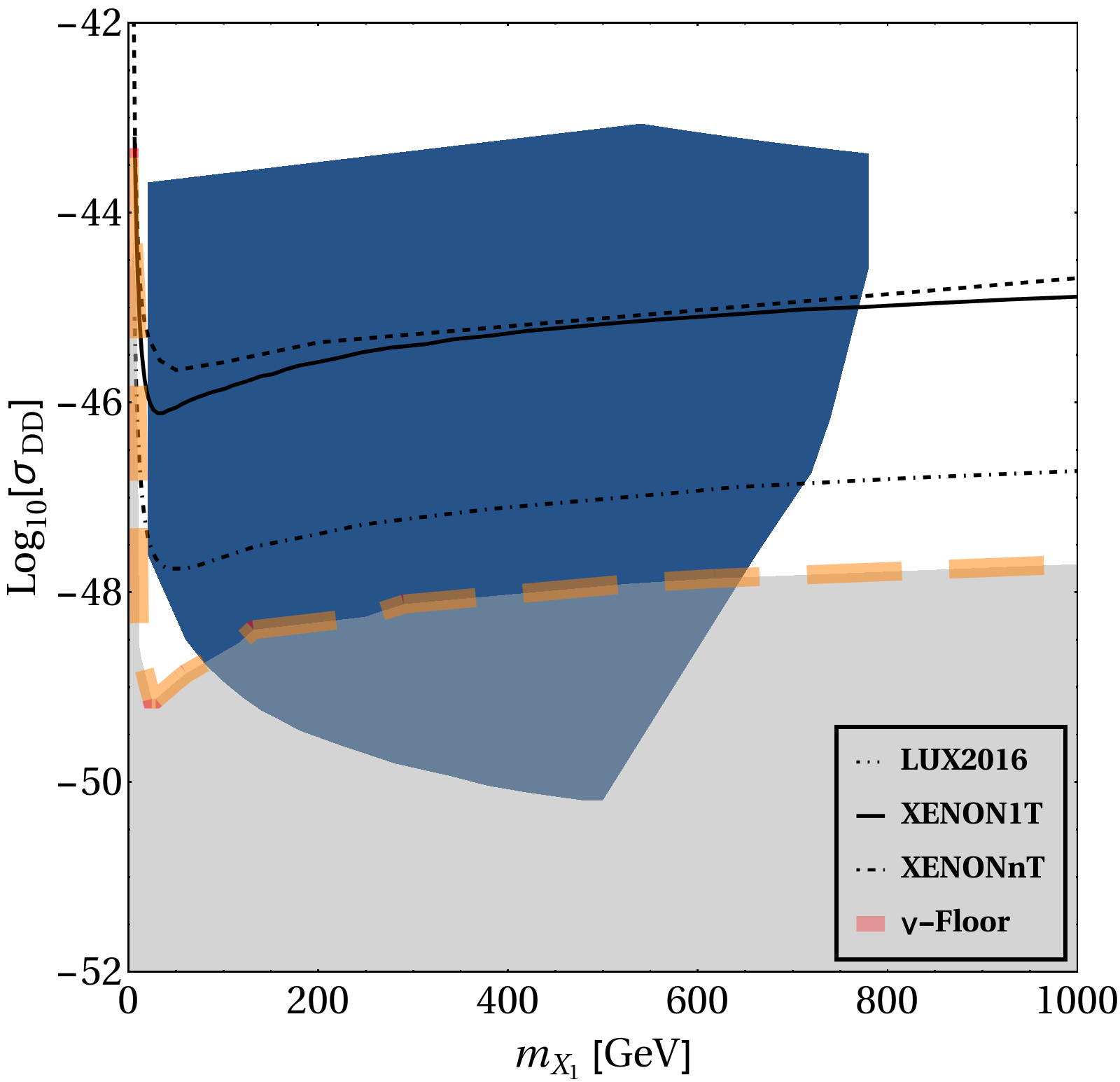}
$$
\caption{Relic density allowed parameter space for degenerate DM scenario ($m_{X_1}=m_{X_2}$) in spin independent direct search cross-section vs. DM mass plane. Limits from LUX 2016, XENON 1T and  XENONnT are shown in black dashed, solid black and black dot dashed lines respectively. The orange thick dashed line shows the background limit from solar, atmospheric and diffuse supernovae neutrinos while the grey shaded region below shows the neutrino floor.}
\label{fig:mx1mx2sig}
\end{figure}

In the breaking pattern of $E(6) \to SU(2)_N \otimes SM \to SM$, we have considered both D-parity conserving and non-conserving cases. For D-parity broken case, we will have one extra $SU(2)_N$ triplet at the TeV scale, and its respective parent multiplet will be present at high scales. We have introduced three intermediate symmetry groups in between $E(6)$ and SM. In principle, these intermediate scales are different. But we have noted, as almost all the fermions (27-dimensional) stick till $SU(2)_N$ breaking scale, beta coefficients for $U(1)_L$ and $U(1)_R$ are quite large ($\sim 115/18$). This results in rapid growth of the respective gauge couplings, making them non-perturbative. Thus, to avoid such catastrophe, we have assumed that $\mathcal{G}_{333}$ is broken directly to $SM\otimes SU(2)_N$, i.e. $M_{I}$ coincides with $M_{I}^{'}$. Hence, in our scenario, $M_{I}^{N}$ and $M_{I}$ are the free parameters. In order to estimate the freedom of choosing $g_N$ different from $g_{2L}$ in D-parity odd scenario, we restrict $M_{I}^{N}$ to be around TeV scale for phenomenological considerations, while unification scale to be within $[10^{19}:10^{16}]$ GeV. The allowed range of $g_{N}$ therefore turns to be $[0.60:0.63]$, as shown in Fig.~\ref{fig:gnuni}.


For D-parity even case, $g_N$ is no more a free parameter and equals to $g_{2L}$. Here  also, we find consistent solutions for unification with $M_{I}$ ranging between $[10^{19}:10^{16}]$ GeV.  The D-parity even case will lead to degenerate mass $\left(m_{X_1}=m_{X_2}\right)$, as the triplet VEV is absent. Thus, we will have a degenerate two component dark matter scenario which is itself an interesting possibility. Relic density allowed parameter space of degenerate DM scenario is shown in direct search plane in association with LUX, XENON1T bounds. It essentially indicates that DM mass has to be less than a TeV in order to satisfy them. However, the DM has to be as heavy as at least a TeV to satisfy constraints from $Z-Z^{'}$ mixing. Hence, the situation is phenomenologically nonviable and D-parity even case with degenerate DM doesn't work for this model. We have to live with only D-parity broken scenario with a single component DM even if the triplet VEV is small. This is one of the main outcomes of this elaborate discussion on the unification which was not indicated in earlier analyses \cite{DiazCruz:2010dc,Bhattacharya:2011tr}. We would also like to add here that the correlations between gauge coupling unification with the intermediate scales associated to $SU(2)_N$, is very specific to our chosen framework. If we fix the particle content of this model and keep the $SU(2)_N$ breaking scale at TeV range, a range of values of $g_N$, $M_I$ and $M_U$ become consistent with the unification picture, as shown in Fig.~\ref{fig:gnuni}. For the low-energy phenomenology of this model, the most important parameter of these three is $g_N$ and thus we have concentrated on the possible range of $g_N$. Therefore, fixing $g_N$ and varying $M_{I,U}$ in the unification-allowed range neither impact nor change our findings in the phenomenological analysis. We would also like to add that a SUSY version of $E(6)$  GUT, where all the 27 fermions and their superpartners are present at the low scale,  is severly constrained by flavour data, FCNC~\cite{King:2005jy} etc. Au contraire, as we are working within a non-supersymmetric framework, these  constraints can be evaded by considering the exotic scalars to be very heavy $\geq 15$ TeV, without altering our conclusion. As the Yukawa couplings are not playing any deterministic role in our analysis, we can assume them to be predominantly diagonal. We skip the detailed formulation of the Yukawa sector in this analysis as we do not aim to address the fermion masses and mixing in this paper. This ensures that our analysis will not be affected by the FCNC and flavour problems.



\subsection{Summary of dark matter phenomenology}
\label{sec:summary-dm}

\begin{figure}[htb!]
\centering
\includegraphics[height=7cm]{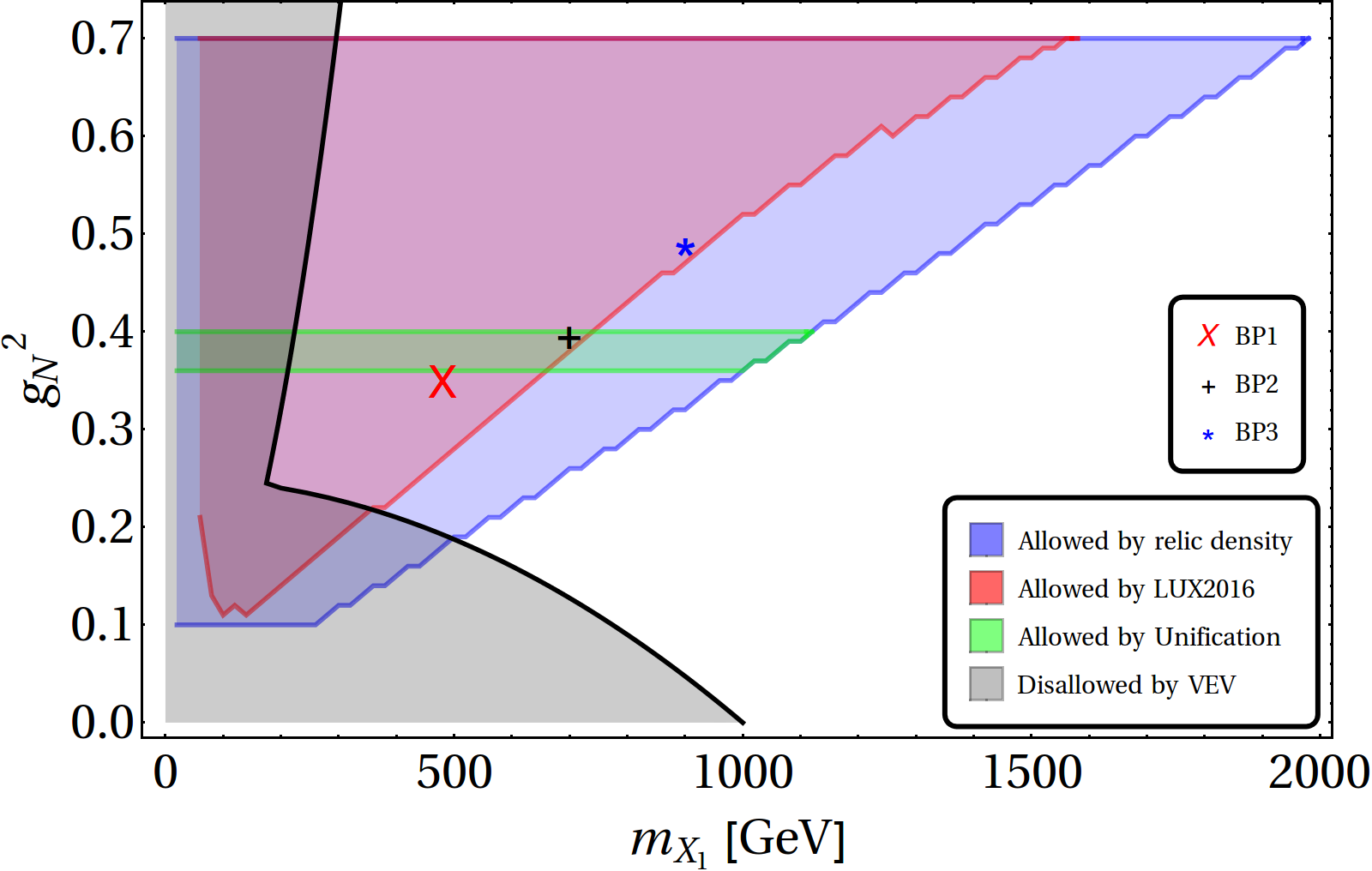}
\caption{Relic density (blue shaded), direct search (red shaded) and unification (green shaded) allowed parameter space in $g_N^2-m_{X_1}$ plane. Three chosen benchmark points BP1, BP2 and BP3 (see Tab.~\ref{tab:BP}) are also shown.} 
\label{fig:m1-gn-final}
\end{figure}

\begin{table}[htb!]
\begin{center}
\begin{tabular}{| c | c | c | c | c | c | c |c| }
\hline
Benchmark &  & $m_{X_{1}}$ & $m_{X_{2}}$ & $m_{h_q}$ & $m$ &  & $\sigma_{DD}$\\ [0.5ex] 
Points & $g_{N}$ & $(GeV)$ & $(GeV)$ & $(GeV)$ & $(GeV)$ & $\Omega h^2$ & $(\rm{cm}^2)$\\ [0.5ex] 
\hline\hline
&  &  & & & & &\\
$BP1$ & 0.59 & 480 & 1000 & 1200 & 600 & 0.09 & $10^{-50}$\\
&  &  &  & & & &\\
\hline
&  &  &  & & & &\\
$BP2$ & 0.63 & 700 & 1000 & 1160 & 860 & 0.1 & $10^{-49}$\\
&  &  & & & & &\\
\hline
&  &  &  & & & &\\
$BP3$ & 0.70 & 900 & 1000 & 1740 & 920 & 0.09 & $10^{-49}$\\
&  &  & & & & &\\
\hline
\end{tabular}
\end{center}
\caption{Three benchmark points (BP1, BP2, BP3) of the model are identified with input parameters $\{g_N,m_{X_1},m_{h_q},m=m_N=m_E \}$. Relic density and direct detection cross-sections for these points are also mentioned. We have set $m_{X_2}$ at 1000 GeV to obey maximum splitting scenario (see Eq.~\ref{eq:maxsplit}).} 
\label{tab:BP}
\end{table}

Before we proceed further, we would like to summarize the outcome of DM phenomenology in the light of relic density, direct search and unification constraints. As we have argued to stick to the D-parity broken scenario, $g_{N}$ is not identical to $SU(2)_L$ coupling, and in principle, a free parameter. Hence DM mass ($m_{X_1}$) is a function of $g_N$ and $SU(2)_N$ symmetry breaking VEVs ($\kappa_2,\delta_{1,2}$). In Fig.~\ref{fig:m1-gn-final}, on top of relic density allowed (blue) and VEV disallowed (grey) parameter space, we incorporate direct detection constraint provided by LUX in (red) region and unification constraint (green). For further phenomenological exploration in context of LHC, we have chosen three benchmark points (BP): BP1 ({\color{red} $\times$}), BP2 ({\color{black} $+$}) and BP3 ({\color{blue} $\star$}). The details of these benchmark points are listed in Tab.~\ref{tab:BP}.  The BPs have been chosen within DM mass ranging between 500 GeV$-$1 TeV. Two of the three gauge couplings $\left(g_N\right)$, chosen for the benchmark points (for BP1 and BP2), fall within the unification window but have negligible co-annihilation contributions due to large mass difference between $m_{X_1}$ and $m_{X_2}$. For BP3, a larger coupling is chosen, with considerable co-annihilation contribution. The choice of $m_{h_q}$ and $m$ follow a specific hierarchy: $m_{h_q}>m_{X_2}>m_N>m_{X_1}$, typically motivated from the collider aspect as we shall discuss in Sec.~\ref{subsec:OSD}.




\section{Collider Phenomenology at LHC}
\label{sec:collider}

Collider signatures of this model yield a lot of interesting possibilities with a number of beyond SM particles, which interact with the SM through the $SU(2)_N$ or Yukawa interactions. In this work, we will particularly highlight two multi-leptonic final states:
\begin{itemize}
\item Opposite sign dilepton plus a single jet and missing energy (OSD: $\ell^{+} \ell^{-}+1~j+{E_T}\!\!\!\!\!\!\!/~~$),
\item Hadronically quiet four lepton and missing energy  (HQ4l: $2\ell^{+} 2\ell^{-}+{E_T}\!\!\!\!\!\!\!/~~$).
\end{itemize}
Leptonic final states are interesting for study at LHC as the SM background contribution is relatively less. Unfortunately no excess has been reported from the existing data yet. In this section, we will show that the benchmark points chosen in Table ~\ref{tab:BP}, satisfy existing data from CMS with center-of-mass-energy $\sqrt{s}=13$ TeV and predict signal events in the aforementioned channels at $\sqrt{s}=14$ TeV. The excess of signal is subject to the final state event selection criteria and we will discuss corresponding SM background contributions in each of the cases accordingly.


\subsection{Simulation methodology} 
\label{subsec:selec_crit} 

We have implemented the model in {\tt CalcHEP}~\cite{Belyaev:2012qa} to generate the parton level events. These events are then fed into {\tt PYTHIA v.6}~\cite{Sjostrand:2006za} for showering and hadronization. We have used {\tt CTEQ6L}~\cite{Placakyte:2011az} parton distribution function with renormalization ($\mu_R$) and factorisation ($\mu_F$) scales set to the sub-process center-of-mass-energy ($\sqrt {\hat s}$).

All the exotic particles produced in the signal will eventually decay into SM leptons, jets and the DM $X_1$. This will give rise to different multilepton plus jet events,  subject to the choice of production processes, along with missing energy. To mimic the experimental environment at LHC, we use the following identification criteria:

\begin{itemize}
\item {\em Lepton ($\ell=e,\mu$)}: They are identified with minimum transverse momentum ($p_T$) of 20 GeV, and with pseudorapidity $|\eta|<$ 2.5 to identify them in the central region of the detector. Two leptons are treated as isolated objects if their mutual separation satisfy $\Delta R=\sqrt{(\Delta \eta)^2+(\Delta \phi)^2}\geq 0.2$. Lepton-jet separation must satisfy $\Delta R \geq 0.4$. $\tau$ leptons are difficult to observe in electromagnetic and muon calorimeters due to their shorter lifetime and so they are not usually classified into lepton category.

\item {\em Jet ($jet$)}: Due to $SU(3)$ interactions, partons form hadrons after emerging out of the collision vertex. These hadrons then cluster to form jets. We have used   {\tt PYCELL}, the cone jet formation algorithm built within {\tt PYHTIA}, to define the clustered hadrons as $jet$s. The detector is assumed to span $|\eta| \le5$ and is segmented in 100-$\eta$ and 70-$ \phi $ bins. The minimum transverse energy $E_T$ of each cell is taken as $0.5$ GeV, while we require $E_T\ge 2 \rm~{GeV}$ for a cell to act as a jet initiator. All the partons within $\Delta R$ = 0.4 from the jet initiator cell are included in the formation of the jet, and we require  $E_T \ge 20 \rm~{GeV} $ for a group to be considered as a jet. For isolation of the jets from the unclustered objects (see below), we also impose a cut of $\Delta R >0.4$.

\item {\em Unclustered Objects}: All the other final state particles, which are not isolated leptons and separated from jets, are considered as unclustered objects. This clearly means all the particles (electron/photon/muon) with $0.5< E_T< 10$GeV and $|\eta|< 5$ and jets with $0.5< E_T< 20$~GeV and $|\eta|< 5$, which leave their presence in the detector, are considered as unclustered objects.

\item {\em Missing energy} (${E_{T}}\!\!\!\!/~$) : Though DMs produced in the decay chain of exotic particles will be missed in the detector, their transverse momentum can be estimated from the momentum imbalance in the associated visible particles (leptons and jets). This missing energy is thus defined as:
\be
E_{T}\!\!\!\!/ = (p_T)_{mis}=-(p_T)_{vis},~
(p_T)_{vis}=\sqrt{(\sum_{\ell,j} p_x)^2+(\sum_{\ell,j} p_y)^2}\,.
\ee
Note that the negative sign in the definition of missing energy do not carry any significance and we will always refer to the absolute value of the visible transverse momentum as missing energy. We also note that missing energy identification takes the unclustered objects, which do not fall into the lepton or $jet$ category as defined by detector sensitivity, into account. 
\end{itemize}

SM background for the corresponding signal events play a crucial role in identifying the signal significance and discovery potential of the underlying model. We generate dominant SM backgrounds using {\tt MADGRAPH} \cite{Alwall:2014hca} and shower them in {\tt PYTHIA} \cite{Sjostrand:2006za}. We have appropriately used $K$ factor to match the next-to-leading-order (NLO) cross-sections for the processes contributing to SM background. The signal cross-section, which we have evaluated at tree level, may vary within 10\% due to the choice of jet energy scale, parton distribution functions, smearing effects of leptons and jets, and different jet formation criteria.  NLO results for the signal events, particularly involving coloured particles like $h_q$, may have larger effects than the detector simulation criteria. 

\subsection{Opposite sign dilepton signal}
\label{subsec:OSD}

\begin{figure}[htb!]
$$
\includegraphics[height=3.5cm]{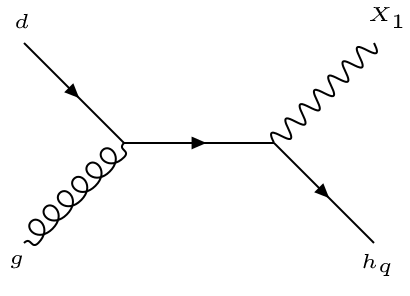}
\includegraphics[height=3.8cm]{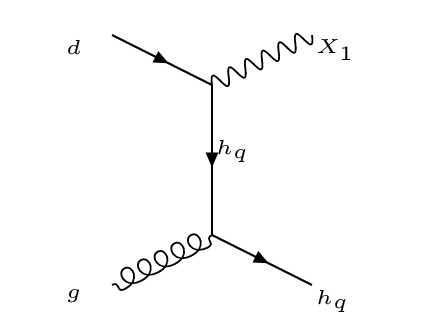}
$$
$$
\includegraphics[height=4.2cm]{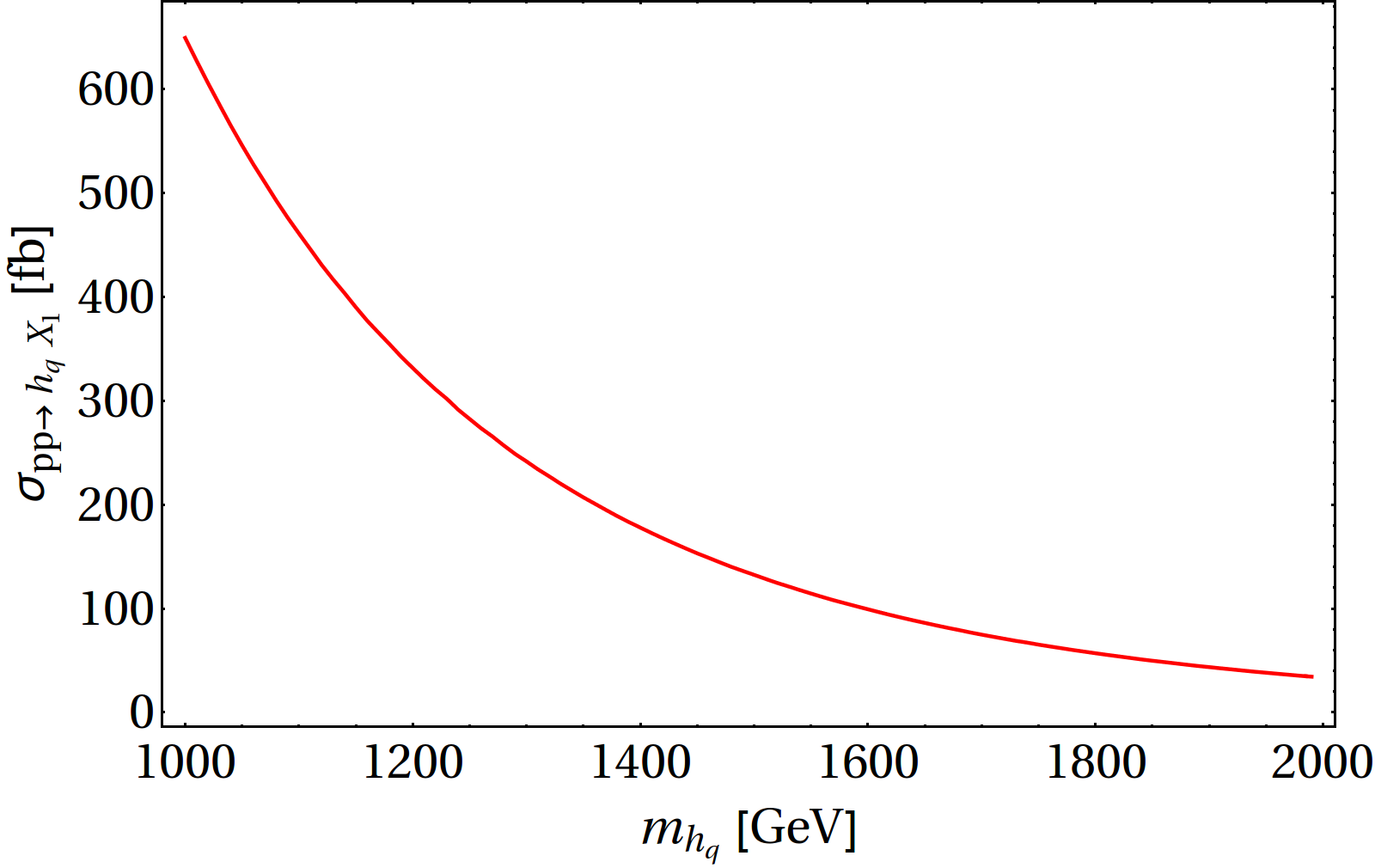} 
$$
\caption{Top panel: Feynman diagrams for producing $pp \to h_q X_1$ at the LHC. Bottom panel: Production cross-section of $pp \to h_q X_1$ is plotted as a function of $m_{h_q}$ (GeV) at LHC for $\sqrt{s}= 14$ TeV, keeping the DM mass fixed at $m_{X_1}=$ 700 GeV and $g_N^2=0.4$.}
\label{fig:prod_hq}
\end{figure}

\begin{figure}[htb!]
$$
\includegraphics[height=3.9cm]{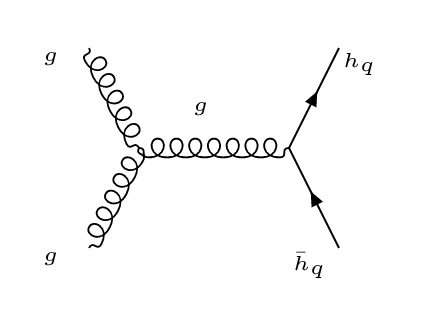} 
\includegraphics[height=3.9cm]{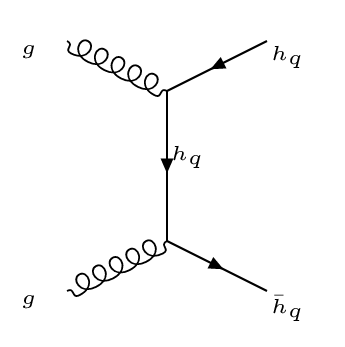}
\includegraphics[height=3.7cm]{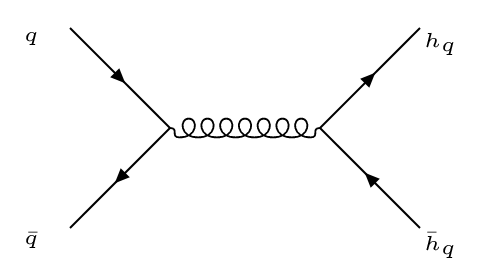}
$$
$$
\includegraphics[height=4.2cm]{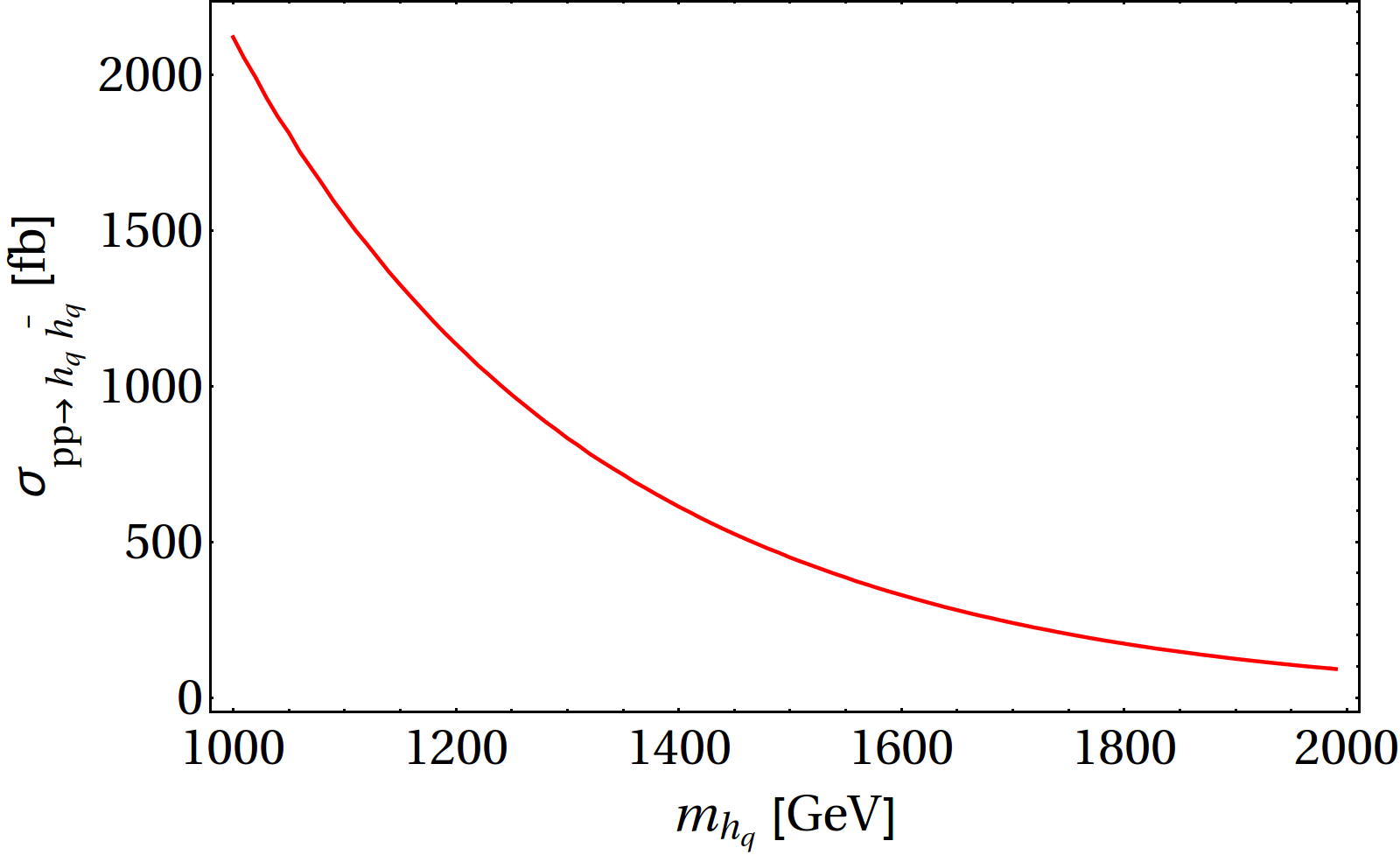} 
$$
\caption{Top panel: Feynman diagram for producing $pp \to h_q \bar{h}_q$ at the LHC. Bottom panel: Production cross-section of $pp \to h_q \bar{h}_q$ is plotted as a function of $m_{h_q}$ (GeV) at LHC for $\sqrt s= 14$ TeV with $g_N^2=0.4$.}
\label{fig:prod_hqhq}
\end{figure}

LHC being a proton-proton collider, exotic particles with colour charge ($h_q$ in our case) can be produced copiously. Opposite sign dilepton (OSD) with $1j+{E_{T}}\!\!\!\!/~$ arises in the model from two different production processes: (i) $pp \to h_q X_1$, through the Feynman graphs shown in the top panel of Fig.~\ref{fig:prod_hq}, and (ii) $pp \to h_q \bar{h_q}$, as shown by the Feynman graphs in the top panel of Fig.~\ref{fig:prod_hqhq}. Variation in production cross-sections of these processes with exotic quark mass $m_{h_q}$ is shown in the bottom panels of Fig.~\ref{fig:prod_hq} and Fig.~\ref{fig:prod_hqhq} respectively. DM mass ($m_{X_1}$) is fixed at 700 GeV and the $SU(2)_N$ gauge coupling is chosen as $g_N^2=0.4$ for these calculations. 

\begin{figure}[htb!]
$$
\includegraphics[height=3.9cm]{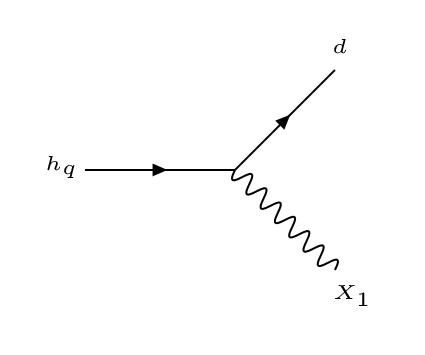}
\includegraphics[height=3.8cm]{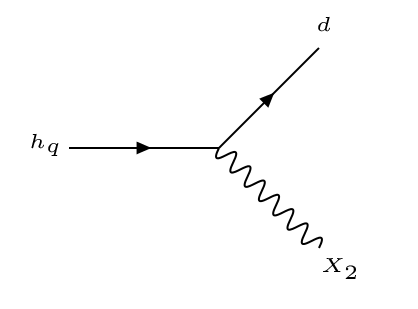}
$$
$$
\includegraphics[height=4.2cm]{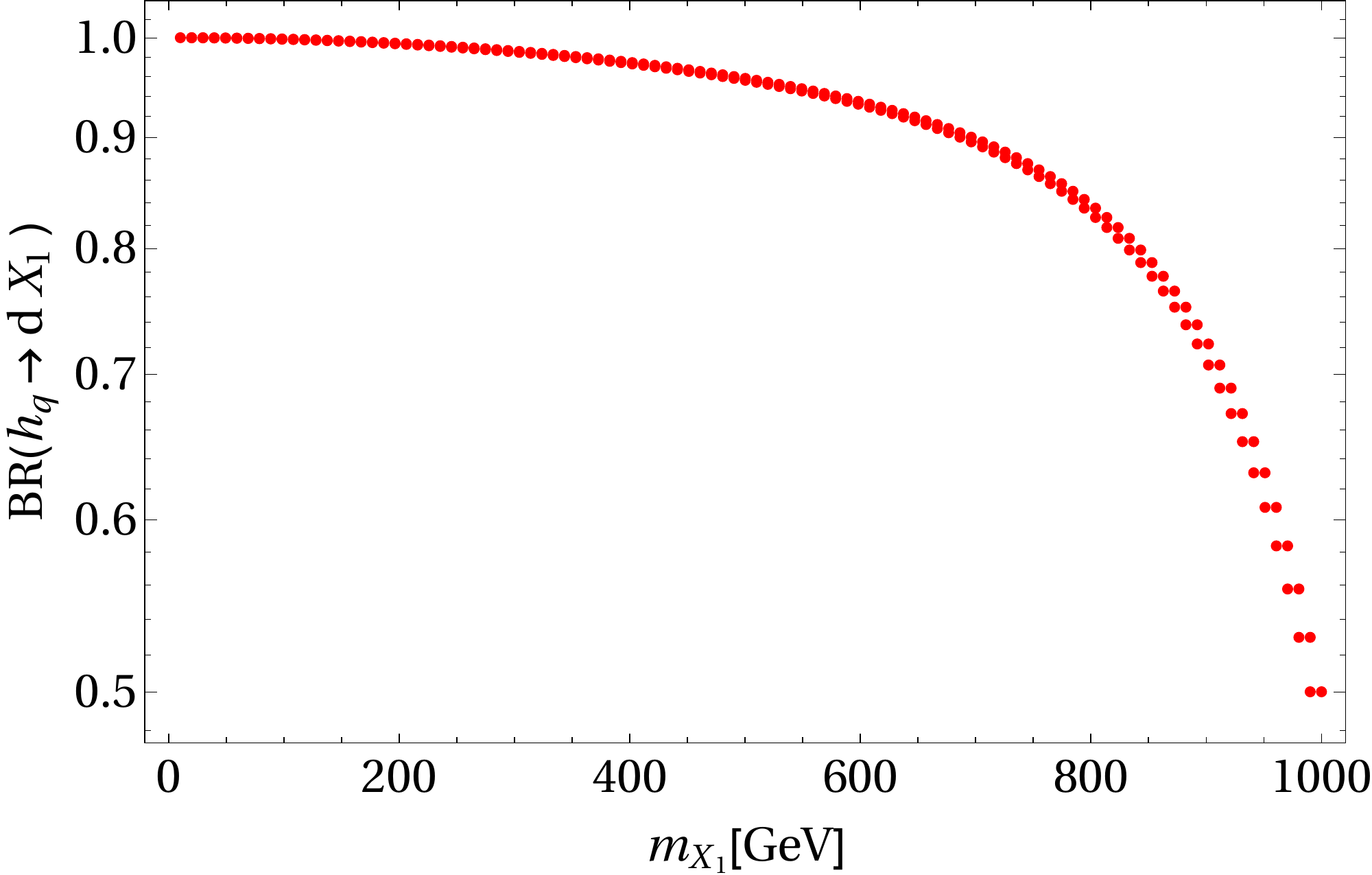}\hspace{1cm}
\includegraphics[height=4.2cm]{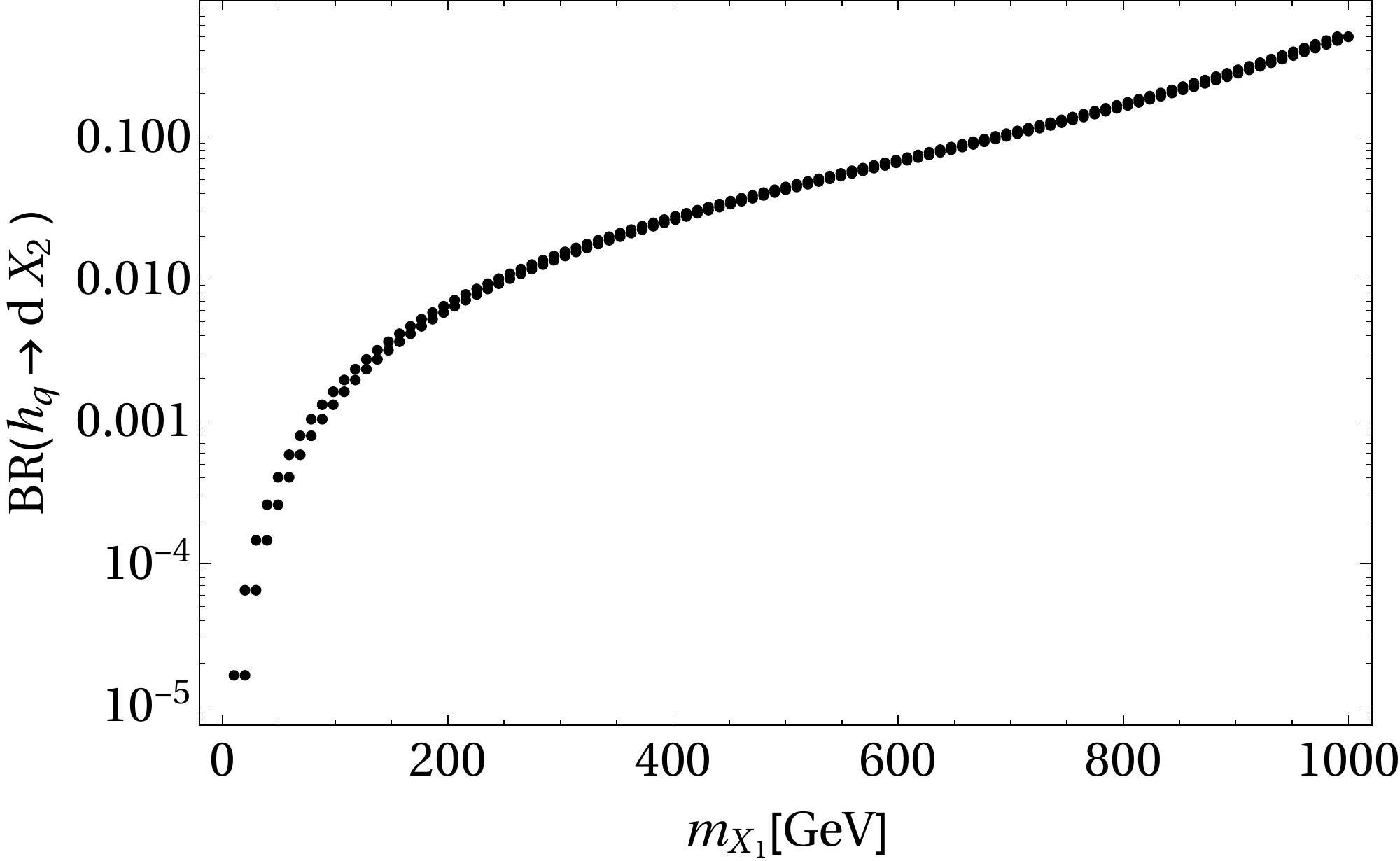}
$$
\caption{Top panel: Feynman diagrams for possible decay modes of $h_q$. Bottom panel: Decay branching fractions of $h_q$ as a function of $m_{X_1}$, Br($h_q \to d X_1$) on the left and Br($h_q \to d X_2$) on the right. We have chosen $m_{h_q}=1160$ GeV and $m_{X_2}=1000$ GeV for illustration.}
\label{fig:decay_hq}
\end{figure}


\begin{figure}[htb!]
\centering
\includegraphics[height=2.4cm]{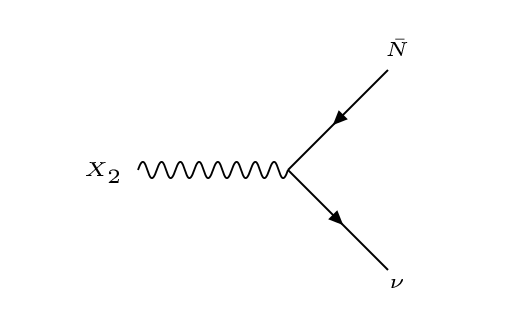}
\includegraphics[height=2.4cm]{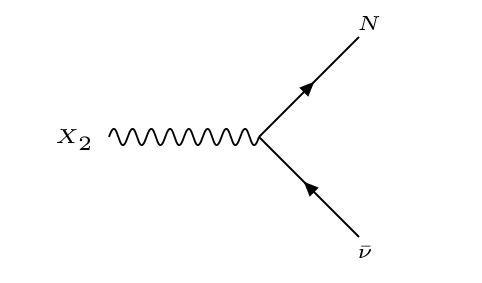}
\includegraphics[height=2.4cm]{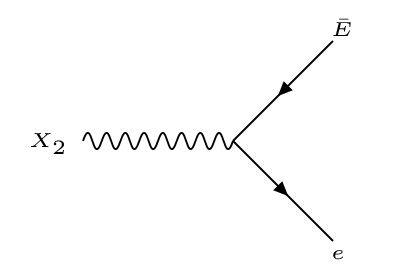}
\includegraphics[height=2.4cm]{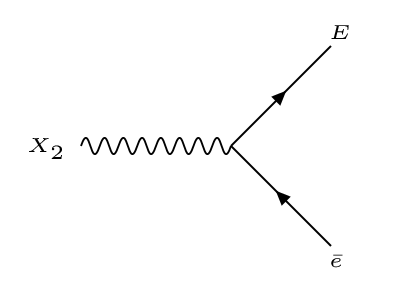}
\caption{Feynman diagrams showing decays of $X_2$  to $N\;\bar{\nu},\;E^+e^-$ and their conjugate final states.}
\label{fig:X2-decay}
\end{figure}

\begin{figure}[htb!]
\centering
\includegraphics[height=2.6cm]{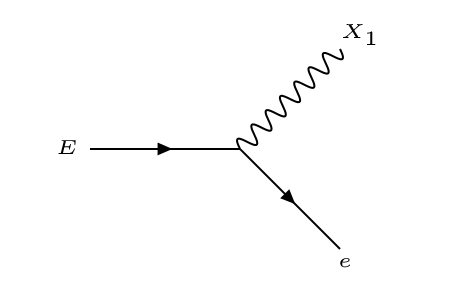}
\includegraphics[height=2.6cm]{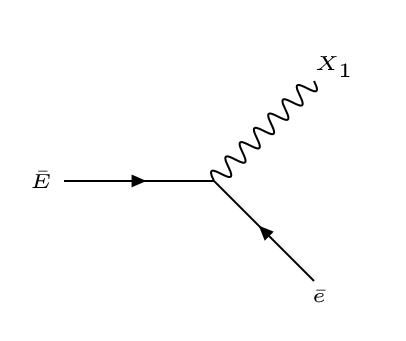}
\includegraphics[height=2.6cm]{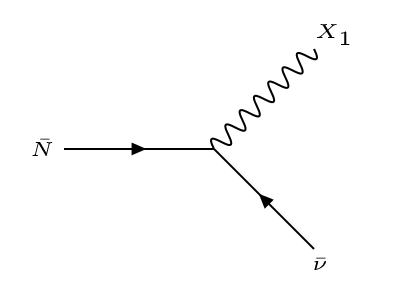}
\includegraphics[height=2.6cm]{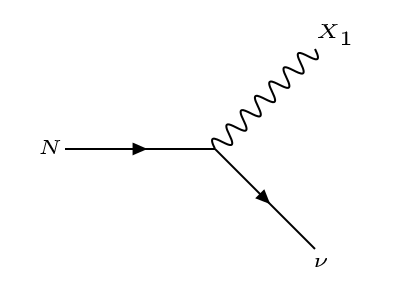}
\caption{Feynman diagrams showing decays of $E$ and $N$  to $X_1$ and SM particles.}
\label{fig:EN-decay}
\end{figure}

Now $h_q$, once produced through $SU(2)_N$ gauge interactions, can only decay via $h_q \to  X_{1,2}~d$ (Fig.~\ref{fig:decay_hq}), depending on the availability of phase space. The branching fractions are plotted with respect to $m_{X_1}$ in the bottom panel of Fig.~\ref{fig:decay_hq}. Given a specific $m_{X_2}$ and $m_{h_q}$, the branching fraction to $X_1~d$ reduces and $X_2~d$ increases with $m_{X_1}$. In the limiting case, when $m_{X_1}=m_{X_2}$, the branching fractions of $h_q$ are same for both the final states. When $h_q$ decays to $d X_1$, the production of $pp \to h_q X_1$ will end in single jet plus missing energy events and $pp \to h_q \bar{h_q}$ will end with two jets plus missing energy events, which are standard signatures of many DM models including those of minimal scalar singlet extensions. SM background for such final states are also huge and hence reducing the background for a possible excess in the signal is difficult if not impossible. However, $h_q\to dX_2$ may yield some interesting possibilities if we choose the following mass hierarchy in the spectrum: 
\be
m_{h_q}>m_{X_2}>m=m_E=m_N>m_{X_1}
\label{eq:hierarchy}\,.
\ee
$X_2$ can then decay to $ N\bar{\nu},\; E^+e^-$ (see Fig.~\ref{fig:X2-decay}). The branching fraction of $X_2$ to neutrino and charged lepton final states are equal in the limit of $m_N=m_E$, which is assumed throughout the analysis. Now, exotic leptons will decay to DM with 100\% branching fraction: $N \to \nu X_1, ~ E \to e X_1$ as shown in Fig.~\ref{fig:EN-decay}. Hence, $pp \to h_q X_1, ~ h_q \to d X_2, ~ X_2 \to E^+e^- $ will lead to opposite sign same flavour dilepton, a soft jet, and missing energy in the final state. Similarly, $pp \to h_q \bar{h_q}, ~ h_q \to d X_2, ~ X_2 \to E^+e^- $ will also mostly yield opposite sign same flavour dilepton, two jets, and missing energy events. In the second case, many of those two jet events qualify for single jet case as the jets are soft due to their production in association with a sufficiently heavy particle $X_2$, which carries away most of the momentum in the decay chain.
We will perform collider analysis only for the selected benchmark points BP1, BP2, and BP3 (Tab.~\ref{tab:BP}). Branching fractions of $h_q$ at those benchmark points are specified in Tab.~\ref{tab:branching}.

\begin{table}
\begin{center}
\begin{tabular}{|c|c|c |c |c|c|}
\hline
Benchmark Points & $m_{X_1}$ & $m_{X_2}$ & $m_{h_q}$ & Br($h_q\rightarrow dX_1$)& Br($h_q\rightarrow dX_2$)  \\ [0.5ex]
\hline

$BP1$ & 480 & 1000 & 1200 & 94.8 & 5.2 \\
\hline
$BP2$ & 700  & 1000 & 1160 & 89.7 & 10.3 \\
\hline
$BP3$ & 900  & 1000 & 1740 & 57.7 & 42.3 \\
\hline
\end{tabular}
\end{center}
\caption{Branching ratio of $h_q \to d X_{1,2}$ for three benchmark points: BP1, BP2 and BP3.}
\label{tab:branching}
\end{table}

\begin{figure}[htb!]
\centering
\includegraphics[height=4.4cm]{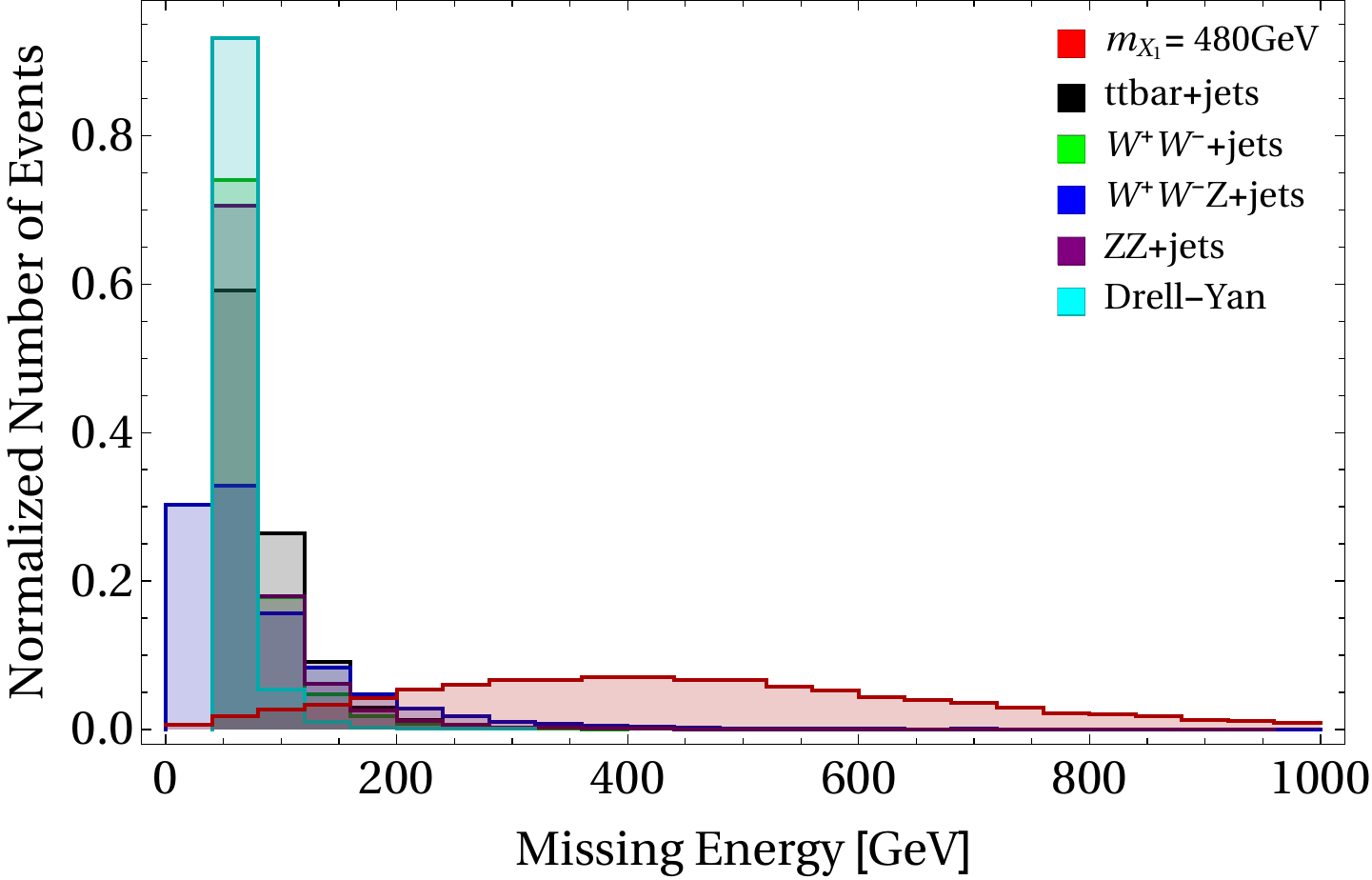}
\includegraphics[height=4.4cm]{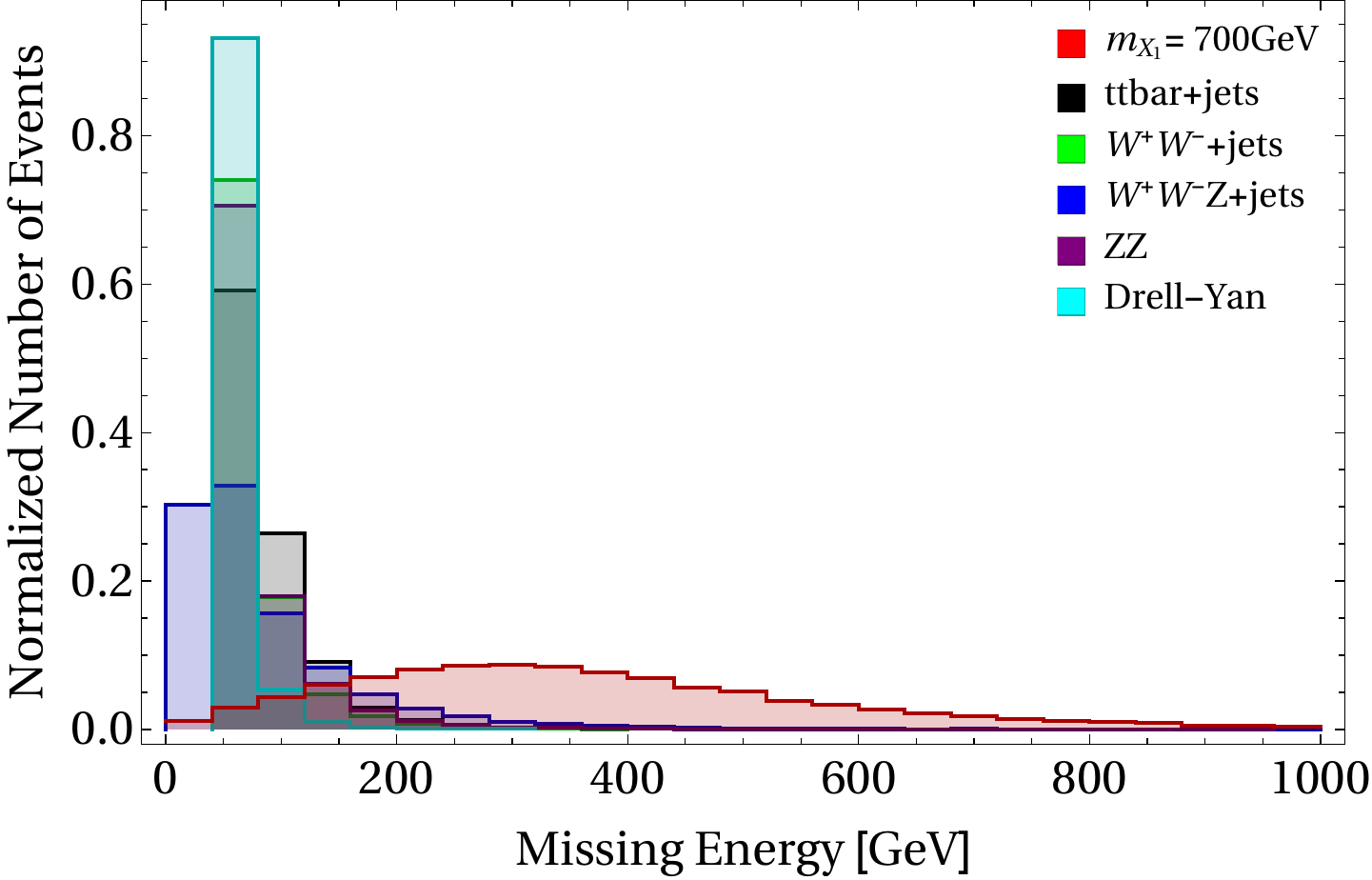}
\includegraphics[height=4.4cm]{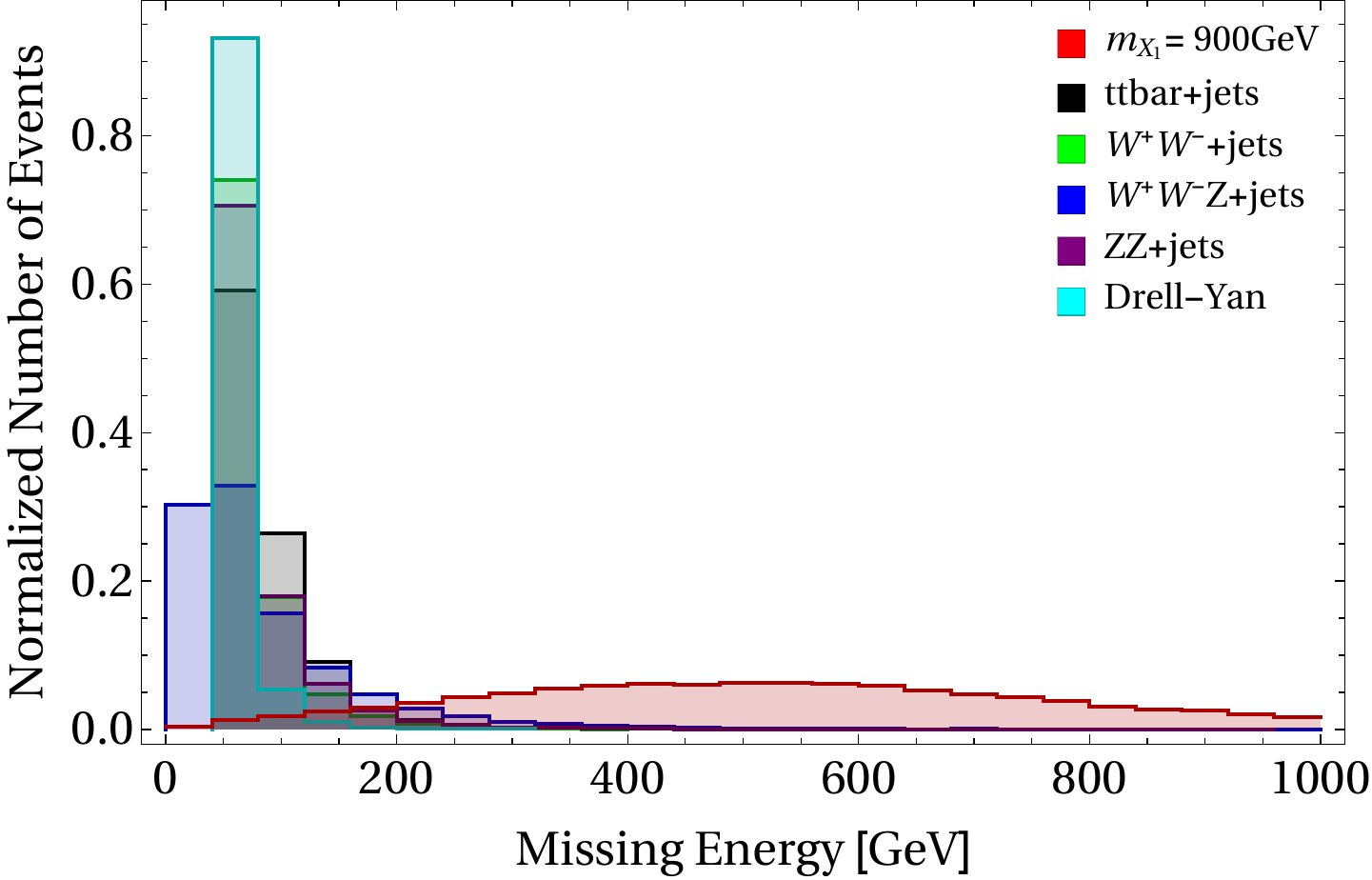}
 \caption{Missing energy distribution for ($\ell^{+}\ell^{-}+1j+{E_{T}}\!\!\!\!/$) events at LHC from $pp \to h_q X_1$ production, for the benchmark points with dominant SM background. Top left: BP1, Top right: BP2, Bottom: BP3.}
 \label{fig:OSD-MET}
\end{figure}

We have identified the dominant SM processes which can mimic OSD signal and consider them as backgrounds. Such processes are $t\bar{t}$+jets,$~WW$+jets,~$WZ$+jets,~$ZZ$+jets, and also the $Drell$-$Yan$. The cross-section of the background is much larger than that of the signal and one needs to judicially choose the cuts on the final state observables to retain the signal and reduce the background. Missing energy ($E_{T}\!\!\!\!/~$) signature of the signal is characteristically different from that of the SM background. For background, $E_{T}\!\!\!\!/~$ dominantly comes from neutrinos and energy mismatch (due to smearing and detector inefficiency).  $E_{T}\!\!\!\!/~$ distributions for the signal for the BPs are plotted along with the dominant SM backgrounds in Fig.~\ref{fig:OSD-MET}. $E_{T}\!\!\!\!/~$ peaks at a larger value for the signal and at a smaller value for the background, thus allowing us to distinguish between the model(signal) from the background. This is done by putting a sufficiently large     
$E_{T}\!\!\!\!/~$.

\begin{figure}[htb!]
$$
\includegraphics[height=5cm]{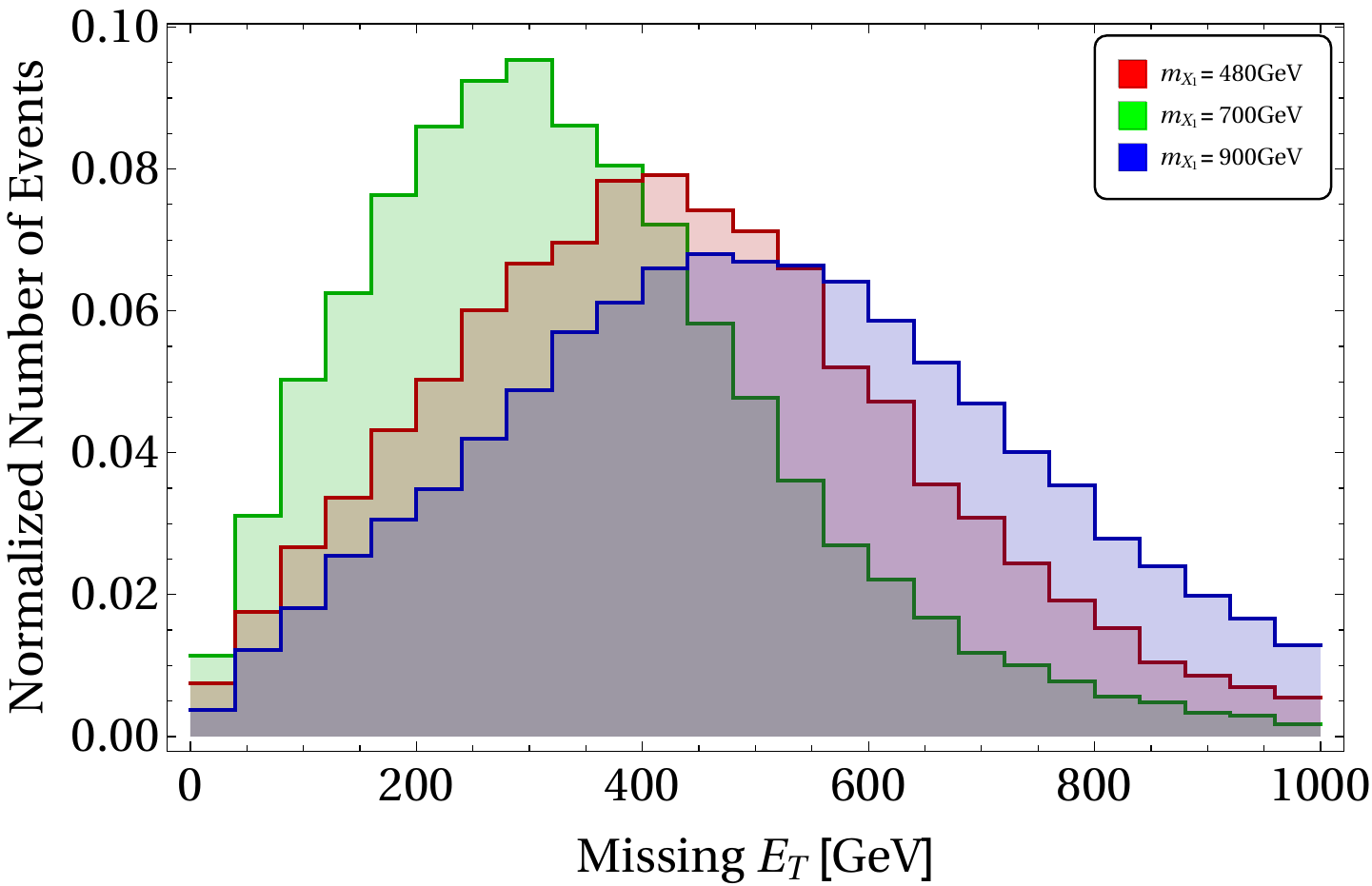}
\includegraphics[height=5cm]{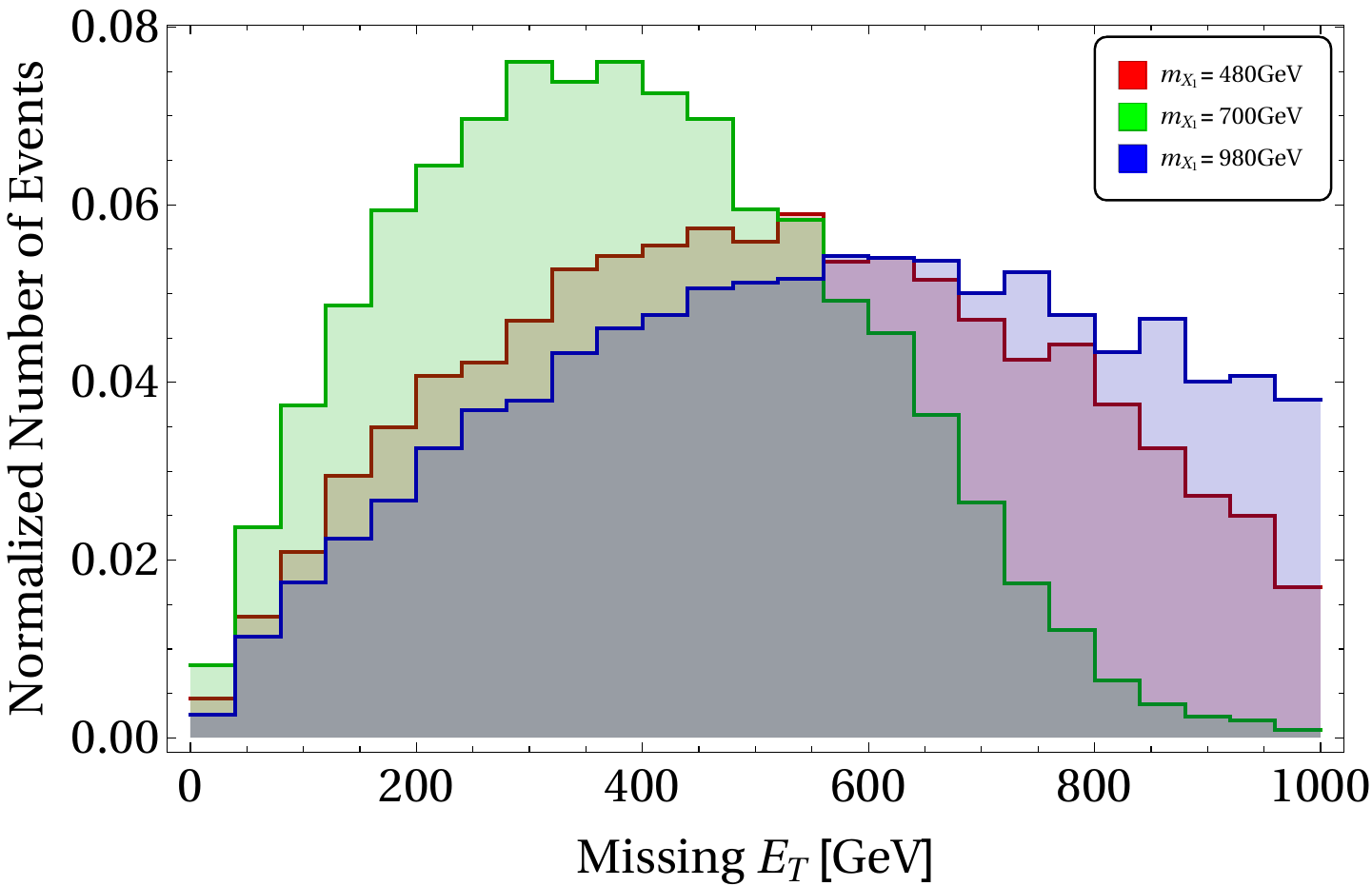}
$$
\caption{Missing energy (${E_{T}}\!\!\!\!/$) distribution in OSD events for benchmark points (BP1, BP2, BP3). On the left, we consider only $pp \to h_q X_1$ production while on the right panel $pp \to h_q \bar{h}_q$ production is considered.}
\label{fig:OSD-MET-sig}
\end{figure}

Another interesting feature emerges from the ${E_{T}}\!\!\!\!/$ distribution when they are compared for the BPs as in Fig.~\ref{fig:OSD-MET-sig}. The peak of the distribution depends on the momentum carried away by the DM. For $pp \to h_qX_1$ (left panel of Fig.~\ref{fig:OSD-MET-sig}), as $m_{h_q}>m_{X_1}$, larger share of the momentum will be carried by the exotic quark $h_q$. Hence, the peak of  distribution will be governed by $\Delta m=m_{h_q}-m_{X_1}$ and will shift to larger value with larger $\Delta m$. $(\Delta m)_{BP3}>(\Delta m)_{BP1}>(\Delta m)_{BP2}$ is clearly therefore reflected in the ${E_{T}}\!\!\!\!/$ distributions. The same is true for the right panel of Fig.~\ref{fig:OSD-MET-sig}, where ${E_{T}}\!\!\!\!/$ distribution is plotted for contributions only from $pp \to h_q\bar{h_q}$. This feature can help to differentiate between the benchmark points with different $\Delta m$. This can also give an idea about the mass of the exotic quark ($m_{h_q}$), given a knowledge of DM mass ($m_{X_1}$). 



\begin{figure}[htb!]
$$
\includegraphics[height=6cm]{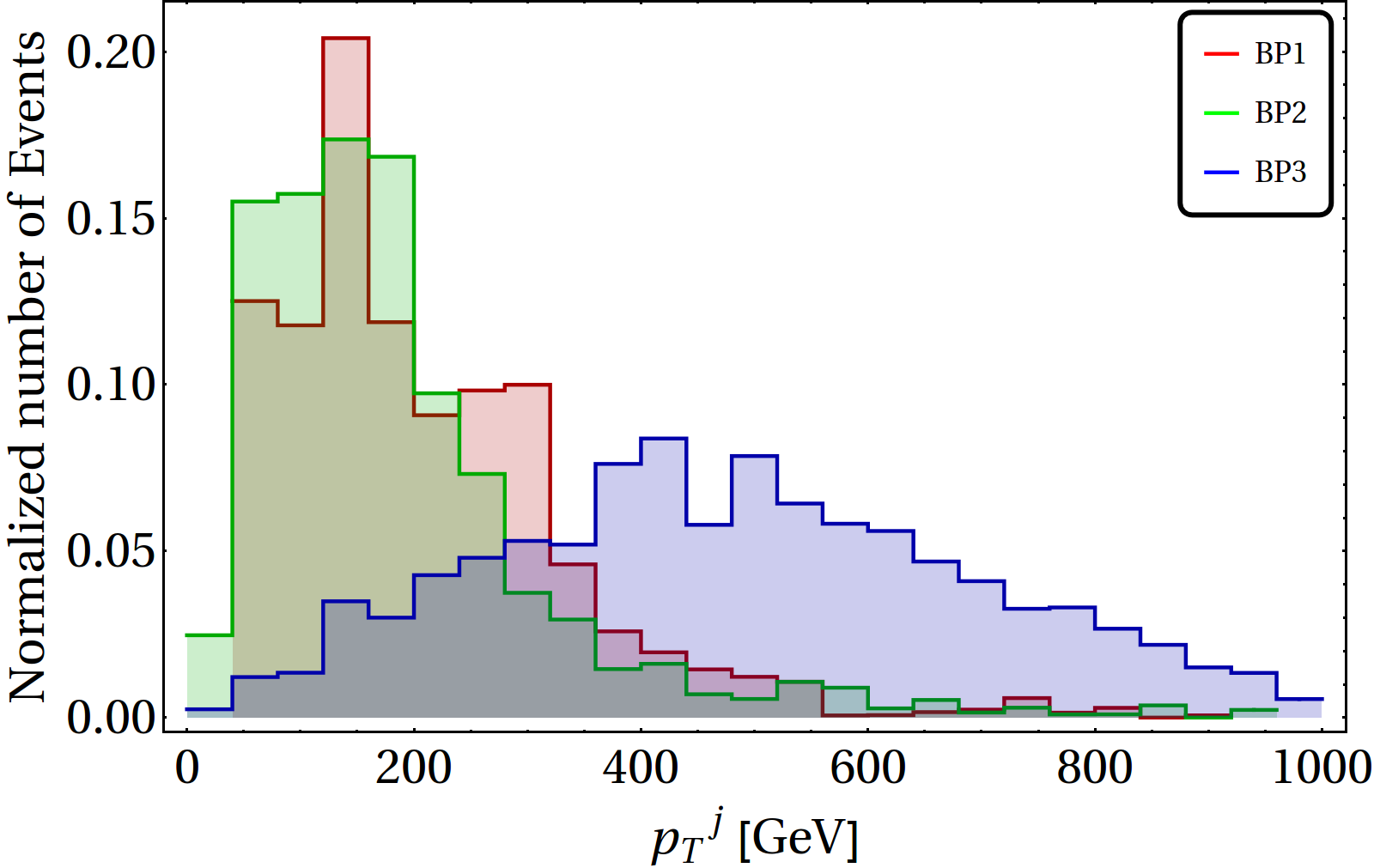}
$$
\caption{Transverse momentum ($p_T$) of jet distribution in OSD ($\ell^{+}\ell^{-}+1j+{E_{T}}\!\!\!\!/$) events for benchmark points BP1, BP2 and BP3.}
\label{fig:pt-dist}
\end{figure}

Cross-sections for OSD events at LHC with $\sqrt{s}=14$ TeV and integrated luminosity $\mathcal L=100~{\rm fb}^{-1}$ are listed in Table \ref{tab:OSD} in terms of actual number of events \footnote{ Actual number of events $N$ for luminosity $\mathcal L$ can be obtained for seed simulation points $N_1$ and final state events $N_2$ as $N=(\sigma_{p}\times N_2\times \mathcal{L})/N_1$, where $\sigma_{p}$ is the corresponding production cross-section. }. On top of the basic cuts (as discussed in \ref{subsec:selec_crit}), we have used following cuts to effectively separate signal from background: 
\begin{itemize}
\item Missing Energy cut (${E_{T}}\!\!\!\!/$): $E_{T}\!\!\!\!/ >$ 100, 200, 300 GeV are used. Large ${E_{T}}\!\!\!\!/$ retains the signal due to large DM masses of the benchmark points and reduces background.
\item Invariant mass cut: Invariant mass is defined as ${m_{\ell\ell}}^2=(p_{\ell^-}+p_{\ell^+})^2$ for opposite sign same flavour dileptons. We require invariant mass not to lie within the Z mass window $\left(|m_z -15|\nleq {m_{\ell\ell}}\nleq |m_z +15|~GeV\right)$ to reduce background events from $Z$ production. 
\item For jets, a moderate cut $p_T>40$ GeV is demanded. This is particularly because of the fact that the jets produced in the signal events carry only a small fraction of the momentum and are mostly soft. The jet $p_T$ distribution for OSD events at the benchmark points are shown in Fig.~\ref{fig:pt-dist}. Larger $p_T$ cut kills a large fraction of signal events.  
\end{itemize}

\begin{table}
\begin{center}
\begin{tabular}{|c| c| c| c|c|c|c|c|}
\hline
Benchmark & $\sigma_{pp\rightarrow h_{q}X_{1}}$ & $\sigma_{pp\rightarrow h_{q}\bar{h}_q}$ & ${E_{T}}\!\!\!\!/$ &  $\sigma^{OSD}_{pp\rightarrow h_{q}X_{1}}$ & $\sigma^{OSD}_{pp\rightarrow h_{q}\bar{h}_q}$ & $N $\\ [0.5ex]

Points & (in~pb) & (in~pb) & (in GeV) &  (in~pb) & (in~pb) & $(100~ \rm{fb}^{-1})$\\ [0.5ex]
\hline\hline	

&  &  & $>100$  & $1.77\times 10^{-3}$ & $8.37\times 10^{-4}$  & 260 \\
$BP1$ & $ 0.331 $ & 1.13 & $>200$  & $1.53\times 10^{-3}$  & $7.26\times 10^{-4}$ & 225 \\
&  &  & $>300$  & $9.70\times 10^{-4}$  & $5.95\times 10^{-4}$  & 156 \\
\hline
&  &  & $>100$ & $1.62\times 10^{-3}$  & $1.58\times 10^{-3}$  & 320 \\
$BP2$ & $0.172$ &  $0.758$ & $>200$  & $1.04\times 10^{-3}$  & $1.23\times 10^{-3}$ & 227 \\
&  &  & $>300$   & $5.45\times 10^{-4}$   & $7.81\times 10^{-4}$  & 132 \\
\hline
&  &  & $>100$ & $2.73\times 10^{-4}$     & $1.42\times 10^{-5}$  & 28 \\
$BP3$ &$ 0.0199$ &  $0.102$ & $>200$   & $2.47\times 10^{-4}$   & $1.36\times 10^{-5}$   & 25 \\
&  &  & $>300$ & $2.10\times 10^{-4}$    & $1.20\times 10^{-5}$  & 22 \\
\hline
\end{tabular}
\end{center}
\caption {OSD ($\ell^+\ell^- + 1 jet + {E_T}\!\!\!\!\!\!\!/ ~$~) events at LHC for chosen benchmark points 
with $p_{T_{\ell}} > 20$, $p_{T_{j}} > 40$ and $\left|m_Z-15\right|\nleq m_{ll}\nleq \left|m_Z+15\right|$ at $\sqrt{s}=14$ TeV. 
Number of events predicted for $\mathcal{L}=100 ~{\rm fb}^{-1}$ luminosity.}  
\label{tab:OSD}
\end{table}

\begin{table}
\begin{center}
\begin{tabular}{|c| c| c| c| c|c|}
\hline
Process & $\sigma_{p}$ (in~pb) & ${E_T}\!\!\!\!\!\!\!/ ~$ (in GeV) & $\sigma^{OSD} $ (in~pb) & $N (100 ~\rm{fb}^{-1})$  \\ [0.5ex] 
\hline\hline
&  & $>100$ & $137.66\times 10^{-3}$ & 13766 \\
$t\bar t+j$ & 809.79 & $>200$ & $<8.09\times 10^{-3}$ & $<$1 \\
&  & $>300$ & $<8.09\times 10^{-3}$ & $<1$ \\
\hline
&  &  $>100$ & 64.82 & 6482 \\
$WW+j$ & 60.58 &  $>200$ & $5.45\times 10^{-2}$ & 545 \\
&  & $>300$ & $1.81\times 10^{-2}$ & 181 \\
\hline
&  & $>100$ & $2.16\times 10^{-4}$ & 21 \\
$WZ+j$ & $0.15$ &  $>200$ & $5.55\times 10^{-5}$ & 5 \\
&  & $>300$ & $2.40\times 10^{-5}$ & 2 \\
\hline
&  &  $>100$ & $2.28\times 10^{-4}$ & 22  \\
$ZZ+j$& 7.63 &  $>200$ & $<7.63\times 10^{-5}$ & $<1$ \\
&  & $>300$ & $<7.63\times 10^{-5}$ & $<1$ \\
\hline
&  &  $>100$ & $<8.79\times 10^{-3}$ & $<$1 \\
$Drell-Yan$&  879.19 &  $>200$ & $<8.79\times 10^{-3}$ & $<1$ \\
&  & $>300$ & $<8.79\times 10^{-3}$ & $<1$ \\
\hline
\end{tabular}
\end{center}
\caption { OSD ($\ell^+\ell^- + 1 jet + {E_T}\!\!\!\!\!\!\!/ ~$~) events for dominant SM background at LHC with $p_{T_{\ell}} > 20$, $p_{T_{j}} > 40$ and $\left|m_Z-15\right|\nleq m_{ll}\nleq \left|m_Z+15\right|$ at $\sqrt{s}=14~$TeV and $\mathcal{L}=100~ {\rm fb}^{-1} $ luminosity. Appropriate $K-$factors are used for different processes to match to the NLO-NLL cross-sections available in literature (see text for details).} 
\label{tab:OSD-back}
\end{table}

\begin{figure}[htb!]
$$
\includegraphics[height=5.5cm]{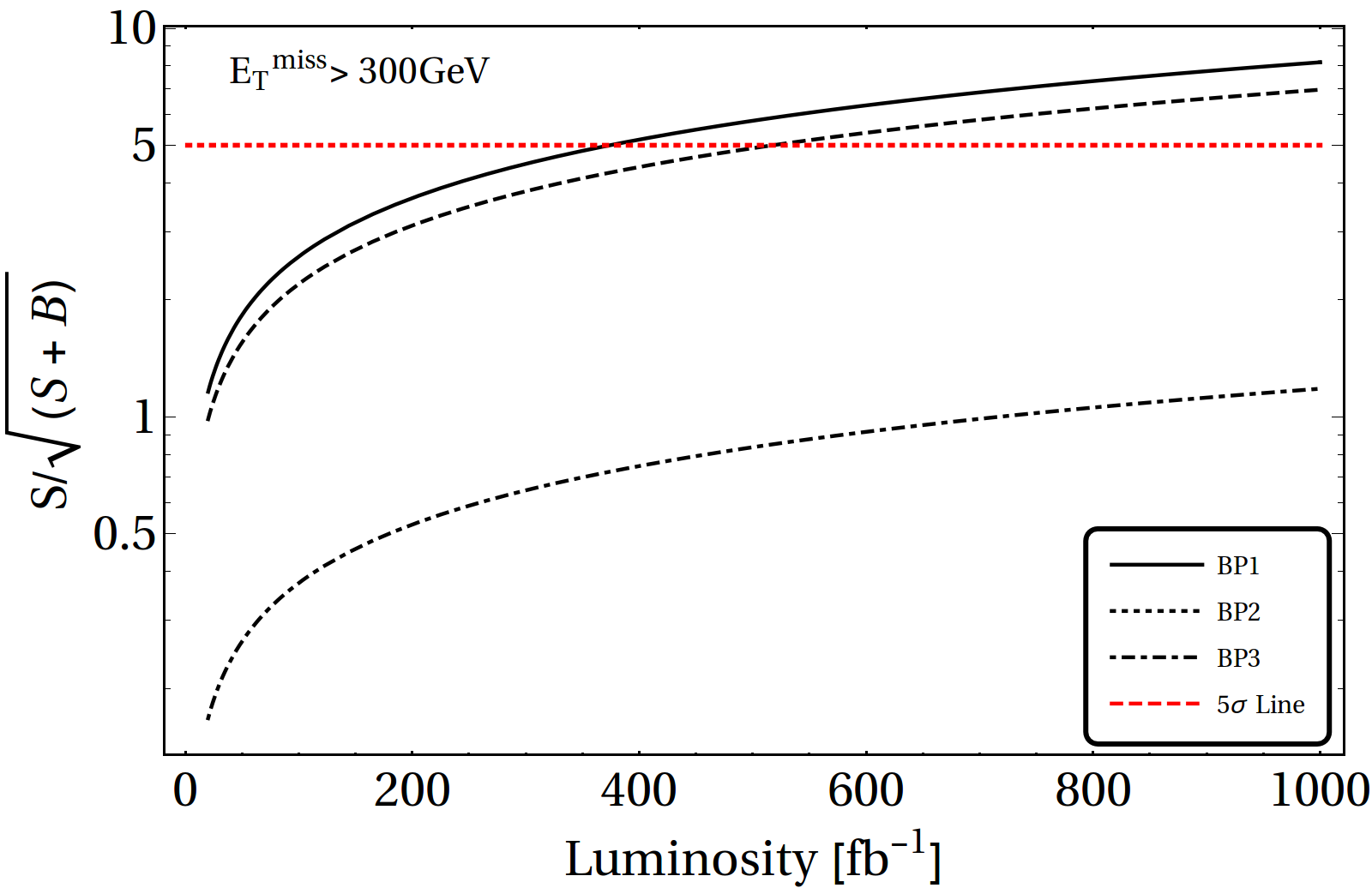}
$$
 \caption{Significance plot of OSD ($\ell^+\ell^- + 1 jet + {E_T}\!\!\!\!\!\!\!/ ~$~) events at the benchmark points with ${E_T}\!\!\!\!\!\!\!/~ > 300$ GeV versus luminosity (in fb$^{-1}$).}
 \label{fig:osd_significance}
\end{figure}

Cross-sections for the dominant SM background processes contributing to OSD events at LHC is tabulated in Table~\ref{tab:OSD-back}. We have multiplied the production cross-sections generated at leading order (LO) by appropriate $K$ factors  to match the NLO cross-section available in the literature. For example, for $t\bar{t}:~K=1.31$, $WWj:~K=1.30$, $WZj:~K=1.27$, $ZZj:~K=1.31$, $Drell$-$Yan: ~K=1.2$~\cite{Alwall:2014hca}. We estimate a limit on the final state event cross-section ($\sigma_e$) as $ \sigma_e< \sigma_{p}/N $ where $N$ events are simulated with production cross section $\sigma_{p}$. A discovery significance for OSD events is shown in Fig.~\ref{fig:osd_significance} in terms of luminosity (in fb$^{-1}$). The main outcome of this analysis is to see an excess in BP1 and BP2 benchmark points in OSD final states at LHC for high luminosity, while BP3 might be a little harder to explore.


\subsubsection{Validation against observed data at LHC}
\label{subsec:validation}

\begin{table}

\small
\setlength\tabcolsep{0pt}
\setlength\thickmuskip{0mu}
\setlength\medmuskip{0mu}

\begin{tabular}{|c |c c c| c c c| c c c||c|c|}
\hline

\multirow{3}{*}{${E_{T}}\!\!\!\!/$ (GeV)} & \multicolumn{3}{c|}{BP1} & \multicolumn{3}{c|}{BP2} & \multicolumn{3}{c||}{BP3} & \multirow{2}{*}{Obs.} & \multirow{2}{*}{SM}  \\
 & & & & & & & & & & data & bck \\

\cline{2-10}

        & $\sigma^{OSD}_{pp\rightarrow h_{q}X_{1}}~$          & $\sigma^{OSD}_{pp\rightarrow h_{q}\bar{h}_q}~$        & $N_{OSD}~$ 
        & $\sigma^{OSD}_{pp\rightarrow h_{q}X_{1}}~$          & $\sigma^{OSD}_{pp\rightarrow h_{q}\bar{h}_q}~$        & $N_{OSD}~$   
        & $\sigma^{OSD}_{pp\rightarrow h_{q}X_{1}}~$          & $\sigma^{OSD}_{pp\rightarrow h_{q}\bar{h}_q}~$        & $N_{OSD}~$ & & \\

\hline
100-150  & $<$0.003         & $<$0.007        & $<$1            & $<$0.001     & $<$0.004      & $<$1        & $<$0.0001         & $<$0.0005           & $<$1  & 28$~$ &  $28.2^{+5.4}_{-4.8}$ \\
\hline
150-225  & $<$0.003         & $<$0.007        & $<$1            & $<$0.001     & $<$0.004      & $<$1        & $<$0.0001         & $<$0.0005           & $<$1  & 6$~$  & $8.7^{+3.2}_{-1.9}$\\
\hline
225-300  & $<$0.003         & $<$0.007        & $<$1            & $<$0.001     & $<$0.004      & $<$1        & $<$0.0001         & $<$0.0005           & $<$1  & 5$~$  & $3.3^{+2.5}_{-1.0}$\\
\hline
$>$300   & $<$0.003         & $<$0.007        & $<$1            & $<$0.001     & $<$0.004      & $<$1        & $<$0.0001         & $<$0.0005           & $<$1  & 6$~$  & $1.9^{+1.4}_{-0.7}$\\
\hline
\end{tabular}
\caption{Results for CMS on-$Z$ search \cite{CMS} at the benchmark points for OSD events associated with 2-3 jets and $H_T>$ 400. The data, signal and SM predictions are made for $\sqrt s=13 $ TeV with $\mathcal L=2.2$ fb$^{-1}$.}
\label{tab:cms on-Z}
\end{table}

We discuss here the competence of the benchmark points chosen for the analysis with the existing data at LHC. OSD events have been searched exhaustively, particularly because supersymmetry (SUSY) provides such a signal so very often. The signal selection criteria however is different from ours and is mainly guided to maximize the SUSY signal efficiency. Searches in OSD channels have been broadly divided into two different regions as described by CMS~\cite{CMS,CMS2}  and  ATLAS~\cite{Aaboud:2016ejt}: (i) on the $Z$-mass window or the `on-$Z$ search', where $|m_z -15|\leq {m_{\ell\ell}}\leq |m_z +15|$ GeV and (ii) out of the $Z$-mass window or `edge search' where ${m_{\ell\ell}}$ does not lie within the Z-mass window. These regions are then further classified into sub-regions. On-Z search is mainly classified by number of associated jets and b-tagged jets. We only validate our benchmark points for {\bf SRA} regions of CMS analysis \cite{CMS}, where the OSD signal is vetoed with 2-3 jets and zero b-jets. In this search, $H_T=\Sigma_{jets}~p_T>400$ GeV is imposed. The event rates of the benchmark points with such criteria are summarised in Table~\ref{tab:cms on-Z}, where the signal region is further divided into different ${E_T}\!\!\!\!\!\!\!/ ~~$ regions. We see that all the benchmark points produce zero events at the luminosity of 2.3 $fb^{-1}$ and therefore are allowed by the observed data. This happens mainly due to the demand of number of associated jets and high $H_T$ cut which the model fails to pass through. Requiring more jets ($N_j \ge 3$) anyway produces zero events for all the benchmark points and hence we do not analyze the other signal region ({\bf SRB}) of CMS analysis. The edge search or off-Z search for OSD event is divided into different invariant mass ($m_{\ell\ell}$) regions and is inclusive of number of associate jets which require a moderate ${E_T}\!\!\!\!\!\!\!/~>150$ GeV cut for $N_j\ge 2$ and ${E_T}\!\!\!\!\!\!\!/~>100$ GeV cut for $N_j\ge 3$. Signal events for off-Z search at the benchmark points are mentioned in Table~\ref{tab:cms off-Z}. Comparing with the observed data we see that all the benchmark points lie comfortably within the limit.

\begin{table}

\small
\setlength\tabcolsep{0pt}
\setlength\thickmuskip{0mu}
\setlength\medmuskip{0mu}
\small

\begin{tabular}{|c |c c c| c c c| c c c||c| c|}
\hline
\multirow{3}{*}{$m_{ll}$(GeV)} & \multicolumn{3}{c|}{BP1} & \multicolumn{3}{c|}{BP2} & \multicolumn{3}{c||}{BP3} & \multirow{2}{*}{Obs.} & \multirow{2}{*}{SM}  \\
 & & & & & & & & & & data &  bck \\
 
\cline{2-10}

   & $\sigma^{OSD}_{pp\rightarrow h_{q}X_{1}}~$        & $\sigma^{OSD}_{pp\rightarrow h_{q}\bar{h}_q}~$                    & $N_{OSD}~$           
   & $\sigma^{OSD}_{pp\rightarrow h_{q}X_{1}}~$        & $\sigma^{OSD}_{pp\rightarrow h_{q}\bar{h}_q}~$                    & $N_{OSD}~$        
   & $\sigma^{OSD}_{pp\rightarrow h_{q}X_{1}}~$        & $\sigma^{OSD}_{pp\rightarrow h_{q}\bar{h}_q}~$                    & $N_{OSD}~$ & & \\

\hline
20-70  & 0.026        & 0.25        & $<$1            & $<$0.001     & 0.844         & 1           & 0.33              & 2.298              & 3 & 132$~$         &  126.7$\pm$ 12.3  \\
\hline 
70-81  & 0.003        & 0.007       & $<$1            & $<$0.001     & 0.032         & $<$1        & 0.01              & 0.064              & $<$1 &  33$~$      & 38.2 $\pm$ 6.2  \\
\hline
81-120 & $<$0.003     & $<$0.007    & $<$1            & $<$0.001     & $<$0.004      & $<$1        & $<$0.0001         & $<$0.000           & $<$1 & 42$~$       & 93.1 $\pm$ 10.5 \\
\hline 
$>$120 & $<$0.003     & $<$0.007    & $<$1            & $<$0.001     & $<$0.004      & $<$1        & $<$0.0001         &  $<$0.0005         &  $<$1 & 141$~$     & 109.9$\pm$ 11.4\\
\hline
\end{tabular}
\caption{OSD events for CMS off-$Z$ search \cite{CMS} at the benchmark points with ${E_T}\!\!\!\!\!\!\!/~~>150$ GeV for $N_j\ge 2$ and  ${E_T}\!\!\!\!\!\!\!/~~>100$ GeV for $N_j\ge 3$. The data, signal and SM background predictions are made for $\sqrt s=13$ TeV with $\mathcal L=2.2$ fb$^{-1}$.}
\label{tab:cms off-Z}
\end{table}

\subsection{Hadronically quiet four lepton signal}
\label{subsec:HQ4l}

 \begin{figure}[htb!]
$$
\includegraphics[height=4.5cm]{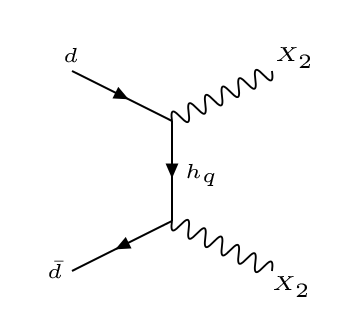}\hspace{1cm} 
\includegraphics[height=4.5cm]{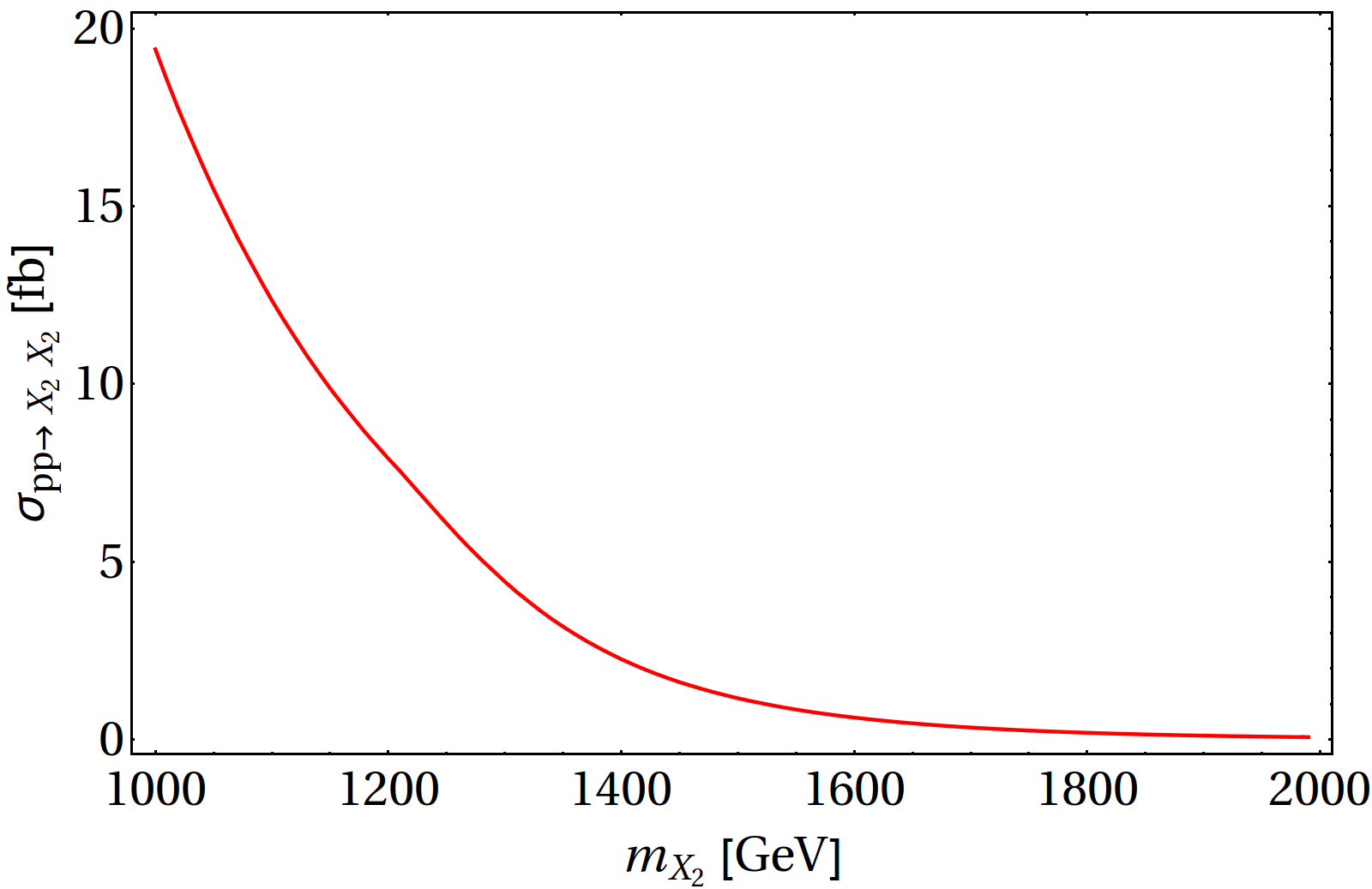} 
$$
\caption{Left: Feynman diagram for producing a pair of $X_2$ at the LHC. Right: Cross-section of $pp\to X_2X_2$ as a function of $m_{X_2}$ at 14 TeV LHC.}
\label{fig:prod_x2x2}
\end{figure}

\begin{figure}[htb!]
\centering
\includegraphics[height=4.4cm]{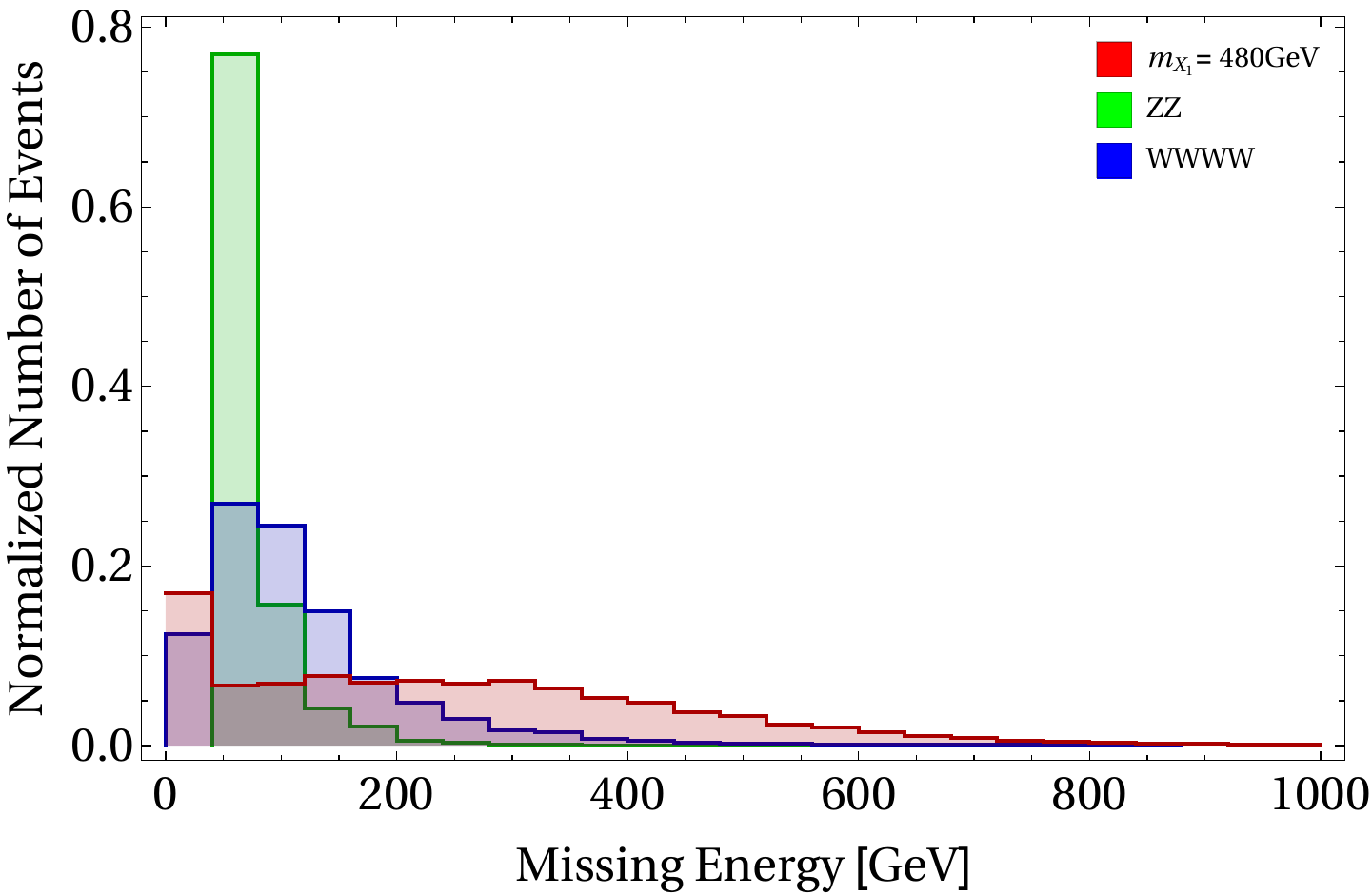}\hspace{1cm}
\includegraphics[height=4.4cm]{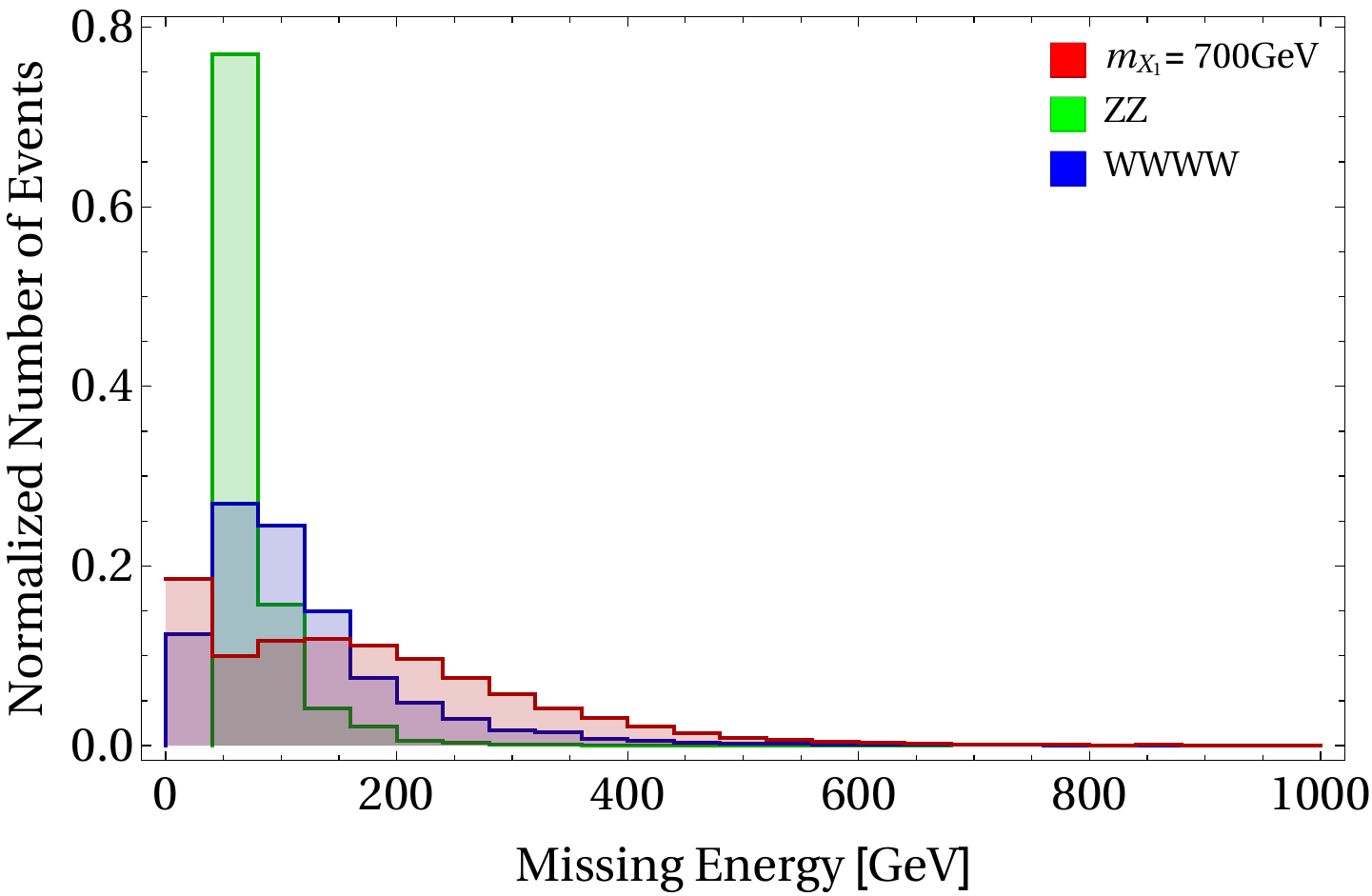}
\includegraphics[height=4.4cm]{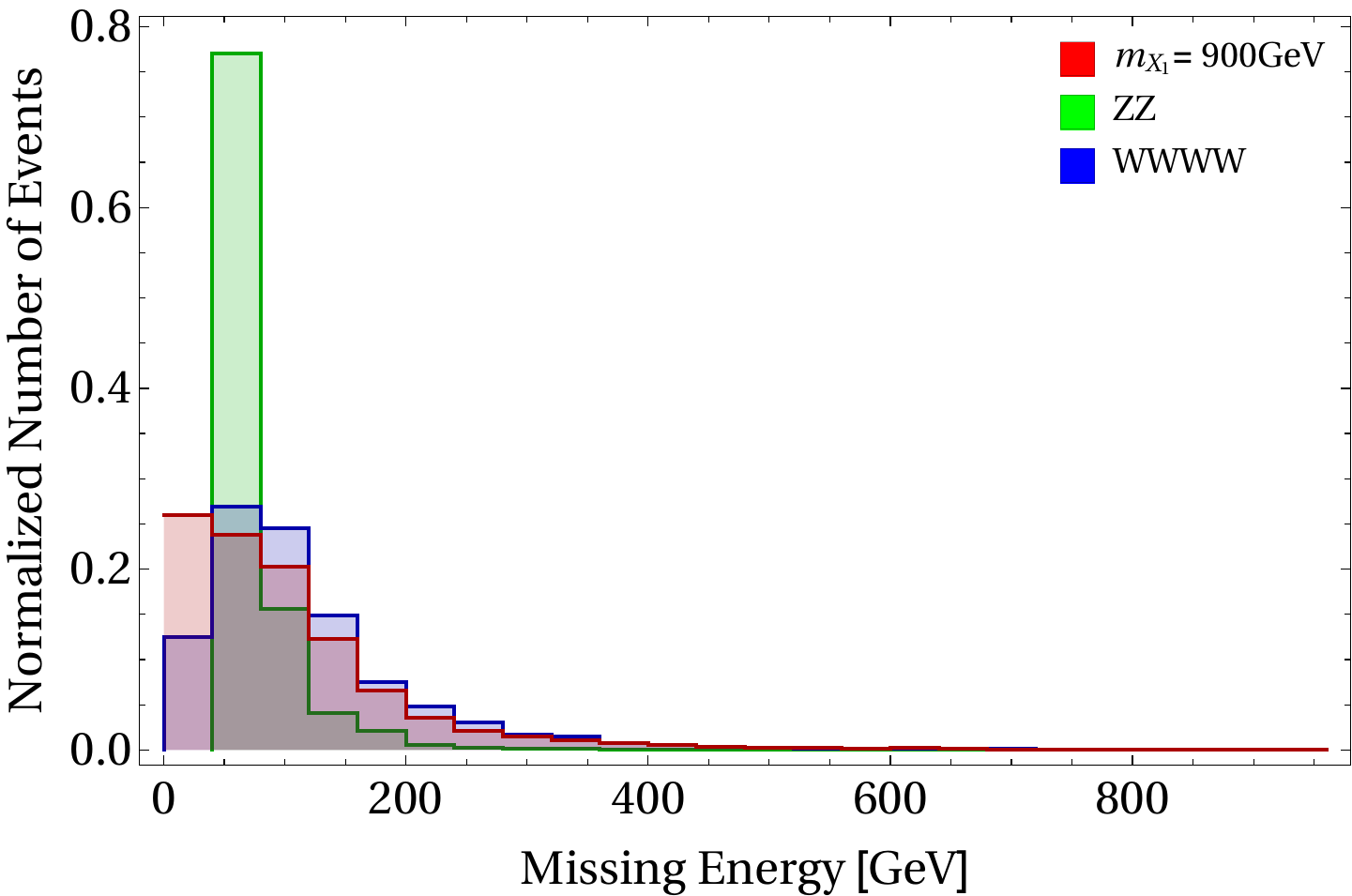}
 \caption{ Missing energy (${E_T}\!\!\!\!\!\!\!/~$~) distribution for HQ4l events at the benchmark points with dominant SM background. Top left: BP1, Top right: BP2, Bottom: BP3.}
 \label{fig:MET-HQ4l}
\end{figure}

One of the unique collider signatures that this model offers, is the hadronically quiet four lepton (HQ4l). This can arise from the production of the heavier $SU(2)_N$ gauge boson $X_2$ following the Feynman graph in left panel of Fig.~\ref{fig:prod_x2x2}. $X_2$ further decays through the exotic leptons following Fig.~\ref{fig:X2-decay} and Fig.~\ref{fig:EN-decay} and produces two pairs of same flavour opposite sign dilepton with large missing energy ($\ell^\pm\ell^{\mp}\ell^{'\pm}\ell^{'\mp}+ {E_T}\!\!\!\!\!\!\!/~$~). The corresponding signal can have any of the following combinations: $e^{\pm}e^{\mp}e^{\pm}e^{\mp},~ e^{\pm}e^{\mp}\mu^{\pm}\mu^{\mp},~ \mu^{\pm}\mu^{\mp}\mu^{\pm}\mu^{\mp}$. We however choose final states inclusive of both the flavours. Hadronically quiet four lepton signature is also an artifact of the mass hierarchy as pointed out in Eq.~\ref{eq:hierarchy} and not necessarily  a generic one for this model. The production cross-section of $pp \to X_2 X_2$ is much smaller as it is an $SU(2)_N$ gauge interaction process. With heavier $m_{X_2}$, the cross-section falls off sharply to a vanishingly small value as shown in the right panel of Fig.~\ref{fig:prod_x2x2}. We again note here that $pp\to h_q \bar{h_q}$ also contributes to HQ4l events, where in both legs $h_q$ decays via $h_q \to d X_2$ with $X_2 \to e^+e^-X_1$ through exotic charged leptons. The signal event rates at the benchmark points are mentioned in Table~\ref{tab:hq4l}.  The lepton selection criteria remains similar to OSD search, while we demand the absence of jets with $p_T>$20 GeV. Although the number of signal events is small in this case, SM background is even more rare. Dominant contributions come from $ZZ$, $WWWW$ and $Zt\bar t$ as estimated in Table~\ref{tab:hq4l-back}. For SM background, we have multiplied the cross-section generated at LO in {\tt Madgraph} by appropriate $K$ factors to match the NLO cross-section available in the literature (for $WWWW:~K=1.74$, $ZZ:~K=1.29$, $Zt\bar t: ~K=1.44$~\cite{Alwall:2014hca}). Again we show that missing energy is one of the main discriminators for separating signal and background. This is shown in Fig.~\ref{fig:MET-HQ4l} for three benchmark points and for the dominant SM backgrounds.


\begin{table}
\begin{center}
\begin{tabular}{|c| c| c| c|c|c|c|c|}
\hline
Benchmark & $\sigma_{pp\rightarrow X_{2}X_{2}}$ & $\sigma_{pp\rightarrow h_{q}\bar{h}_q}$ &  &  $\sigma^{HQ4l}_{pp\rightarrow X_{2}X_{2}}$ & $\sigma^{HQ4l}_{pp\rightarrow h_{q}\bar{h_q}}$ & $N$\\ [0.5ex] 

Points & $(in~pb)$ & $(in~pb)$ & $E_{T}^{miss}$ &  $(in~pb)$ & $(in~pb)$ & $(100 fb^{-1})$\\ [0.5ex] 
\hline\hline	
&  &  & $>100$  & $1.88\times 10^{-3}$ & $3.32\times 10^{-4}$  & 221 \\
$BP1$ & $ 0.0193 $ & 1.13 & $>200$  & $1.53\times 10^{-3}$  & $2.82\times 10^{-4}$  & 181 \\
&  &  & $>300$  & $1.07\times 10^{-3}$  & $2.01\times 10^{-4}$  & 127 \\
\hline
&  &  & $>100$ & $2.27\times 10^{-3}$  & $5.99\times 10^{-4}$    & 286 \\
$BP2$ & $0.0264$ &  $0.758$ & $>200$  & $1.36\times 10^{-3}$  & $4.11\times 10^{-4}$  & 177 \\
&  &  & $>300$   & $5.43\times 10^{-4}$   & $2.15\times 10^{-4}$    & 75 \\
\hline
&  &  & $>100$ & $1.73\times 10^{-4}$     & $2.61\times 10^{-4}$  & 43 \\
$BP3$ &$ 0.0205 $ &  $0.102$ & $>200$   & $3.93\times 10^{-5}$   & $2.47\times 10^{-4}$   & 27 \\
&  &  & $>300$ & $1.56\times 10^{-5}$    & $2.23\times 10^{-4}$  & 23 \\
\hline
\end{tabular}
\end{center}
\caption {Hadronically quiet four lepton ($\ell^+\ell^- \ell^{'+}\ell^{'-}+ {E_T}\!\!\!\!\!\!\!/ \: \;$~) events at LHC for chosen benchmark points 
with $p_{T_{\ell}} > 20$ and $\left|m_Z-15\right|\nleq m_{ll}\nleq \left|m_Z+15\right|$ for all opposite sign same flavour dilepton at $\sqrt{s}=14~TeV$  
and $\mathcal{L}= 100 fb^{-1}$ luminosity.}  
\label{tab:hq4l}
\end{table}

\begin{table}[htb!]
\begin{center}
\begin{tabular}{|c| c| c| c|c|}
\hline
Process & $\sigma_{production} (in~pb)$ & $E_{T}^{miss}$  & $\sigma_{HQ4l} (in~pb)$ & $N(100 fb^{-1})$  \\ [0.5ex] 
\hline\hline
       &                                & $>100$          & $1.68\times 10^{-5}$    & 1  \\
$ZZ$   & $0.024$                        & $>200$          & $<1.2\times10^{-6}$     & $<1$ \\
       &                                & $>300$          & $<1.2\times10^{-6}$     & $<1$ \\
\hline
       &                                &  $>100$         & $<7.8\times10^{-12}$     & $<1$ \\
$W^{+}W^{-}W^{+}W^{-}$  & $1.17\times 10^{-6}$            &  $>200$                 & $<7.8\times10^{-12}$ & $<1$ \\
       &                & $>300$        & $<7.8\times10^{-12}$ & $<1$ \\
\hline
&  &  $>100$ & $1.8\times10^{-4}$ & 18 \\
$Zt\bar t$ & 0.90 &  $>200$ & $4.5\times10^{-5}$ & 4 \\
&  & $>300$ & $<4.5\times10^{-5}$ & $<1$ \\
\hline
\end{tabular}
\end{center}
\caption {HQ4l ($\ell^+\ell^-\ell^{'+}\ell^{'-}+{E_T}\!\!\!\!\!\!\!/ \: \;$~) events for dominant SM background at LHC with $p_{T_{\ell}} > 20$ and $\left|m_Z-15\right|\nleq m_{ll}\nleq \left|m_Z+15\right|$ for all opposite sign same flavour dilepton at $\sqrt{s}=14~TeV$  and $\mathcal{L}= 100 fb^{-1} $ luminosity. Appropriate $K-$factors are used for different processes to match to the NLO-NLL cross-sections available in literature (see text for details).} 
\label{tab:hq4l-back}
\end{table}

\begin{figure}[htb!]
$$
\includegraphics[height=4.4cm]{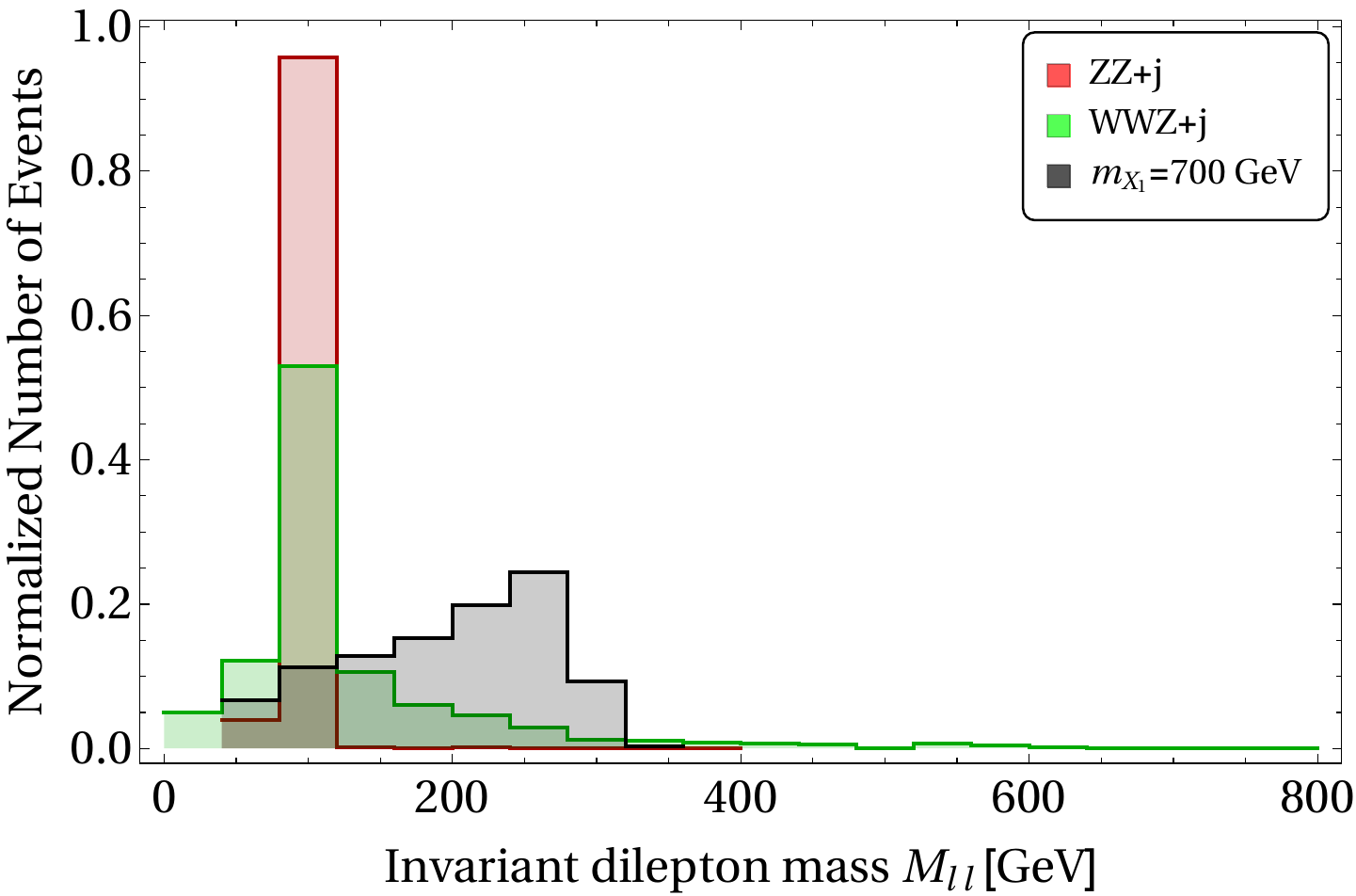}\hspace{1cm}
\includegraphics[height=4.4cm]{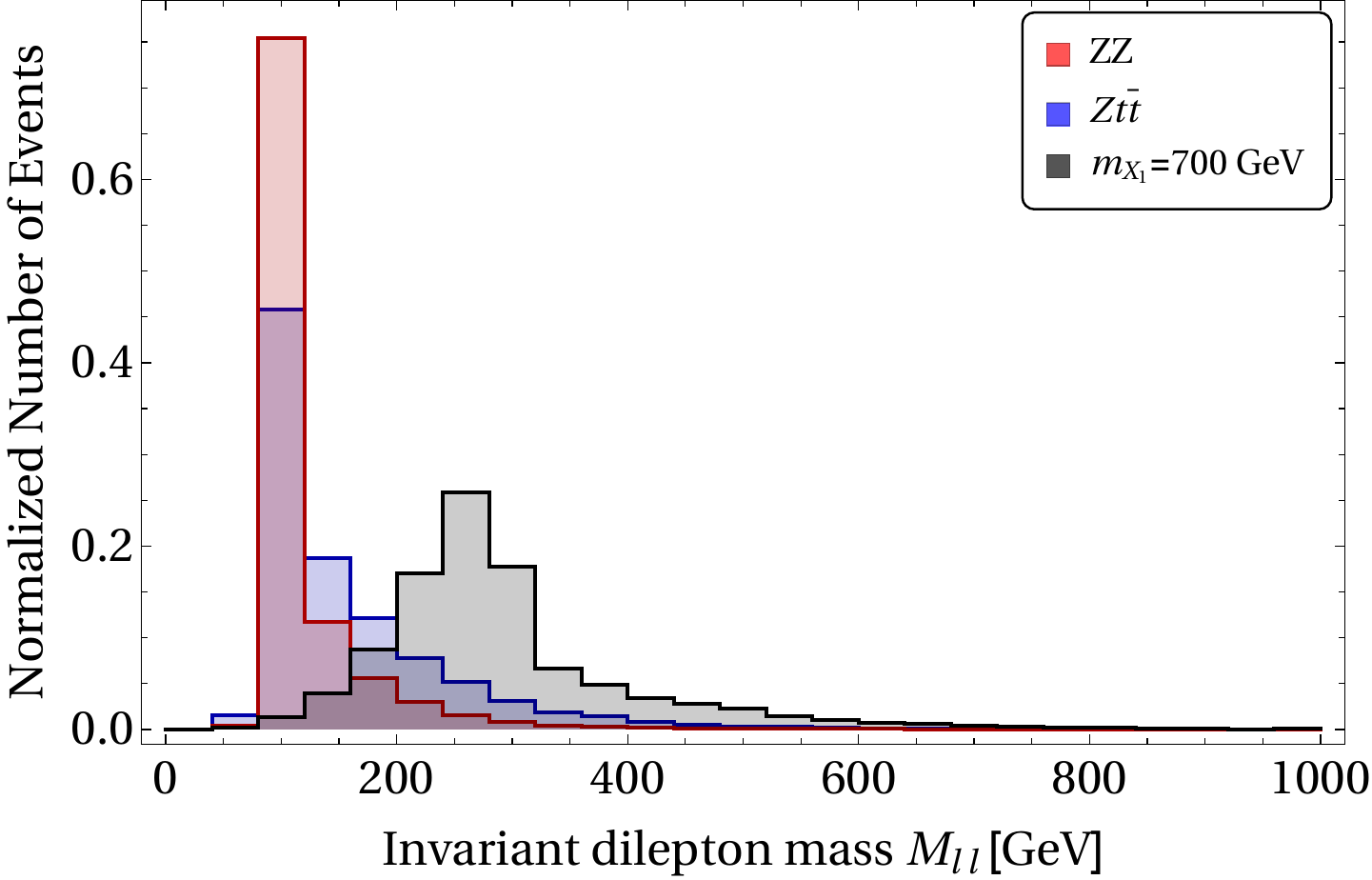}
$$
\caption{Invariant mass $(m_{ll})$ distribution for OSD (left) and HQ4l channel (right) for BP2 compared with $Z$ background.}
\label{fig:mll}
\end{figure}

We have also checked the efficiency of invariant mass ($m_{\ell\ell}$) cut employed in the analysis and explicitly demonstrated the $m_{\ell\ell}$ distributions in Fig.~\ref{fig:mll}. In the left panel, we plot $m_{\ell\ell}$ for OSD ($\ell^+\ell^- + 1 jet + {E_T}\!\!\!\!\!\!\!/~$~) events for signal from BP2 and dominant SM background from $ZZ$+jets,\;$W^+W^-Z$+jets. On the right-panel we plot $m_{\ell\ell}$ of any two same flavour opposite sign dilepton pair of HQ4l events for signal at BP2 with dominant SM backgrounds from $ZZ$,$Zt\bar t$. For OSD, both the backgrounds $ZZ$+jets,\;$W^+W^-Z$+jets have $m_{\ell\ell}$ peaks around $Z$ mass window and reduce drastically with the invariant mass cut.  For HQ4l, we can form dilepton invariant mass $m_{\ell\ell}$ depending upon same flavour or two different flavour leptons in the final state respectively. The $m_{\ell\ell}$ distribution for HQ4l events are shown for highest $m_{\ell\ell}$ obtained in the events. As can easily be seen that for $ZZ$ events, the distribution peaks at Z-mass, while signal events for BP2 peaks at a larger value. We therefore, employ $\left|m_Z-15\right|\nleq m_{ll}\nleq \left|m_Z+15\right|$ cut for all the opposite sign same flavour invariant mass pairs formed in HQ4l events, which retains the signal to a large extent while the background drops significantly.  

\begin{figure}[htb!]
$$
\includegraphics[height=5.5cm]{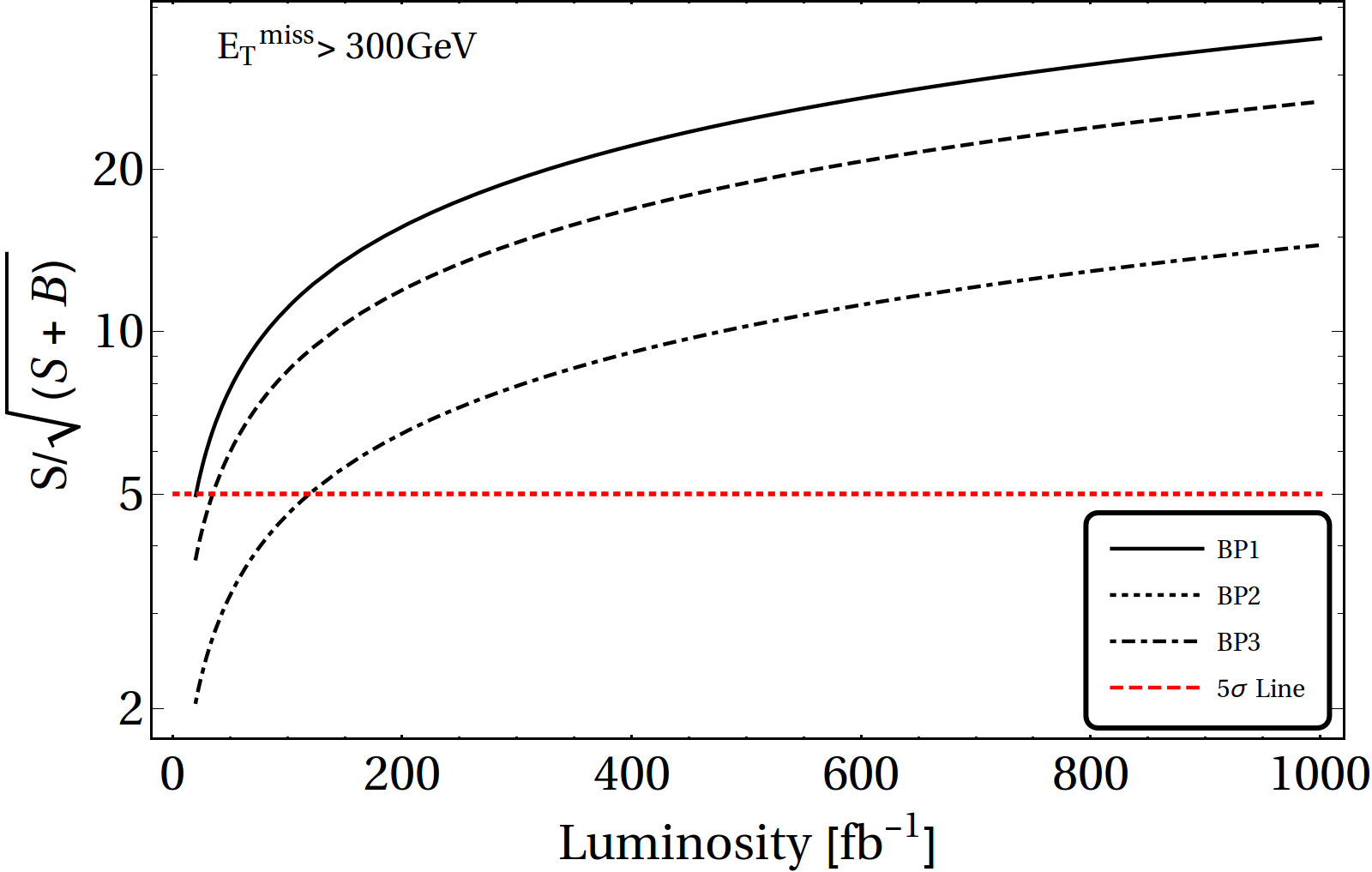}
$$
 \caption{Significance plot for hadronically quiet four lepton final state (HQ4l) events for benchmark points at LHC for ${E_T}\!\!\!\!\!\!\!/~ > 300$ GeV with luminosity (in fb$^{-1}$).}
 \label{fig:hq4l_significance}
\end{figure}

It is clear that hadronically quiet four lepton channel provides a smoking gun signature of the model as the significance plot suggests in Fig.~\ref{fig:hq4l_significance}. Such a signature is not often studied and should be analyzed with existing data for a possible excess. SUSY with R-parity violation can yield four lepton final states, but doesn't offer a large missing energy as we have here. On the other hand, SUSY with $R$- parity conservation with lighter third generation squarks may have four lepton final state and missing energy, but infected by the presence of multiple $b$ jets. Hence, four lepton final state in vector boson DM model discussed here can be disentangled from that of SUSY. The existing multilepton search criteria \cite{CMS2}, due to large b-tagged jets yields zero events for signal at the benchmark points and hence allowed by the data as shown in Table~\ref{tab:cms off-Z hq4l}. Lastly, we note that four leptons can also arise with two soft jets and missing energy in our model from the $pp\to h_q \bar{h}_q$ production, and its subsequent decays through heavier gauge bosons. However, the signal cross-section is smaller (for example, BP1: $\sigma^{4l+2j}=0.17$ fb with ${E_T}\!\!\!\!\!\!\!/~ > 300$ GeV) than the HQ4l events and the SM background is also larger. Hence, the benchmark points can easily satisfy non-observation of data in this channel and we refrain from providing a detailed analysis on that. 

A few non-supersymmetric scenarios, like left-right symmetric and Type-II seesaw models~\cite{Bambhaniya:2013yca}, contain  $SU(2)$ triplet scalars with hyper charge $+2$ (\cite{Bambhaniya:2013wza},\cite{Bambhaniya:2015wna}). Such multiplets contain doubly charged scalars. Through the pair production of such scalars and their dominant leptonic decay modes, we can have HQ4l final states. This resembles our signal but without the large missing energy. Also, unlike our scenario here, one can construct the dilepton (same signed) invariant mass, and that distribution  will peak around the mass of that doubly charged scalar. This sharp invariant mass reconstruction is not possible for our scenario and that can differentiate our model from such models containing doubly charged scalars.

 \begin{table}[htb!]
\begin{center}
\begin{adjustbox}{width=1\textwidth}
\small
\begin{tabular}{|c| c| c| c| c| c |c| c|c|c|c|}
\hline
SR  & ${E_{T}}\!\!\!\!/$ (GeV) & ${H_{T}}\!\!\!\!/$ (GeV)& BP & $\sigma^{multilep}(fb)$ & $N(2.3 fb^{-1})$           & SM prediction  & Obs. data \\ [0.5ex] 

\hline\hline
SR1 & 50-150                   & 60-400           & BP1   & $<$0.007      & $<$1                           &     & \\
    &                          &                  & BP2   & 0.009         & $<$1                           & $19.26^{+4.81}_{-4.80}$   & 18 \\
    &                          &                  & BP3   & 0.026         & $<$1                          &     & \\
\hline
SR2 & 150-300                  & 60-400           & BP1   & $<$0.007      & $<$1                           &     &  \\
    &                          &                  & BP2   & $<$0.004      & $<$1                           & $1.16^{+0.31}_{-0.20}$    & 4 \\
    &                          &                  & BP3   & $<$0.0005     & $<$1                           &     &   \\
\hline
SR3 & 50-150                   & 400-600          & BP1   & $<$0.007      & $<$1                           &     &    \\
    &                          &                  & BP2   & $<$0.004      & $<$1                           & $1.20^{+0.47}_{-0.40}$   &  3 \\
    &                          &                  & BP3   & $<$0.0005     & $<$1                           &     &  \\
\hline
SR4  & 150-300                 & 400-600          & BP1   & $<$0.007      & $<$1                           &     &    \\
     &                         &                  & BP2   & $<$0.004      & $<$1                       & $0.29^{+0.44}_{0.40}$   & 0  \\
     &                         &                  & BP3   & $<$0.0005     & $<$1                   &     &    \\
\hline
\end{tabular}
\end{adjustbox}
\end{center}
\caption{Multilepton ($\ge 3$ electrons or muons) events for CMS off-$Z$ search \cite{CMS2} at the benchmark points for $\sqrt s=13$ TeV with $\mathcal L=2.3$ fb$^{-1}$.}
\label{tab:cms off-Z hq4l}
\end{table}

\section{Summary and Conclusions}
\label{sec:summary}

In this paper we have analyzed the phenomenology of a vector boson DM with $SU(2)_N$ extension of SM, in detail. The model had already been proposed in literature and is extremely relevant for current studies as it can easily evade the ever growing direct search constraints from non-observation of DM. We elaborate on relic density and direct search outcome of the DM taking care of the freedom in choosing the $SU(2)_N$ gauge coupling through elaborate parameter space scan. Several new features have emerged from this analysis including the crucial effects of co-annihilation and non-viability of a degenerate DM scenario that could emerge in absence of a scalar triplet.

We have also explicitly demonstrated the unification of the low scale model into gauge group $E(6)$ as was proposed earlier, with consistent intermediate symmetries and breaking scheme. The breaking adopted here suggests that in D-parity conserved case, $SU(2)_N$ and $SU(2)_L$ couplings are equal ($g_{N}=g_L$) with two degenerate vector boson DM, while for D-parity broken scenario we may have non-universality through $g_{N}\neq g_{L}$ and a single component DM emerges. The spread in the gauge coupling for D-odd case is determined by explicit calculation and the freedom is utilized for phenomenological analysis. 

One of the crucial construct of the model is that the $SU(2)_N$ charges do not contribute to hypercharge and therefore the gauge bosons $X_{1,2,3}$ are neutral. The stability of the DM ($X_1$) is achieved by a modified $R=(-1)^{3B+L+2J}$ charge, through $L=T_{3N}+P$. The phenomenology alters completely with a change in choosing the $R$ symmetry and will be discussed elsewhere. Apart from unification, the other important constraint on the model comes from small $X_3(Z^{'})$-$Z$ mixing which demands $m_{X_3}=m_{Z^{'}}\geq 1$ TeV. This, in turn, constrains the lower limit of the DM mass for specific $g_N$.

Relic density constraint of the model correlates coupling and mass of the DM to the exotic quark mass $m_{h_q}$ and exotic lepton mass $m$. Dominant annihilation processes are $t$-channel diagrams through exotic fermion exchange. This allows a large range of DM masses above $\sim$200 GeV, lower than that is discarded by VEV constraints. 

Direct search interaction for this model is mainly $s$-channel process and mediates only via exotic quarks. Hence, except for those regions where $m_{h_q}\sim m_{DM}$, the model is very loosely constrained by spin independent direct search bounds. Co-annihilation of DM with $X_2$ helps the DM to evade direct search constraints and the detection may go beyond the reach of XENONnT sensitivity. This is simply because co-annihilation only contributes to relic density and does not take part in direct search  due to kinematical constraints. We also analyzed degenerate DM scenario in D-parity even case, where the DM mass has to be at least $m_{X_3}=m_{Z^{'}}\geq 1$ TeV, as three of the gauge bosons (including the DM) become degenerate. But this situation is ruled out because of relic density constraints.

The exotic particles present in the model, can be produced at LHC and their subsequent decays yield multi-lepton signatures with large missing energy. Opposite sign dilepton and hadronically quiet four lepton channels are discussed in context of the present LHC data through detailed simulations. We note that the presence of only one soft jet in two lepton channel can help to distinguish this model from SM background events like $t \bar t$ and also from other supersymmetric signals. Large missing energy cuts are shown to be effective to reduce the backgrounds and a possibility of seeing such a signal at 14 TeV emerges, albeit with large luminosity. A strong $H_T$ cut, as provided in the current LHC analysis of the data, washes away signal cross-section of this model. We therefore propose a minimal or no $H_T$ cut at all for seeing opposite sign dilepton signature of this particular scenario. However, a more crucial signal appears in the form hadronically quiet four lepton, that has not been analyzed before. SM background for this channel is tiny, which can be further minimized by employing a high missing energy and invariant mass cut within $Z$-window. The model therefore can be tested in hadronically quiet four lepton events at future runs of LHC at 14 TeV even with a moderate integrated luminosity.  

One of the important outcomes of this analysis is therefore to show that, collider search provides more sensitivity in unraveling the DM model while direct search may get delayed, even can be submerged into neutrino floor. This is in sharp contrast to many other DM scenarios. This is attributed to the $t$-channel DM annihilation and dominant $s$-channel direct search along with co-annihilation processes.

A comparative analysis on different vector boson DM frameworks will be helpful to classify and distinguish the phenomenological outcome with respect to future observations. We will discuss some of those features in our next analysis.

\acknowledgments 

SB would like to acknowledge discussions with Qing-Hong Cao at Argonne National Lab and M. Zakeri of U. C. Riverside; appreciate WHEPP workshop held at IIT Kanpur where part of the work was initiated and acknowledge support from DST INSPIRE faculty award (IFA-13-PH-57 at IIT Guwahati). BB would like to thank Dr. Subhadeep Mondal of HRI, Allahabad for helping out with the collider simulations, Dr. Flip Tanedo of UC Riverside for his hand written notes on DM direct detection and Prof. Biswarup Mukhopadhyay of HRI, Allahabad for his critical comments on the collider part. BB would also like to acknowledge the hospitality of IIT Kanpur during his visit, where a part of the work was completed. J.C. is supported by the Department of Science and Technology, Government of India, under the Grant IFA12-PH-34 (INSPIRE Faculty Award); and the Science and Engineering Research Board, Government of India, under the agreement SERB/PHY/2016348.

\appendix
\section{Scalar Potential}
\label{sec:potential}

The scalar potential of the model is given by :

\begin{align}
V & = \mu_1^{2}~Tr\left(\phi_{13}^{\dagger}\phi_{13}\right)+\mu_2^{2}\left(\phi_{2}^{\dagger}\phi_{2}\right)+\mu_{\chi}^{2}\left(\chi\chi^{\dagger}\right)+\mu_{\Delta}^{2}~Tr\left(\Delta^{\dagger}\Delta\right)+\left(\mu_3^{2}~det\Delta+h.c.\right)\nonumber \\ &+\left(\mu_{22}\tilde\chi\phi_{13}^{\dagger}\tilde\phi_2+\mu_{12}\chi\Delta\tilde\chi^{\dagger}+\mu_{23}\tilde\chi\Delta\chi^{\dagger}+h.c.\right)+\frac{1}{2}\lambda_1\left(Tr\left(\phi_{13}^{\dagger}\phi_{13}\right)\right)^{2}+\frac{1}{2}\lambda_2\left(\phi_2^{\dagger}\phi_2\right)^2 \nonumber \\&+\frac{1}{2}\lambda_{3}~Tr\left(\phi_{13}^{\dagger}\phi_{13}\phi_{13}^{\dagger}\phi_{13}\right)+\frac{1}{2}\lambda_{4}\left(\chi\chi^{\dagger}\right)^2+\frac{1}{2}\lambda_5\left[Tr\left(\Delta^{\dagger}\Delta\right)\right]^2\nonumber \\&+\frac{1}{4}\lambda_6~Tr\left(\Delta^{\dagger}\Delta-\Delta\Delta^{\dagger}\right)^2+ \tilde{\lambda_1}\chi\phi_{13}^{\dagger}\phi_{13}\chi^{\dagger}+\tilde{\lambda_2}\chi\tilde\phi_{13}^{\dagger}\tilde\phi_{13}\chi^{\dagger}\nonumber +\tilde{\lambda_3}\phi_2^{\dagger}\phi_{13}\phi_{13}^{\dagger}\phi_2\\&+\tilde{\lambda_4}\phi_2^{\dagger}\tilde\phi_{13}\tilde\phi_{13}^{\dagger}\phi_2+\tilde{\lambda_5}\left(\phi_2^{\dagger}\phi_2\right)\left(\chi^{\dagger}\chi\right)\nonumber +\tilde{\lambda_6}\left(\chi\chi^{\dagger}\right)~Tr\left(\Delta^{\dagger}\Delta\right)\nonumber \\&+\tilde{\lambda_7}\chi\left(\Delta^{\dagger}\Delta-\Delta\Delta^{\dagger}\right)\chi^{\dagger}+\tilde{\lambda_8}\left(\phi_2^{\dagger}\phi_2\right)Tr\left(\Delta^{\dagger}\Delta\right)+\tilde{\lambda_9} Tr\left(\phi_{13}^{\dagger}\phi_{13}\right)Tr\left(\Delta^{\dagger}\Delta\right)\nonumber \\&+\tilde{\lambda_{10}}~Tr\left(\phi_{13}\left(\Delta^{\dagger}\Delta-\Delta\Delta^{\dagger}\right)\phi_{13}^{\dagger}\right),
\end{align}

where 

\begin{align}
\phi_2=\begin{pmatrix}
   \phi_2^{+} \\
   \phi_2^{0}
  \end{pmatrix},
\tilde\phi_2=\begin{pmatrix}
   \bar\phi_2^{0} \\
   -\phi_2^{-}
  \end{pmatrix},
  \phi_{13}=\quad 
  \begin{pmatrix}
\phi_1^0 & \phi_3^0 \\
\phi_1^- & \phi_3^-
\end{pmatrix},\nonumber \\
\tilde\phi_{13}=\quad 
\begin{pmatrix}
\phi_3^{+} & -\phi_1^{+} \\
-\phi_3^{0} & \bar\phi_1^{0}
\end{pmatrix},~
\chi=(\chi_1^0\;\;\chi_2^0),~
\tilde\chi= (\bar\chi_2^{0}\;\; -\bar\chi_1^{0}).
\end{align}

In above potential $\mu_3^{2}$, $\mu_{23}$ terms explicitly break $L$ softly to $\left(-1\right)^L$.

\section{The Yukawa Couplings}
\label{sec:yukawa}

The allowed Yukawa couplings of the model ~\cite{Bhattacharya:2011tr} for quarks are:
\be
\left(d\phi_1^{0}-u\phi_1^{-}\right)d^{c}-\left(d\phi_3^{0}-u\phi_3^{-}\right)h_q^{c},\left(u\phi_2^{0}-d\phi_2^{+}\right)u^{c},\left(h_q^{c}\chi_2^{0}-d_c\chi_1^{0}\right)h_q.
\ee
The allowed Yukawa coupling for leptons are:
\be
\left(N\phi_3^{-}-\nu\phi_1^{-}-E\phi_3^{0}+e\phi_1^{0}\right)e^{c},
\left(E\phi_2^{+}-N\phi_2^{0}\right)n^{c}-\left(e\phi_2^{+}-\nu\phi_2^{0}\right)\nu^{c},
\ee

\be
\left(EE^{c}-NN^{c}\right)\chi_2^{0}-\left(eE^{c}-\nu N^{c}\right)\chi_1^{0},
\left(E^{c}\phi_1^{-}-N^{c}\phi_1^{0}\right)n^{c}-\left(E^{c}\phi_2^{-}-N^{c}\phi_2^{0}\right)\nu^{c},
\ee

\be
n^{c}n^{c}\Delta_1^{0}+\left(n^{c}\nu^{c}+\nu^{c}n^{c}\right)\Delta_2^{0}/\sqrt{2}-\nu^{c}\nu^{c}\Delta_3^{0}.
\ee

\end{document}